\documentclass{article}
\usepackage{epsfig,latexsym,graphics,eufrak, subfigure, cite}
\usepackage{feynarts}
\newcommand{\be}{\begin{equation}}
\newcommand{\ee}{\end{equation}}
\newcommand{\ba}{\begin{eqnarray}}
\newcommand{\ea}{\end{eqnarray}}
\newcommand{\bd}{\begin{displaymath}}
\newcommand{\ed}{\end{displaymath}}
\newcommand{\sla}[1]{/\!\!\!\!#1}

\begin{document}

\vspace*{-2cm}

\begin{flushright} 
\small{MPP-2008-53}
\end{flushright}

\vspace{1cm}

{

\centering

{ \bf \Large Squark--anti-squark pair production at the LHC: \\[0.3cm] 
the electroweak contribution}

\vspace{1.cm}

{\sc \large   W.~Hollik} {\large and}  { \large \sc E.~Mirabella}

\vspace{0.2cm}
{\it Max-Planck-Institut f\"ur Physik
     (Werner-Heisenberg-Institut) \\ 
F\"ohringer Ring 6, 
D-80805 M\"unchen, Germany }

}

\vspace{0.3cm}

\begin{abstract}
\noindent
We present the complete NLO electroweak contribution of
$\mathcal{O}(\alpha^2_s \alpha)$ to the production of diagonal
squark--anti-squark pairs in proton--proton collisions. 
Compared to the lowest-order $\mathcal{O}(\alpha_s\alpha +  \alpha^2)$
electroweak terms, the NLO contributions are also significant.
We discuss the LO and NLO electroweak effects in cross sections
and distributions at the LHC for the production of squarks
different from top squarks, in various supersymmetric benchmark
scenarios. LO and NLO can add up to $10\%$ in cross sections and
$25\%$ in distributions.
\end{abstract}

\vspace{0.3cm}

\section{Introduction}
The exploration of electroweak symmetry breaking will be one of the main
tasks at the Large Hadron Collider (LHC). Experiments are expected to either
verify the Higgs mechanism of the Standard Model (SM), or to detect signals
of physics beyond the SM. The concept of supersymmetry~\cite{Wess, Wess2} provides a 
promising alternative version of the Higgs mechanism where symmetry breaking
occurs without introducing new scalar couplings that potentially can 
become strong,
thus stabilizing the electroweak scale.
The realization of supersymmetry in terms of 
the Minimal Supersymmetric Standard Model (MSSM)~\cite{MSSM, MSSM2, MSSM3}, 
has come up as the most promising extension of the SM, 
a predictive framework that allows to make
precise predictions to be investigated 
by indirect and direct experimental studies.
The indirect access through virtual effects in 
electroweak precision data~\cite{PrecisionObservables}
provides an overall fit~\cite{Ellis:2007fu, Buchmueller:2007zk}
with a quality at least as good as in the SM, in specific observables 
like  $g-2$ of the muon~\cite{MuonEX, MuonEX2} even better, 
and yields bounds on the
light Higgs boson mass with less tension than in the SM~\cite{SMfit,SMfit2, SMfit3}. \\

If supersymmetry (SUSY) is realized at the TeV scale or below, 
it will be accessible to
direct experimental studies at the LHC through the production of SUSY
particles.
In particular, colored particles like squarks and gluinos
will be copiously produced, and the hadronic production of
squark-- anti-squark pairs is expected to play an important role 
for SUSY hunting. 
The cross section is in the range from
$0.5~\mbox{to}~10$ pb for masses of squarks and gluinos below 1 TeV  
and can be measured with a statistical uncertainty 
of the order of a few percent even in the low luminosity regime. 
Moreover, squark cascade decays    % $\tilde{q} \to 
into $q \tilde{\chi}^0_1$ 
lead to a signature with missing $E_T$  % $\sla{E}_T$ 
plus jets and possibly leptons 
that is well suited to detect MSSM signals~\cite{Plehn:2005cq,Randall}. 
The number of hard jets allows the distinction 
between gluino and squark decays. Finally with the help of 
decay chains one can reconstruct the mass of the squarks up to 2 TeV 
with a resolution better than $10~\%$~\cite{Tricomi, Tricomi2, Tricomi3}. \\

The first prediction of the 
cross section for hadronic production of squark pairs  
in the early 1980's was done at lowest order $\mathcal{O}(\alpha_s^2)$
in supersymmetric QCD~\cite{Tree, Tree2, Tree3, Tree4, Tree5}. 
QCD contributions at NLO, 
$\mathcal{O}(\alpha_s^3)$, for the processes
$
%\label{Eq:Process}
%\sum_{Q,a}
P P \to \tilde{Q}^a \tilde{Q}^{b *} X,\; 
P P \to \tilde{Q}^a \tilde{Q}^{b} X\; (Q\neq t)
$
were calculated more than ten years later~\cite{Beenakker1996,Beenakker19962}. 
They increase the cross section by
typically $20~\mbox{to}~30~\%$, and they 
substantially reduce the dependence on the factorization 
and renormalization scale.
NLO QCD corrections to the production of top-squark pairs, 
performed in~\cite{Beenakker1997}, are also positive 
and can increase the cross section by 40--50\%. \\

Besides the QCD-based production mechanisms,
there are also partonic processes of electroweak 
origin, like  diagonal and non-diagonal
squark pair production from $q\bar{q}$ annihilation~\cite{Drees,Fuks}. 
They proceed 
through $s$-channel photon and $Z$ exchange, and also through  
neutralino/chargino exchange in the $t$-channel (if 
$\tilde{Q}$ is different from $\tilde{t}$), yielding terms of
$\mathcal{O}(\alpha^2)$ and $\mathcal{O}(\alpha_s \alpha)$.
Due to interference between the tree-level QCD 
and electroweak amplitudes for $\tilde{Q}\neq \tilde{t}$,
the electroweak contributions can also become sizable, 
reaching values up to 20\%~\cite{Drees}. \\

For reliable predictions, electroweak contributions at NLO
have to be taken into account as well. 
In the case of top-squark pair production~\cite{StopEW2,StopEW,Beccaria:2007dt},
they were found to be significant, with effects up to 20\%.
In general, NLO electroweak (EW) contributions consist of loop contributions to
the tree-level amplitudes for $q\bar{q}$ annihilation and gluon fusion,
together with real photon and gluon bremsstrahlung processes, yielding
an involved struture of interference terms in  $q\bar{q}$ annihilation.
Moreover, photon--gluon induced parton processes also contribute owing to
the non-zero photon distribution in the proton.   
In this paper we present the NLO electroweak
contributions, of $\mathcal{O}(\alpha^2_s\alpha)$, to the production
of diagonal squark--anti-squark pairs different from top- and bottom-squarks,
\be
\label{Eq:Process}
 P~P \to \tilde{Q}^a~\tilde{Q}^{a*}\, X\quad (\tilde{Q}\neq \tilde{t},
 \tilde{b}) \, .
\ee
They show significant differences to top-squark production, based on the
following pecularities.
\begin{itemize}
\item 
In leading order $\mathcal{O}(\alpha^2_s)$,
the squark pair $\tilde{Q}^a~\tilde{Q}^{a*}$  can be produced via 
annhilation of a $Q~\overline{Q}$ pair 
through amplitudes that involve also the exchange of a gluino in the $t$-channel, 
thus enhancing the relative weight of the 
annihilation channel in~(\ref{Eq:Process}).
\item 
Electroweak tree 
diagrams with $t$-channel neutralino and chargino exchange are part of the 
amplitudes for $Q~\overline{Q} \to \tilde{Q}^a \tilde{Q}^{a*}$  
and  $Q'~\overline{Q'} \to \tilde{Q}^a \tilde{Q}^{a *}$, where $Q'$
is the isospin partner of $Q$ in a quark doublet, yielding EW--QCD interference
already at the tree-level.
\item  
At $\mathcal{O}(\alpha_s^2 \alpha)$ many types of interferences occur between 
amplitudes of $\mathcal{O}(\alpha_s\alpha)$ and  $\mathcal{O}(\alpha_s)$
as well as between $\mathcal{O}(\alpha_s^2)$ and $\mathcal{O}(\alpha)$   
amplitudes.
\end{itemize}
\noindent
These features make the calculation of the EW contributions 
of $\mathcal{O}(\alpha^2_s\alpha)$ 
to the processes (\ref{Eq:Process}) 
more involved  than in the case of $\tilde{t}\, \tilde{t}^*$ production
where no $t$-channel diagrams occur at lowest order.
Our analysis shows that the EW effects of NLO
can reach the same size as the tree-level EW contributions of
$\mathcal{O}(\alpha_s \alpha)$ and  $\mathcal{O}(\alpha^2)$, which we will
include in our discussion as well. \\

The case of $\tilde{b}\tilde{b}^{*}$ production will not be treated here. 
Owing to $b$-tagging, bottom-squarks can 
be experimentally distinguished from the squarks of the first two
generations~\cite{Tricomi3,MassR1,MassR3}. Moreover, in the case of $\tilde{b}
\tilde{b}^{*}$ production the partonic process $b \bar{b} \to \tilde{b}
\tilde{b}^{*}$  exhibits specific features, like 
mixing between left- and right-handed $b$-squarks,
mixing angle renormalization~\cite{RzehakHollik2}, non-negligible
Higgs-boson contributions and
enhanced Yukawa couplings for large vaues of $\tan \beta$ with the need of 
resummation~\cite{ResumYukawa}; 
other peculiarities for massive initial-state partons 
are the proper counting of the orders of the perturbative 
expansion~\cite{Olness:1987ep, Dicus:1998hs} and the 
appropriate choice of the factorization scale~\cite{Maltoni:2003pn}. 
A dedicated extra analysis for $b$-squark final states thus seems 
appropriate. \\

The outline of the paper is as follows. 
In Section~\ref{Sec:3level} we 
briefly summarize the various tree-level contributions 
to the processes~(\ref{Eq:Process}). 
Section~\ref{S:ObigAlfa} describes the structure of the
NLO terms of EW origin that contribute at $\mathcal{O}(\alpha^2_s\alpha)$
and the strategy of the calculation. Evaluation
of the EW effects and their analysis for the LHC  
are presented in Section~\ref{Sec:Numerics} 
and summarized in Section~\ref{Sec:Conclusions}.
A list of Feynman diagrams and counter terms with
specification of renormalization, and 
technical details for
the calculation of singular integrals  
are collected in the Appendix.

\section{ Tree-level contributions to squark pair production}
\label{Sec:3level}
In this section we list the lowest-order cross sections for the process
(\ref{Eq:Process}) arising from tree-level amplitudes at order
$\mathcal{O}(\alpha_s^2)$, $\mathcal{O}(\alpha_s \alpha)$ and  $\mathcal{O}(\alpha^2)$.
We will use the convention $d \sigma^{a,b}_X$ to  denote
the cross section for a partonic process $X$
at a given order $\mathcal{O}(\alpha_s^a \alpha^b)$ 
in the strong and electroweak coupling constants.
The parton luminosities for getting to the hadronic cross section 
are given by the convolution
\be
\label{Eq:Lumi}
\frac{dL_{i j}}{d\tau}(\tau)= \frac{1}{1+\delta_{ij}}~
\int_{\tau}^1 \frac{dx}{x} \left[ f_i(x)f_j\left(\frac{\tau}{x} \right)+ 
f_j(x)f_i\left(\frac{\tau}{x}\right)\right] ,
\ee
where
$f_{i}(x)$ is the momentum distribution of the parton 
$i$ in the proton (PDF).

\subsection{Squark pair production at leading order}
The leading-order contribution to the process (\ref{Eq:Process}) is 
QCD based, of $\mathcal{O}(\alpha_s^2)$. 
In the notation mentioned above, 
the differential cross section reads as follows,  
\be
\label{Eq:DiffXSec}
d \sigma^{\rm LO}_{PP\to \tilde{Q}^a\tilde{Q}^{a *}}(S)= 
\sum_{q}~\int_{\tau_0}^{1}~d\tau\;  \frac{dL_{q\overline{q}}}{d\tau}(\tau)\, 
d \sigma^{2,0}_{q\overline{q}\to \tilde{Q}^a\tilde{Q}^{a *}}({s}) + 
\int_{\tau_0}^{1}~d\tau\; \frac{dL_{gg}}{d\tau}(\tau)\, 
d \sigma^{2,0}_{gg\to \tilde{Q}^a\tilde{Q}^{a *}}({s}).
\ee
The sum runs over the quarks $q=u,~d,~c,~s$.~~S  and 
${s}=\tau S$ are the squared CM energies of the hadronic process~(\ref{Eq:Process}) 
and of the partonic subprocess, respectively. Moreover, 
with the squark mass $m_{\tilde{Q},a}$, the threshold value $\tau_0$ is determined by
$\tau_0= 4m^2_{\tilde{Q},a}/S$.

\medskip
$~d\sigma^{2,0}_{q\overline{q}\to\tilde{Q}^a\tilde{Q}^{a *}}$  and 
$~d\sigma^{2,0}_{gg\to\tilde{Q}^a \tilde{Q}^{a *}}$ denote 
the $\mathcal{O}(\alpha_s^2)$ differential cross sections
for the partonic processes
\ba
\label{Eq:PartProQ}
q(p_1)~\overline{q}(p_2) &\to& ~\tilde{Q}^a(k_1)~\tilde{Q}^{a *}(k_2), \\
\label{Eq:PartProG}
g(p_1)~g(p_2) &\to& ~\tilde{Q}^a(k_1)~\tilde{Q}^{a *}(k_2) ,
\ea
respectively,
which are obtained from the Feynman diagrams in
Fig.~\ref{Fig:TREE} of Appendix~\ref{SSec:FeynmanDiagrams}. 
Explicit expressions for these leading-order cross sections  can be found in
Refs.~\cite{Tree,Tree4,Tree5}.
Owing to  flavour conservation in SUSY QCD,
the diagram with the exchange of a gluino in the $t$ channel contributes only if $q=Q$.

\subsection{Tree-level electroweak contributions 
 of {\boldmath{$\mathcal{O}(\alpha_s\alpha)$}} and 
 {\boldmath{$\mathcal{O}(\alpha^2)$}} }
The $\mathcal{O}(\alpha_s \alpha)$ and $\mathcal{O}(\alpha^2)$ 
contributions to the process~(\ref{Eq:Process}), 
involving electroweak terms, can be written as follows,
\be
\label{Eq:MainFormulaTREE}
d \sigma^{\rm ew,LO}_{PP\to\tilde{Q}^a\tilde{Q}^{a *}X} = \sum_{q}\, \int_{\tau_0}^1 d\tau\,
                                        \bigg  \{
                                        \frac{dL_{q \overline{q}}}{d\tau}(\tau) \left(
                                         d\sigma^{1,1}_{q
                                        \overline{q} \to
                                        \tilde{Q}^a\tilde{Q}^{a *} }+
d \sigma^{0,2}_{q \overline{q} \to \tilde{Q}^a\tilde{Q}^{a *}} \right) +
                                        \frac{dL_{\gamma g}} {d\tau}(\tau)\,
                                       d \sigma^{1,1}_{\gamma g \to \tilde{Q}^a\tilde{Q}^{a *} } \bigg \}.
\ee 
The parton cross section
$d \sigma^{0,2}_{q\overline{q}\to\tilde{Q}^a\tilde{Q}^{a *}}$ is obtained squaring the tree-level
electroweak diagrams depicted in Fig.~\ref{Fig:TREEwkQ} of Appendix~\ref{SSec:FeynmanDiagrams}. 
The diagram with $t$-channel neutralino exchange contributes only if $q = Q$, and 
the diagram with chargino exchange appears only if  $q' =Q$, $q'$ 
being the SU(2) partner of the quark $q$, since we
treat the CKM matrix as unity.
$d \sigma^{1,1}_{q\overline{q}\to\tilde{Q}^a\tilde{Q}^{a *}}$ originates  from interference between the aforementioned 
tree-level electroweak diagrams and the tree-level QCD graphs of Fig.~\ref{Fig:TREE}.
Analytical expressions for
these cross sections can be found in Ref.~\cite{Drees}. \\

As a new element at $\mathcal{O}(\alpha_s\alpha)$, 
photon--gluon fusion occurs
as a further partonic process,
\be
\gamma(p_1)~g(p_2) \to ~\tilde{Q}^a(k_1)~\tilde{Q}^{a *}(k_2). 
\label{Eq:GaGLU}
\ee
The corresponding cross section, with  $t=(p_1-k_1)^2$,
\be
d \sigma^{1,1}_{\gamma g \to\tilde{Q}^a\tilde{Q}^{a *} } = \frac{dt}{16 \pi s^2}\; \overline{\sum} \;
|\mathcal{M}^{0}_{\gamma g \to \tilde{Q}^a\tilde{Q}^{a *}}|^2, 
\ee
contains the spin- and color-averaged squared tree-amplitudes from the diagrams in 
Fig.~\ref{Fig:GaGl},
\bd
\overline{\sum}\; | \mathcal{M}^{0}_{\gamma g \to \tilde{Q}^a\tilde{Q}^{a *}}|^2 = 
16\pi^2\, \alpha \alpha_s e_{\tilde{Q}} \;
\frac{m_{\tilde{Q},a}^4(m_{\tilde{Q},a}^4+ s^2 )+ t u(tu -2 m_{\tilde{Q},a}^4 )}
     {(t-m_{\tilde{Q},a}^2)^2\, (u-m_{\tilde{Q},a}^2)^2}
\ed
with $u=(p_1-k_2)^2$, and the electric charge $e_{\tilde{Q}}$ of the squark $\tilde{Q}$. 
The presence of photons in the proton follows from including NLO QED effects in the evolution equations 
for the PDFs. The photon PDF is part of the publicly available PDF set of~\cite{QEDpdf}; together with
the gluon PDF, the $\gamma g$ luminosity entering~(\ref{Eq:MainFormulaTREE}) 
is built according to Eq.~(\ref{Eq:Lumi}).

\section{Virtual and real {\boldmath{$\mathcal{O}(\alpha_s^2\alpha)$}} corrections}
\label{S:ObigAlfa}
In this section we describe the computation of the 
$\mathcal{O}(\alpha_s^2 \alpha)$ corrections to the process (\ref{Eq:Process}) 
arising from loops and from photon/gluon bresmsstrahlung.  The corresponding contributions to the
hadronic cross section are expressed in obvious notation,
\ba
\label{Eq:MainFormula}
d \sigma^{\rm ew, NLO}_{PP\to\tilde{Q}^a\tilde{Q}^{a*}X} 
                                       &=&  \int_{\tau_0}^1 d\tau\;  
                                        \frac{dL_{gg}}{d\tau}(\tau)\, \left(d  \sigma^{2,1}_{gg \to \tilde{Q}^a\tilde{Q}^{a *} }+ 
                                                                   d \sigma^{2,1}_{gg \to \tilde{Q}^a\tilde{Q}^{a *}\gamma }  \right) \nonumber \\
                                    &+&    \sum_{q}\, \Bigg \{ \int_{\tau_0}^1 d\tau\, 
                                        \frac{dL_{q \overline{q}}}{d\tau}(\tau) \left(d \sigma^{2,1}_{q \overline{q} \to \tilde{Q}^a\tilde{Q}^{a*} }+
                                       d \sigma^{2,1}_{q \overline{q} \to \tilde{Q}^a\tilde{Q}^{a*}\gamma } +
                                       d \sigma^{2,1}_{q \overline{q} \to
                                       \tilde{Q}^a\tilde{Q}^{a*}g } \right)
                                       \nonumber \\                                       
                                     &+& \int_{\tau_0}^1 d\tau\, \left[ 
                                 \frac{dL_{qg}}{d\tau}(\tau)  d
                                 \sigma^{2,1}_{qg \to
                                 \tilde{Q}^a\tilde{Q}^{a*}q }  +  
                                 \frac{dL_{\overline{q}g}}{d\tau}(\tau)  d
                                 \sigma^{2,1}_{\overline{q}g \to
                                 \tilde{Q}^a\tilde{Q}^{a*}\overline{q} } \right]
                                       \Bigg \} \, . 
\ea 
Other bremsstrahlung contributions to the hadronic cross section are of the type
\be
\label{Eq:ProOut}
\gamma(p_1)~          q (p_2) \to ~\tilde{Q}^a(k_1)~\tilde{Q}^{a *}(k_2)~          q (k_3),~~~~
\gamma(p_1)~\overline{q}(p_2) \to ~\tilde{Q}^a(k_1)~\tilde{Q}^{a *}(k_2)~\overline{q}(k_3).
\ee
We will not consider this class of processes here; they are further
suppressed by an additional  factor $\alpha_s$ with 
respect to process~(\ref{Eq:GaGLU}), and thus negligible.\\

Diagrams and  corresponding amplitudes are generated using 
\verb|FeynArts|~\cite{FeynArts,FeynArts2}. The algebraic treatment and numerical evaluation
of loop integrals is performed with support of \verb|FormCalc| and~\verb|LoopTools|~\cite{FormCalc,FormCalc2}.
IR singularities are regularized by a small photon mass, while  quark masses
are kept as regulators for the collinear singularities.  \\

\subsection{Gluon fusion with electroweak loops}
\label{S:loopGluonCorrection}
The first class of corrections entering Eq.~(\ref{Eq:MainFormula}) are the $\mathcal{O}(\alpha^2_s \alpha)$ 
electroweak virtual contributions to $gg$ fusion~(\ref{Eq:PartProG}), given by the partonic cross section
\be
\label{Eq:ggvirtual}
d \sigma^{2,1}_{gg \to \tilde{Q}^{a}\tilde{Q}^{a*}} = 
\frac{d t }{16 \pi s^2}\;  \overline{\sum}\; 2\, \mathfrak{Re}\, \{ \mathcal{M}^{0}_{gg \to \tilde{Q}^{a}\tilde{Q}^{a*}}\,
\mathcal{M}^{1,\mbox{\tiny ew}}_{gg \to \tilde{Q}^{a}\tilde{Q}^{a*}}  \},
\ee
$\mathcal{M}^{0}$ is the tree level $gg$ amplitude (Fig.~\ref{Fig:TREE}), and
$\mathcal{M}^{1,\mbox{\tiny ew}}$  is the one-loop amplitude with EW insertions in the QCD-based $gg$ tree diagrams.
These loop diagrams 
do not depend on the flavour of the final squark and thus they are identical to those listed in~\cite{StopEW}
for the particular case of $\tilde{Q}=\tilde{t}$.   
We therefore do not repeat them here. \\

In order to get rid of the UV divergences we have to include the proper counterterms for one-loop renormalization.
Their explicit expressions in terms of
the renormalization constants can be found in Appendix~\ref{S:AppCTT}. 
In the case of the gluon-fusion subprocess we have to  
renormalize the squark sector only.
We use the on-shell scheme~\cite{StopEW,RzehakHollik,RzehakHollik2}, where the independent parameters 
for a squark isospin doublet\footnote{Due to SU(2) invariance, mass renormalization 
of the different squarks from the same SU(2) doublet is correlated
and has to be performed simultaneously.} 
are chosen
to be the masses of the two up-squarks, the mass of one of the two down-squarks,
and the two mixing angles (which, however are irrelevant for the light-quark squarks where mixing can be neglected).
The actual expressions for the renormalization constants  are also given
in Appendix~\ref{S:AppCTT}.\\

Notice that part of the virtual corrections to squark pair production are loop diagrams 
for the gluon-gluon-$H^0$ vertex, with the heavy neutral MSSM Higgs boson $H^0$. 
These terms become resonant when $m_{H^0} \ge  2 m_{\tilde{Q},a}$ and
have to be considered a contribution to the process of $H^0$ production 
via gluon fusion with the subsequent decay $H^0 \to \tilde{Q}^a\tilde{Q}^{a*}$, rather than
an electroweak loop correction.
We will not consider scenarios in which such  resonances occur.

\subsection{Gluon fusion with real photon emission}
\label{S:RealGluonCorrection}
The IR singularities arising from virtual photons in~(\ref{Eq:ggvirtual})
are cancelled by including bremsstrahlung of real photons at $\mathcal{O}(\alpha_s^2\alpha)$, 
\be
\label{Eq:PartonGluon_Gamma}
g(p_1)~g(p_2) \to ~\tilde{Q}^{a}(k_1)~\tilde{Q}^{a *}(k_2)~\gamma(k_3) \, ,
\ee
according to the diagrams depicted in Fig.~\ref{Fig:2in3gamma}.
The integral over the photon phase space is IR divergent in the soft-photon region, 
{\it i.e.}~for $k^0_3 \to 0$, and cancels the corresponding virtual singularities
when added to the virtual contributions according to Eq.~(\ref{Eq:MainFormula}). 
%according to Bloch-Nordsieck~\cite{Singularities}.
\\

For the technical treatment of photon-momentum integration and isolation of divergences
we apply two different procedures: the methods of dipole subtraction 
and of phase space slicing. 
In the dipole subtraction approach, one has to add and subtract 
an auxiliary function to the differential cross section
that matches pointwise the singuilarities 
and is easy enough to be integrated analytically; the integral over the subtracted cross section is 
convergent and can be done numerically.
Due to the universality of the soft singularities general expression for these functions are available. 
In particular we use 
the expressions given in Ref.~\cite{Dipole}. Although the formulae
quoted in this reference apply to  processes
involving fermions only, 
they can be generalized to processes with charged bosons owing to the
universal structure of the IR singularities.\\

The phase space slicing technique restricts 
the phase space integration to the region with a minimum photon energy 
$\Delta E = \delta_{s} {\sqrt s} / 2 $.
The integration over this region is thus convergent and
can be performed numerically.
The complementary integral over 
the singular region with $k_3 < \Delta E$ 
can be done analytically in the eikonal approximation~\cite{Fadin}, which is a good approximation if 
the cut $\delta_{s}$ is sufficiently small.
More details are given in Appendix~\ref{S:AppPSS}. 
Comparison between the two methods provides a non trivial check of the computation. 
As illustrated in Fig.~\ref{Fig:METODI}, 
the two methods yield results which are in good numerical agreement.

\subsection{{\boldmath{$q\bar{q}$}}  annihilation with electroweak and QCD loops}
\label{S:loopQuarkCorrection}
The structure of the parton processes of $q\bar{q}$ annihilation 
at higher order is more involved and requires a simultaneous treatment
of electroweak and QCD loops.
The virtual contributions of one-loop order to the partonic cross section
is given by the interference of tree-level and loop amplitudes,
\be
d \sigma^{2,1}_{qq \to \tilde{Q}^{a}\tilde{Q}^{a*}} = 
\frac{dt}{16 \pi s^2} \; \overline{\sum}\;  \left \{
2\,  \mathfrak{Re}\{ \mathcal{M}^{0,\mbox{\tiny qcd}*}_{q \overline{q} \to \tilde{Q}^{a}\tilde{Q}^{a*}}
\mathcal{M}^{1,\mbox{\tiny ew}}_{q \overline{q} \to \tilde{Q}^{a}\tilde{Q}^{a*}}  \} +
2\,  \mathfrak{Re}\{ \mathcal{M}^{0,\mbox{\tiny ew}*}_{q \overline{q} \to \tilde{Q}^{a}\tilde{Q}^{a*}}
\mathcal{M}^{1,\mbox{\tiny qcd}}_{q \overline{q} \to \tilde{Q}^{a}\tilde{Q}^{a*}}  \}   \right \} ,
\ee
where $\mathcal{M}^{0,\mbox{\tiny qcd}}$ ($\mathcal{M}^{0,\mbox{\tiny ew}}$) is the amplitude related to the tree-level QCD (EW) diagrams 
depicted in Fig.~\ref{Fig:TREE}~(\ref{Fig:TREEwkQ}). $\mathcal{M}^{1,\mbox{\tiny ew}}$ is the one-loop amplitude arising from the  
EW corrections to the QCD tree-level diagrams and the QCD corrections to the EW  tree-level diagrams.  
Finally, $\mathcal{M}^{1,\mbox{\tiny qcd}}$ is the one-loop amplitude corresponding 
to the QCD corrections to the  QCD tree-level diagrams.\\

The diagrams entering $\mathcal{M}^{1,\mbox{\tiny ew}}$ are displayed in
%Figs.~\ref{Fig:Q22weak0},\ref{Fig:Q22weak1},\ref{Fig:Q22weak2} 
Figs.~\ref{Fig:Q22weak0}--\ref{Fig:Q22weak2} 
of Appendix~\ref{SSec:FeynmanDiagrams}. 
They also contain the diagrams with counterterm insertions required for renormalization and cancellation
of UV divergences.
The counterterms and the necessary renormalization constants can be found in Appendix~\ref{S:AppCTT}.
Besides squark renormalization, also quark renormalization is needed.\\

$\mathcal{M}^{1,\mbox{\tiny qcd}}$ can be obtained from the Feynman diagrams in Fig.~\ref{Fig:Q22strong} of Appendix~\ref{SSec:FeynmanDiagrams},
including the proper counterterms. 
Besides renormalization of the squark sector, we have to renormalize also the gluino mass, 
the strong coupling $g_s$, and the quark--squark--gluino  
coupling $\hat{g}_s$, which is related to $g_s$ via supersymmetry.
%Gluino mass is renormalized in the on shell scheme while the 
The strong coupling constant is renormalized in the $\overline{\mbox{MS}}$ scheme, modified  according to Ref.~\cite{Beenakker1996}
in order to decouple heavy particles (top, gluino, squarks) from the running of $\alpha_s$.
For the non-standard loop contributions,
this procedure is  equivalent to the ``zero-momentum subtraction''  used in Ref.~\cite{Berge:2007}. 
Since dimensional regularization violates supersymmetry in higher orders, a finite difference between 
$\hat{g}_s$ and $g_s$ is encountered at one-loop order. Supersymmetry is restored by shifting the renormalization constant 
for $\hat{g}_s$ by the corresponding finite amount, which means an unsymmetric renormalization of $\hat{g}_s$ and $g_s$.
More details and the specification of the counterterms can be found in Appendix~\ref{S:AppCTT}.

\subsection{{\boldmath{$q\bar{q}$}} annihilation with real photon emission}
\label{S:RealQuarkCorrection1}
The diagrams in Fig.~\ref{Fig:2in3gammaQ} of Appendix~\ref{SSec:FeynmanDiagrams}
constitute the generic amplitude for photon bremsstrahlung at 
$\mathcal{O}(\alpha_s^2\alpha)$ in the $q\bar{q}$ annihilation channel,
\be
\label{Eq:PartonQuark_Gamma}
q(p_1)~\overline{q}(p_2) \to ~\tilde{Q}^{a}(k_1)~\tilde{Q}^{a *}(k_2)~\gamma(k_3) \, .
\ee
The corresponding cross section is singular both in the IR soft-photon region and in the collinear region 
({\it e.g.}~whenever $k_3 p_i \to 0$). Although IR singularities cancel in sufficiently inclusive observables, 
collinear singularities from initial-state radiation 
remain and have to be absorbed via factorization in the PDFs.

The extraction of the singularities has been performed using the two different
methods described in section~\ref{S:RealGluonCorrection}. 
In phase space slicing, in this case, we have to introduce a further collinear cutoff  $\delta_c$ on the angle 
between the photon and the radiating quark/antiquark.
For sufficiently small $\delta_c$,
the integral over the singular region can be performed analytically.
Explicit expressions can be found in Appendix~\ref{S:AppPSS}. 
In Fig.~\ref{Fig:METODI} we visualize the comparison between the two methods
in the specific case of the partonic process $u \overline{u} \to \tilde{u}^{L} \tilde{u}^{L*}\gamma$
as an example.

\subsection{{\boldmath{$q\bar{q}$}} annihilation with real gluon emission}
\label{S:RealQuarkCorrection2}
Finally, we have to take into account the class of $q\bar{q}$ annihilation processes with real gluon bremsstrahlung,
\be
\label{Eq:PartonQuark_Gluon}
q(p_1)~\overline{q}(p_2) \to ~\tilde{Q}^{a}(k_1)~\tilde{Q}^{a *}(k_2)~g(k_3)\, ,
\ee
from either EW-based (Fig.~\ref{Fig:Q23strong}~a) or QCD-based Born diagrams (Fig.~\ref{Fig:Q23strong}~b).
This  class contributes to the cross section at $\mathcal{O}(\alpha_s^2 \alpha)$ 
through interference between the graphs of Fig.~\ref{Fig:Q23strong}~a and  Fig.~\ref{Fig:Q23strong}~b.
The cross section exhibits singularities  
when the gluon becomes soft or collinear to the initial-state quark/antiquark.  
The soft singularities cancel against those from the virtual photon/gluon contributions in $q\bar{q}$ annihilation,
when added along Eq.~(\ref{Eq:MainFormula}),
while remaining collinear singularities have to be absorbed in the PDFs by factorization. 
IR and collinear singularities can be treated by mass regularization.

For applying the phase space slicing method, 
the eikonal current has to be modified due to colour correlations after the emission of the soft gluon
(see Refs.~\cite{DipoleCatani,DipoleCatani2,DipoleCatani3} and Appendix~\ref{S:AppPSS} for details).
Colour correlation has to be taken into account also 
when using the Dipole Subtraction Method; we modified the formulae of
Ref.~\cite{Dipole} accordingly, 
following the guidelines of Ref.~\cite{DipoleCatani}. 
In Fig.~\ref{Fig:METODI} we illustrate the comparison between the two methods 
also for gluon radiation, with good numerical agreement.

\subsection{{\boldmath{$q(\bar{q})\, g$}} fusion}
\label{S:QuarkGluonFusion}
A last class of partonic processes
at the considered order is given by 
(anti-)quark-gluon fusion,
\ba
q(p_1)~g(p_2) \to ~\tilde{Q}^{a}(k_1)~\tilde{Q}^{a *}(k_2)~q(k_3)\, ,  \nonumber \\
\label{Eq:PartonQuark_Gluon2}
\overline{q}(p_1)~g(p_2) \to ~\tilde{Q}^{a}(k_1)~\tilde{Q}^{a *}(k_2)~\overline{q}(k_3)\, .
\ea
This IR finite class contributes at $\mathcal{O}(\alpha_s^2 \alpha)$ through the interference
between the diagrams of Fig.~\ref{Fig:QGfusion}~a and  Fig.~\ref{Fig:QGfusion}~b.
Mass singularities arise when the incoming gluon and outgoing
(anti-)quark are collinear. These collinear divergences are again
absorbed into the PDFs. Their extraction has been performed using the two methods described 
above in section~\ref{S:RealGluonCorrection}. Explicit expressions for the cross
section in the collinear region can be found in
Appendix~\ref{S:AppPSS}. The actual expression of the subtraction function
used in the Dipole Subtraction method is obtained
from  the formulae in Ref.~\cite{Diener}. Since those formulae are given there 
for the case of photon--quark splitting  we have to consistently redo
the color algebra. 
In Fig.~\ref{Fig:METODI} we show the agreement between the two methods  
for the example   $ug \to \tilde{u}^{L} \tilde{u}^{L*} u$.

In specific cases of SUSY parameters, when kinematically allowed,
the internal-state gauginos can be on-shell.
The poles are regularized by introducing the width 
of the corresponding gluino, neutralino, or chargino.
Potential problems related to gauge invariance~\cite{Kurihara} 
do not occur here.

\subsection{Factorization of initial-state collinear singularities}
The $\mathcal{O}(\alpha_s^2 \alpha)$ corrections to partonic cross sections
contain universal initial-state 
collinear singularities
that can be absorbed into the PDFs choosing a factorization scheme where
%Using the $\overline{\mbox{MS}}$ scheme for the collinear 
%singularities of relative order $\mathcal{O}(\alpha_s)$ and the DIS scheme
%for the collinear 
singularities of relative order $\mathcal{O}(\alpha)$,
the lowest order PDF $f_i(x)$ for parton $i\, (= q,\, \bar{q}$)
%used in Eq.~(\ref{Eq:MainFormula}) 
is related to the experimentally accessible distribution $f_i(x,\mu_F)$ 
via \cite{BaurKeller,Tobi}
\ba
\label{Eq:PDFfac}
f_i(x) &=&   f_i(x,\mu_F) \left \{ 1 + \frac{\alpha e_i^2 + \alpha_s C_F}{\pi} \left [-1 + \ln \delta_s + \ln^2 \delta_s 
+  \left ( \ln \delta_s +\frac{3}{4} \right)\ln \left( \frac{m_i^2}{\mu_F^2} \right) \right] 
             +\frac{1}{4} \frac{\alpha e^2_i}{\pi} h(\delta_s) \right \} \nonumber \\
       &+&   \int_x^{1-\delta_s} \frac{dz}{z} f_i \left( \frac{x}{z},\mu_F \right) 
                            \Bigg \{ P_{ii}(z)\frac{\alpha e_i^2 + C_F \alpha_s}{2 \pi}   
                            \Bigg [ \ln \left( \frac{m_i^2(1-z)^2}{\mu_F^2}  \right) + 1 \Bigg] -
             \frac{\alpha e^2_i}{2\pi} H (z) \Bigg \} \nonumber \\
       &+&  \int_x^1 \frac{dz}{z} f_g \left( \frac{x}{z},\mu_F \right)
       P_{ig}(z)\frac{\alpha_s T_F}{2 \pi} \ln \left(\frac{m^2_i}{\mu^2_F} \right)  ,   
\ea 
with the factorization scale $\mu_F$, $C_F = \frac{4}{3}$,$T_F=\frac{1}{2}$ and the electric charge $e_i$. 
The splitting functions $P_{ii}$, $P_{ig}$ are  defined in the usual way,
\be
\label{EQ:splittingfunction}
P_{ii}(z) = \frac{1+z^2}{1-z}~~~~~~P_{ig}(z)= z^2 + (1-z)^2  \, ,
\ee
and the functions $h$ and $H$ are given by
\be
h(\delta_s) = 9 + \frac{2 \pi^2}{3} +3 \ln \delta_s -2 \ln^2 \delta_s, ~~~~
H(z)        = P_{ii}(z)\ln \left( \frac{1-z}{z} \right)-\frac{3}{2}\frac{1}{1-z}+2z+3. 
\ee
%We do not consider  the contributions of the processes~(\ref{Eq:ProOut}) and so in the factorization procedure 
%we consistently neglect the contributions  arising from the splitting of a (anti)quark into a photon and a (anti)quark. 

The actual effect of the factorization of the initial collinear singularities is to
substitute $f_i(x)$ by $f_i(x,\mu_F)$ in the definition of the
quark--antiquark luminosity (\ref{Eq:Lumi})
and thus to obtain a further  $\mathcal{O}(\alpha_s^2\alpha)$
contribution to be added to Eq.~(\ref{Eq:MainFormula}).
This contribution reads:
\ba
\label{Eq:ExtraCont}
&&\sum_q  \; \int_{\tau_0}^1 d\tau\,\frac{dL_{q \overline{q}}}{d\tau}(\tau)\Bigg \{ 
 \left( \frac{2\alpha e^2_{q}}{\pi} \kappa^{\mbox{\tiny soft}}_{q} + \frac{ \alpha
     e^2_{q}}{2\pi}h(\delta_s)\right )  d\sigma^{2,0}_{q\overline{q}\to \tilde{Q}^a\tilde{Q}^{a*}}(s) 
     +  \frac{2\alpha_s C_F}{\pi}\kappa^{\mbox{\tiny soft}}_{q}  d\sigma^{1,1}_{q\overline{q}\to
       \tilde{Q}^a\tilde{Q}^{a*}}(s) \nonumber   \\
&& + \;  \int_{x_0}^{1- \delta_s}dz \left [ \left( \frac{\alpha e^2_{q}}{\pi} \kappa^{\mbox{\tiny coll}}_{q}(z) - \frac{ \alpha
     e^2_{q}}{\pi}H(z)\right )  d\sigma^{2,0}_{q\overline{q}\to
   \tilde{Q}^a\tilde{Q}^{a*}}(zs) +  \frac{\alpha_s C_F}{\pi} \kappa^{\mbox{\tiny coll}}_{q}(z) d\sigma^{1,1}_{q\overline{q}\to
       \tilde{Q}^a\tilde{Q}^{a*}}(zs) \right ] \Bigg\} \nonumber \\
&& + \; \sum_q \; \int_{\tau_0}^1 d\tau\, \left [ \left( \frac{dL_{qg}}{d\tau}(\tau) +
     \frac{dL_{\overline{q}g}}{d\tau}(\tau) \right)
\int_{x_0}^{1} dz \, 
P_{qg}(z)\frac{\alpha_s T_F}{2 \pi} \ln \left(\frac{m^2_q}{\mu^2_F} \right)
d\sigma^{1,1}_{q\overline{q} \to \tilde{Q}^a\tilde{Q}^{a*} }(zs) \right ] \, ,
\ea
where $x_0 = (4 m^2_{\tilde{Q},a})/ s$, while  $\kappa^{\mbox{\tiny soft}}_{q}$ and $\kappa^{\mbox{\tiny coll}}_{q}(z)$ are defined as
\ba
\kappa^{\mbox{\tiny soft}}_{q} &=&  \ln \delta_s + \ln^2 \delta_s 
 + \left ( \ln \delta_s +\frac{3}{4} \right)\ln \left(
  \frac{m_q^2} {\mu_F^2}     \right) -1,   \\
\kappa^{\mbox{\tiny coll}}_{q}(z) &=&  P_{qq}(z)\Bigg [ \ln \left( \frac{m_q^2(1-z)^2}{\mu_F^2}  \right) +1 \Bigg].   \nonumber     
\ea
The singularities in $\kappa^{\mbox{\tiny coll}}_{q}(z)$ cancel in the sum of the real
corrections and of the contribution~(\ref{Eq:ExtraCont}), as can be easily checked using the analytic 
expressions of $d \sigma^{2,1}_{q \overline{q}\to
  \tilde{Q}^a\tilde{Q}^{a*}\gamma}$  and $d \sigma^{2,1}_{q \overline{q}\to
  \tilde{Q}^a\tilde{Q}^{a*}g}$ in the collinear regions [see Eqs.~(\ref{Eq:CollinearGamma}) 
and~(\ref{Eq:CollinearGluon}) of Appendix~\ref{S:AppPSS}]. The remaining singularities of the real corrections
are exactly cancelled against those in $\kappa^{\mbox{\tiny soft}}_{q}$ and in
the virtual corrections.
The mass singularities in the last line of Eq.~(\ref{Eq:ExtraCont}) are
cancelled by those of 
$d\sigma^{2,1}_{qg \to\tilde{Q}^a\tilde{Q}^{a*}q}$ and 
$d\sigma^{2,1}_{\overline{q}g \to\tilde{Q}^a\tilde{Q}^{a*}\overline{q}}$, as
can be inferred from the their analytic expressions in the collinear region 
[Eq.~(\ref{Eq:CollinearQuark}) of Appendix \ref{S:AppPSS}]. \\

\begin{table}[t!]
\begin{center} 
\vspace*{0.4cm}
\begin{tabular}{c|c|c|c|c}
\hline
\hline
 parameter               & SPS1a$'$     & SPS5         & SU1          &  SU4           \\
\hline
$M_{1/2}$                 & $250$~GeV    &  $300$~GeV   &  $350$~GeV   &  $160$~GeV     \\
%\hline
$M_0$                     & $70$~GeV    & $150$~GeV    &  $70$~GeV  &  $200$~GeV   \\
%\hline
$A_0$                     &  $-300$~GeV &  $-1000$~GeV &   $0$        &  $-400$~GeV           \\
%\hline
$\mbox{sign}(\mu) $       &  $+$      & $+$        &  $+$       &  $+$         \\
%\hline
$\tan \beta(M_Z)$& $10.37$        &  $5$      &   $10$       &  $10$         \\
\hline
\hline
\end{tabular}
\vspace*{-0.4cm}
\end{center}
\caption{Input parameters in the four benchmark scenarios.}
\label{TAB:parameters}
\end{table}

For the calculation of hadronic observables we use the MRST2004qed 
parton distribution functions~\cite{QEDpdf}. 
Factorization and renormalization scales are chosen as equal, 
$\mu_R = \mu_F = m_{\tilde{Q},a}$. 
%According to the discussion 
%in Ref.~\cite{Diener}, this set of PDF is defined in NLO QED in the DIS scheme 
%while the factorization of $\mathcal{O}(\alpha_s)$ corrections  
%is performed in the $\overline{\mbox{MS}}$ scheme.  

\section{Numerical Analysis}
\label{Sec:Numerics}
For the numerical evaluation and for illustration of the EW effects, we choose 
four different benchmark scenarios: the point SPS1a$'$ suggested by the SPA convention~\cite{SPA}, the snowmass point 
SPS5~\cite{SNOWmass} characterized by light stops,
and two of the points chosen for detector simulation in the ATLAS ``Computing System Commissioning''
exercise~\cite{AtlasCCS}: the point SU1 in the coannihilation region, and the
point SU4 characterized by light SUSY particles. The input parameters
$M_{1/2}$, $M_0$, $A_0$, defined at the GUT scale, and $\tan\beta$
are put together in Table~\ref{TAB:parameters}.
The MSSM input for the actual calculation is obtained  with the help of  
the program \verb|SPheno|~\cite{SPheno}, together with
the program \verb|SuSpect|~\cite{SuSpect} as a cross check.
The pole masses of the squarks of the first generation obtained with the two different codes are shown
in Table~\ref{TAB:squarkmasses}. Since the quarks of the first two
generations are treated as massless, same-chirality and same-isospin
squarks are degenerate, therefore we do not show the masses of the
squarks belonging to the second generation.
The difference  between the masses provided by the two 
codes is below $1\%$. The different inputs given by the two codes
give rise to a differences in  the total cross section of the 
order of $2-3\%$. The standard model parameters are taken from
Ref.~\cite{PDG}. \\

\begin{table}[tbh!]
\begin{center}
\begin{tabular}{c|c|c|c|c}
\hline
\hline
                         & SPS1a$'$     & SPS5         & SU1          &  SU4           \\
\hline
$m_{\tilde{u},R}$          & $548.1$    &  $660.3$   
                         & $739.7$    &  $412.6$     \\
                         & ($545.6$)   &  ($657.4$)   
                         & ($736.3$)    &  ($411.2$)     \\
\hline
$m_{\tilde{u},L}$          & $565.3$    & $681.5$    
                         & $765.6$  &  $420.3$   \\
                         & ($562.0$)  & ($677.5$)    
                         & ($760.7$)    &  ($418.6$)   \\
\hline
$m_{\tilde{d},R}$          & $547.9$    &  $659.2$   
                         & $738.0$    &  $413.9$     \\
                         & ($545.4$)    &  ($656.9$)   
                         & ($734.6$)    &  ($412.5$)     \\
\hline
$m_{\tilde{d},L}$          & $570.7$    & $685.5$    
                         &  $769.6$  &  $427.5$   \\
                         & ($567.5$)    & ($681.8$)    
                         & ($764.7$)  &  ($425.8$)   \\
\hline
\hline
\end{tabular}
\end{center}
\caption{Pole  masses (in GeV) of the
 squarks of the first generation in the various SUSY scenarios. 
They are obtained using {\tt SPheno}~\cite{SPheno}; those computed with
{\tt SuSpect}~\cite{SuSpect} are quoted inside the brackets for comparison.}
\label{TAB:squarkmasses}
\end{table}

\noindent
We introduce the following conventions:
\begin{itemize}
\item We will refer to the sum of $\mathcal{O}(\alpha_s\alpha)$, $\mathcal{O}(\alpha^2)$ and $\mathcal{O}(\alpha_s^2\alpha)$ contributions 
as ``the EW contribution''.
\item We will use the quantity $\delta$ to denote the relative EW contribution, defined as
$ \delta = ( \mathcal{O}_{\mbox{\tiny NLO}} -\mathcal{O}_{\mbox{\tiny LO}} ) / \mathcal{O}_{\mbox{\tiny LO}}$,
where $\mathcal{O}$ is a generic observable and $\mathcal{O}_{\mbox{\tiny NLO}}$ is the sum of the LO in Eq.~(\ref{Eq:DiffXSec})
and the EW contribution. 
\end{itemize}

\subsection{Different squark species}
\label{Sec:squarkspecies}
Electroweak interactions depend on the hypercharge of the squarks, hence the production cross sections are flavour
and chirality dependent.
In this subsection we will study the production of 
four squark species, focusing on the SPS1a$'$ point. 
Since the masses of the light quarks can be neglected, 
the weak eigenstates of the squarks are also the mass eigenstates; 
thus, in the following, the two squarks of a given  flavour are distinguished 
by means of their chiralities, $\tilde{Q}^a=\tilde{Q}^L,\tilde{Q}^R$. \\

\noindent
{\bf Dependence on squark flavour and chirality} \\

\noindent
In Tab.~\ref{TAB:res1} we show the integrated hadronic cross section
for the diagonal pair production of $\tilde{u}^{L}$, $\tilde{u}^R$,
$\tilde{d}^{L}$ and $\tilde{c}^L$.   
In the case of the
production of the squarks of the first generation there is a
cancellation beetween $\mathcal{O}(\alpha_s \alpha)$  
and $\mathcal{O}(\alpha^2)$ contributions. The overall  $\mathcal{O}(\alpha_s \alpha+ \alpha^2)$
correction is negative and of the same order of magnitude as
the $\mathcal{O}(\alpha_s^2\alpha)$ one. Since they have the same sign
their effect is enhanced.
In the
case of $\tilde{c}^L$ production
the situation is different:
 $\mathcal{O}(\alpha_s \alpha)$,  
$\mathcal{O}(\alpha^2)$, and $\mathcal{O}(\alpha^2_s\alpha)$ corrections
are positive, $\mathcal{O}(\alpha^2_s\alpha)$ contribution being the most
important ones (see also the discussion below). 

As a general remark, the EW effects are always larger for left-handed squarks.
For a given chirality and generation, the EW contributions are more important in 
in the case of up-type squarks.  For comparison we
also estimate the corresponding NLO QCD corrections  using the code \verb|PROSPINO|~\cite{Beenakker19962};
they are positive, weakly dependent on the flavour of the produced squarks,  and of the order of $47-48\%$. \\

%\noindent 
%In Figures~\ref{Fig:IC_SQ}--\ref{Fig:YY_SQ} we display
%various differential cross sections for each case $\tilde{u}^{L}$, $\tilde{u}^R$
%$\tilde{d}^{L}$ and $\tilde{c}^L$.  \\

% CUMULATIVE INVARIANT MASS, TOTAL CROSS SECTION
Fig.~\ref{Fig:IC_SQ} shows the relative EW contribution (right part) in 
the ``cumulative invariant mass distribution'' $\sigma (M_{\mbox{\tiny inv}})$, that is the cross section integrated up to 
the value $M_{\mbox{\tiny{inv}}}$ of the squark--antisquark invariant mass. A common feature is that in the low invariant mass region the NLO EW contribution 
is positive, rather steeply decreasing as the invariant mass increases, reaching the plateau at $M_{\mbox{\tiny inv}} \ge 2000$~GeV
which corresponds to the total cross section. The left part of Fig.~\ref{Fig:IC_SQ}
shows the relative size of the individual contributions arising from the various channels.
The contribution from the gluon fusion channel is always positive and dominates at lower values of $M_{\mbox{\tiny inv}}$, wheras the 
$q \overline{q}$ annihilation channel part is negative. \\
Looking at the relative contributions of the different channels in the
high invariant mass region, which corresponds to the total cross section, one can
understand the origin of the different behaviour of the NLO EW
corrections in the case  of $u^Lu^{L*}$ and $c^L c^{L*}$ production.
For up-squark pairs, the $\mathcal{O}(\alpha_s
\alpha)$ and $\mathcal{O}(\alpha_s^2\alpha)$ terms are dominated by the 
the $q\bar{q}$ annihilation channels, which yield a negative contribution; 
for charm-squark production, however, the  
$\mathcal{O}(\alpha_s\alpha)$ [$\mathcal{O}(\alpha_s^2\alpha)$]
corrections are dominated by 
the $q\gamma$ fusion [$gg$ fusion] channel and thus positive
This shows 
the key role played by the partonic processes
$
Q~\overline{Q},\, Q'~\overline{Q'} \to \tilde{Q}^a \tilde{Q}^{a*}$, 
where $Q$ and $Q'$ belong to the same isospin doublet.
Indeed, in the case of $u^Lu^{L*}$ production their
contribution is negative and the largest out of the
$q \bar{q}$ annihilation channels.
In $c^Lc^{L*}$ production they are suppressed by the PDFs
of the charm and strange quarks and hence the 
contributions from the  $q \bar{q}$ annihilation channels
are negligible. As a result the overall contribution to 
total cross section is
negative at the level of 5\% for the left-handed up-squarks, while for
the left-handed charm-squarks it is of the same order of magnitude but
positive. \\
\renewcommand{\arraystretch}{0.5}
\begin{table}[t!]
\begin{center} 
\vspace*{0.4cm}
\begin{tabular}{c|c|c|c|c}
\hline
\hline
&&&& \\
                   & $\tilde{u}^R \tilde{u}^{R*}$
                   & $\tilde{u}^L \tilde{u}^{L*}$        
                   & $\tilde{d}^L \tilde{d}^{L*}$          
                   & $\tilde{c}^L \tilde{c}^{L*}$  \\
&&&& \\
\hline
&&&& \\  
$\mathcal{O}(\alpha_s^2)$ &   $(36.83 \pm 0.03)\cdot10^{-2}$ 
                          &   $(31.31 \pm 0.01)\cdot 10^{-2}$
                          &   $(25.89   \pm   0.01) \cdot 10^{-2}$
                          &  $(22.65 \pm 0.01) \cdot 10^{-2}$ \\
&&&& \\  
$\mathcal{O}(\alpha_s \alpha)$ & $(-9.00 \pm 0.01) \cdot 10^{-3}$
                               &  $(-3.54 \pm 0.01)\cdot10^{-2}$
                               &  $(-3.83   \pm   0.01) \cdot 10^{-2}$
                               &  $(~2.82 \pm 0.01) \cdot 10^{-3}$ \\
&&&& \\  
$\mathcal{O}(\alpha^2)$ & $(~2.42 \pm 0.01) \cdot 10^{-3}$ 
                        &  $(~2.39 \pm 0.01) \cdot 10^{-2}$
                        & $(~3.20   \pm   0.01) \cdot 10^{-2}$
                        &$(~2.11 \pm 0.01) \cdot 10^{-3}$ \\
&&&& \\
$\mathcal{O}(\alpha_s^2\alpha)$  &   $(-3.09 \pm 0.05) \cdot 10^{-3}$ 
                                 &  $(-1.05 \pm 0.01) \cdot 10^{-2}$
                                 &  $(-7.82   \pm   0.07) \cdot10^{-3}$
                                 &  $(~5.89 \pm 0.01) \cdot 10^{-3}$ \\
&&&& \\
$\delta(\%)$   &   $-2.6$
               &  $-7.0$  
               &  $ -5.5$ 
               &  $4.8$  \\
&&&& \\ 
\hline
\hline
\end{tabular}
\vspace*{-0.4cm}
\end{center}
\caption{Total cross section for the diagonal pair production of
different squark species in the SPS1a$'$ scenario. Beside the LO contribution, of
$\mathcal{O}(\alpha_s^2)$, we show the yields of the different orders contributing to the NLO EW corrections. Cross
sections are given in pb. $\delta$ is defined according to Sec.~\ref{Sec:Numerics}.}
\label{TAB:res1}
\end{table}
%This feature was observed also in \cite{Drees}
%for the tree-level QCD--EW interference contribution of
%$\mathcal{O}(\alpha_s\alpha)$. 
%
\indent
The contribution of the $g \gamma$ channel is independent on the squark chirality, determined
only by the electric charge of the produced squarks, which makes
the $g \gamma$ channel contribution
for up-squark pair production four times bigger than that for
down-squarks. Owing to the mass degeneracy between same-chirality and
same-isospin squarks the $gg$ fusion
channel is independent on the generation of the produced squark. \\

% IM DISTRIBUTION
The invariant mass distribution itself is displayed in  Fig.~\ref{Fig:IM_SQ} for the various squark species, showing also the breakdown 
into the individual channels.
For each squark species, the EW contributions are positive in the low invariant mass region  
and become negative for larger values of $M_{\mbox{\tiny{inv}}}$, reaching the level of 15\% for $\tilde{u}^L$ squarks.\\

% PT DISTRIBUTION
Fig.~\ref{Fig:PT_SQ} contains 
the transverse momentum distribution of the squarks.
Again, the EW effects are more pronounced for left-handed chirality
yielding more than 30\% negative contributions for large $p_T$. As new feature, the LO EW contribution can be positive for low $p_T$,
especially for the $\tilde{d}^L$ case, originating from the PDF-enhanced parton process $u\bar{u} \rightarrow \tilde{d}^L\tilde{d}^{L *}$
through $t$-channel chargino exchange. This positive part is practically compensated by the NLO $\mathcal{O}(\alpha_s^2\alpha)$
contributions in the $q\bar{q}$ annihilation channel. \\

% YY DISTRIBUTION 
%In Fig.~\ref{Fig:YY_SQ} we show the squark rapidity distribution and the various parts of the EW contributions.
%For right-handed squarks, the EW effects are rather small; for left-handed squarks their sum
%is  typically of the order 5\%. Again NLO EW corrections change sign
%when moving from the production of the squarks of the first generation
%to the processes involving squarks of the second generation.\\

%\newpage
\noindent
{\bf Dependence on the squark masses} \\

\noindent
To study the dependence of the NLO EW contributions on the mass of the squarks,
we vary $m_{\tilde{u},R}$, setting $m_{\tilde{d},R} = m_{\tilde{u},R}$
and $m_{\tilde{u},R} = m_{\tilde{u},L}(1+\varepsilon)$ with $\varepsilon = 0.03$, which is the
value at the  SPS1a$'$ point. The values are also taken for the other generations as well as 
for the sleptons. The other parameters are kept as in SPS1a$'$.
Each parameter point was checked to satisfy the bounds on SUSY particles from LEP~\cite{LEP, LEP2} and 
Tevatron~\cite{GluinoSearchLast}, and the bound on the mass of the
light Higgs boson $h^0$, which has been computed using \verb|FeynHiggs 2.5.1|~\cite{FeynHiggs,FeynHiggs2, FeynHiggs3}. 
Moreover, each point fullfills the condition
$ |\Delta \rho| < 0.025$,
where $ \Delta \rho $ is the dominant squark contribution to the electroweak $\rho$ parameter. \\

The  relative EW contributions are shown  
in Fig.~\ref{Fig:Scan2} for the total cross section, for each of the various squark types.
The quantity $\xi$ displayed in the right panel is the fraction of each 
the $gg$ fusion and the $q\bar{q}$ annihilation channel 
in the total cross section, at leading order $\mathcal{O}(\alpha_s^2)$.  
The $q \bar{q}$ channel becomes more and more important as $m_{\tilde{u},R}$ increases.
This feature, already pointed out in Ref.~\cite{Tree}, is a consequence of the 
$t$-channel gluino exchange diagrams.
The increasing importance of $q\bar{q}$ annihilation allows 
a better understanding of the particular role of the NLO corrections to
the $q\bar{q}$ channel with increasing squark masses.
Especially for left-handed up- and down-squarks, the NLO EW contributions become more important than
the LO ones, with effects of more than 20\%. In the charm-squark
production case $q \bar{q}$ channel is subleading with respect to the
$gg$ and $g\gamma$ fusion channels due to the aforementioned
suppression of charm and strange PDFs.  
The total sum of the EW contributions is shown in the right panel
of  Fig.~\ref{Fig:Scan2}. For illustration, we also give an
estimate of the formal statistical uncertainty 
$\delta_{\mbox{\tiny stat}} = (L\, \sigma^{\rm NLO})^{-1/2}$, 
assuming a  luminosity $L=100~\mbox{fb}^{-1}$.

\subsection{Different SUSY scenarios}
\label{SSec:Disc2}
Here we discuss the electroweak effects 
in the different SUSY scenarios mentioned above.
As a concrete example, we consider the production of $\tilde{u}^L$ squarks,
with the corresponding masses listed in
Table~\ref{TAB:squarkmasses}. \\

In Table~\ref{TAB:res2} we show the total cross section for the
aforementioned production process. The LO contribution
and the different orders entering the NLO EW corrections are shown
separately. As one can see the absolute value of the different contributions
decreases as the mass of $m_{\tilde{u},L}$ increases, while the
relative yield of the NLO EW corrections increases with the mass of
the produced squarks. In the case of the SU1 scenario  NLO EW
corrections are negative and of the order of 10\%. The corresponding
NLO QCD corrections 
are estimated  using the code \verb|PROSPINO|~\cite{Beenakker19962};
they are  of the order of $45-50\%$.\\

\renewcommand{\arraystretch}{0.5}
\begin{table}[t!]
\begin{center} 
\vspace*{0.4cm}
\begin{tabular}{c|c|c|c}
\hline
\hline
&&& \\
                   & SPS5
                   & SU1        
                   & SU4         \\
&&& \\
\hline
&&& \\  
$\mathcal{O}(\alpha_s^2)$ &   $(10.62 \pm 0.01)\cdot10^{-2}$ 
                          &   $(51.77 \pm 0.02)\cdot 10^{-3}$
                          &   $(16.14   \pm   0.01) \cdot 10^{-1}$\\
&&& \\  
$\mathcal{O}(\alpha_s \alpha)$ & $(-1.37 \pm 0.01)\cdot 10^{-2}$
                               &  $(-7.22 \pm 0.01)\cdot10^{-3}$
                               &  $(-1.45   \pm   0.01) \cdot 10^{-1}$\\
&&& \\  
$\mathcal{O}(\alpha^2)$ & $(~9.11 \pm 0.01) \cdot 10^{-3}$ 
                        &  $(~4.73 \pm 0.01) \cdot 10^{-3}$
                        & $(~10.16   \pm   0.01) \cdot 10^{-2}$\\

&&& \\
$\mathcal{O}(\alpha_s^2\alpha)$ & $(-4.83 \pm 0.03) \cdot 10^{-3}$ 
                        &  $(-2.75 \pm 0.02) \cdot 10^{-3}$
                        & $(-2.61   \pm   0.01) \cdot 10^{-2}$\\
&&& \\
$\delta(\%)$   &   $-8.9$
           &  $-10.1$  
           &  $ -4.3$ \\
&&& \\ 
\hline
\hline
\end{tabular}
\vspace*{-0.4cm}
\end{center}
\caption{Same as Tab.~\ref{TAB:res1} but focusing on $\tilde{u}^L \tilde{u}^{L*}$ production in
  different SUSY scenarios. }
\label{TAB:res2}
\end{table}

Fig.~\ref{Fig:IC} contains the cumulative invariant mass distribution, again with the individual  
and the total EW contributions, which show a similar behaviour for all the chosen scenarios.
Also the differential invariant mass distribution, displayed in Fig.~\ref{Fig:IM}, has similar
qualitative features in all scenarios. At low values, the gluon fusion part dominates and   
renders the total EW contribution positive. At larger values, the contributions from $q\overline{q}$ annihilation
turn the EW contribution to the negative region; thereby the NLO part is always of about the same size
as  the LO part. \\

In Fig.~\ref{Fig:PT} we show the transverse momentum distribution in the various cases. Again, 
their shape depends only weakly on the scenario. \\
%This also applies to the rapidity distribution, which is illustrated 
%in Fig.~\ref{Fig:YY}. The EW contributions
%are largest in the low rapidity region and are dominated by the negative contributions 
%arising from the $q \overline{q}$ channel at both tree level and NLO. \\

This general situation is only slightly changed when kinematical cuts are imposed, as we find 
from repeating our analysis for an exemplary set of    
cuts on the transverse momentum and on the rapidity of the two squarks,
\bd
p_T > 150~\mbox{GeV}, ~~~|y| < 2.5.
\ed
%As can be inferred from Fig~\ref{Fig:YY}, t
The cut on the rapidity is not effective because the NLO EW contributions to the rapidity distribution are 
very small
for $|y|> 2.5$. More important is the cut on the transverse momentum. It excludes the kinematical region where the largest part of the gluon channel 
contribution comes from.
Moreover, this cut suppresses also the contribution of the $g \gamma$ channel and enhances the influence of the 
$q \overline{q}$ channel by
excluding the region with a positive $p_T$ distribution.
As a result, the negative EW contribution to the total cross section is larger than without cuts, as one can see 
from Fig.~\ref{Fig:ICcut}. \\

%\newpage
\noindent
{\bf Dependence  on the gluino mass} \\

\noindent
Finally we study the dependence of the EW contribution as a function of the mass of the gluino $m_{\tilde{g}}$, 
with the other parameters kept fixed according to the SPS1a$'$ point. 
Again, the parameter range is in accordance with 
the phenomenological constraints described in the previous subsection~\ref{Sec:squarkspecies}. 
At LO, 
the gluon fusion channel does not depend on the gluino mass, while the $q \overline{q}$ annihilation channel 
contribution decreases with increasing $m_{\tilde{g}}$, as displayed in Fig.~\ref{Fig:Scan3}.
In the low $m_{\tilde{g}}$ region the two channel contribute equally to the production cross section, 
while gluon fusion becomes dominant as the mass of the gluino increases.
The relative EW contributions from the various channels are flat, adding up
to a total EW contribution from  $-7~\mbox{to}~-3\%$ for gluino masses between 500 and 2000 GeV.
Thereby, in $q\bar{q}$ annihilation, both 
the tree-level term $\mathcal{O}(\alpha_s\alpha+\alpha^2)$ and the NLO corrections $\mathcal{O}(\alpha_s^2\alpha)$,
are practically of the same size.

\section{Conclusions}
\label{Sec:Conclusions}
We have computed the $\mathcal{O}(\alpha^2_s \alpha)$ NLO electroweak contributions 
to the production of flavour-diagonal squark--anti-squark pairs in proton--proton collisions,
in combination with the electroweak LO tree-level contributions of $\mathcal{O}(\alpha_s \alpha +\alpha^2)$. \\

We have performed an explicit study  of the electroweak contributions for 
each case of the four squark species in the first SU(2) doublet,
with a numerical analysis for the LHC.
The electroweak effects can give rise to sizeable modifications in cross sections and distributions, 
in particular for left-handed squarks.
Thereby, the NLO terms are significant and 
have to be considered together with the tree-level contributions.
They  show a strong dependence on the squark masses, increasing 
their relative influence with the mass of the squarks. \\

Moreover, we have investigated several SUSY benchmark scenarios 
and found that the behaviour of the electroweak contributions is only weakly dependent
on the scenario. Also the gluino-mass dependence is weak.
In summary, the electroweak contributions in squark-pair production can reach 20--25\% 
in size and are thus significant; about  half is carried by the NLO
contributions. As a final remark we would like to mention that the
NNLO QCD contributions, of $\mathcal{O}(\alpha_s^4)$, can be expected
to be of similar size.

\subsection*{Acknowledgments} 
We thank  Stefan~Dittmaier and  Stefano~Pozzorini for useful discussions,
and Maike~Trenkel for cross checking part of the results.

\newpage

\section*{Appendix}
\appendix
%
%
%
%
%%%%% APPENDICE COI DIAGRAMMI DI FEYNMAN %%%%%%%%%%%%%%%%%%% 
%
\addcontentsline{toc}{section}{Appendix}
\section{Feynman diagrams}
\label{SSec:FeynmanDiagrams}
In this Appendix generic diagrams for the various contributions to the
different channels are shown. 
We choose the up-squark case as a specific example.
In the following we will use the label $S^0$ ($S$) to denote all the neutral
(charged) 
Higgs bosons, while $V^0 = \gamma,Z$.
%
%%%%% PRIMA FIGURA
% 
\begin{figure}[htbp]
% tree level QCD
\begin{center}
\vskip -0.1cm
\input{DIAG/TREE1.tex}
\caption{Tree-level QCD diagrams for $q\overline{q}\to \tilde{u}^a\tilde{u}^{a*}$ and
for  $gg \to \tilde{u}^a\tilde{u}^{a*}$.}
\label{Fig:TREE}
% tree level EW
\vskip 0.2cm
\input{DIAG/TREE2.tex}
\caption{Tree-level EW diagrams for $q\overline{q}\to \tilde{u}^a\tilde{u}^{a*}$.}
\label{Fig:TREEwkQ}
% g gamma
\vskip 0.2cm
\input{DIAG/TREE3.tex}
\caption{Lowest-order  diagrams for photon--gluon fusion $\gamma g \to \tilde{u}^a\tilde{u}^{a*}$.}
\end{center}
\label{Fig:GaGl}
% gamma from gluon 
\vskip 0.2cm
\input{DIAG/G23gamma.tex}
\caption{Tree-level diagrams for real photon emission in $g g \to \tilde{u}^a\tilde{u}^{a*}\gamma$. }
\label{Fig:2in3gamma} 
\end{figure}
%
%%%%% SECONDA  FIGURA virtual q qbar EW
% 
\begin{figure}[htbp]
\input{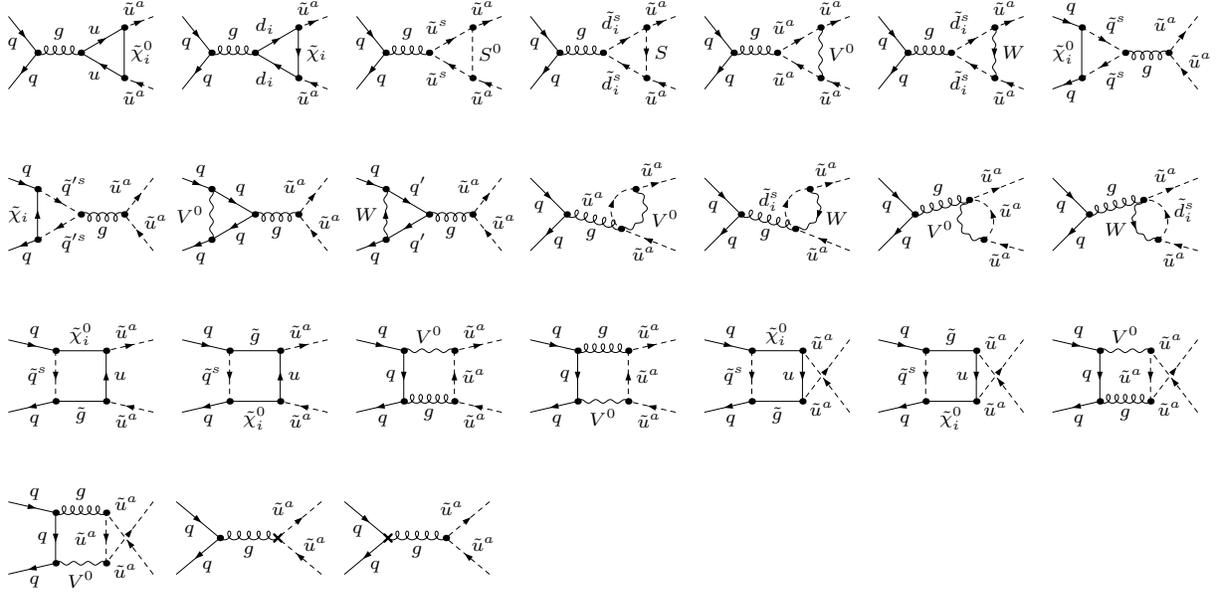}
\caption{One-loop EW diagrams for $q \overline{q}\to \tilde{u}^a
  \tilde{u}^{a*}$. 
The diagrams with  counter terms can be computed
according to the Feynman rules in appendix~\ref{S:AppCTT}. 
The renormalization constants in the counter terms have to be evaluated at 
$\mathcal{O}(\alpha)$. }
\label{Fig:Q22weak0}
\end{figure}
%
%%%%% TERZA FIGURA virtual d dbar EW
% 
\begin{figure}[htbp]
\input{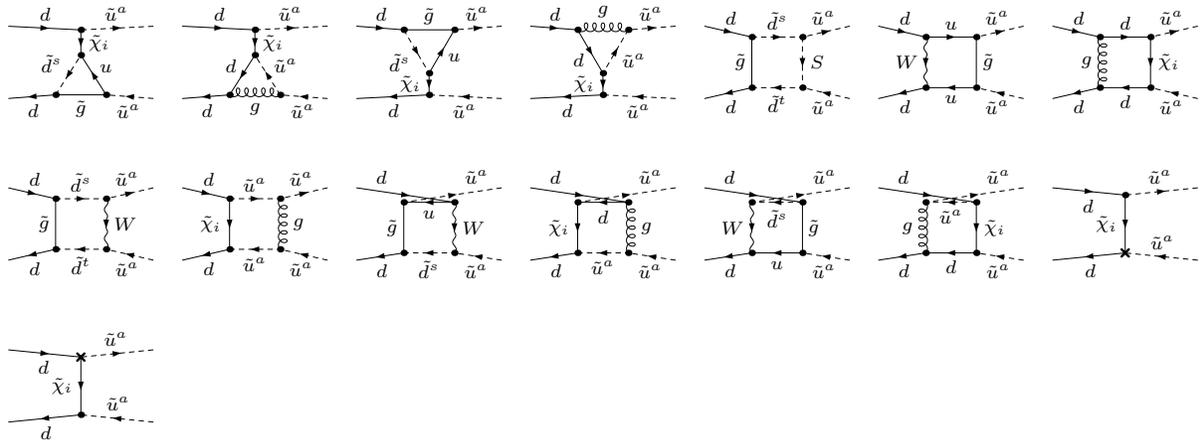}
\caption{One-loop EW diagrams that enter only in the case of the process  
$d \overline{d} \to \tilde{u}^a\tilde{u}^{a*}$. The diagrams containing the counter terms  can be computed 
according to the Feynman rules in Appendix~\ref{S:AppCTT}. 
The renormalization constants in the counter terms have to be evaluated at $\mathcal{O}(\alpha_s)$.}
\label{Fig:Q22weak1}
\end{figure}
%
%%%%% QUARTA FIGURA virtual u ubar EW
% 
\begin{figure}[htbp]
\input{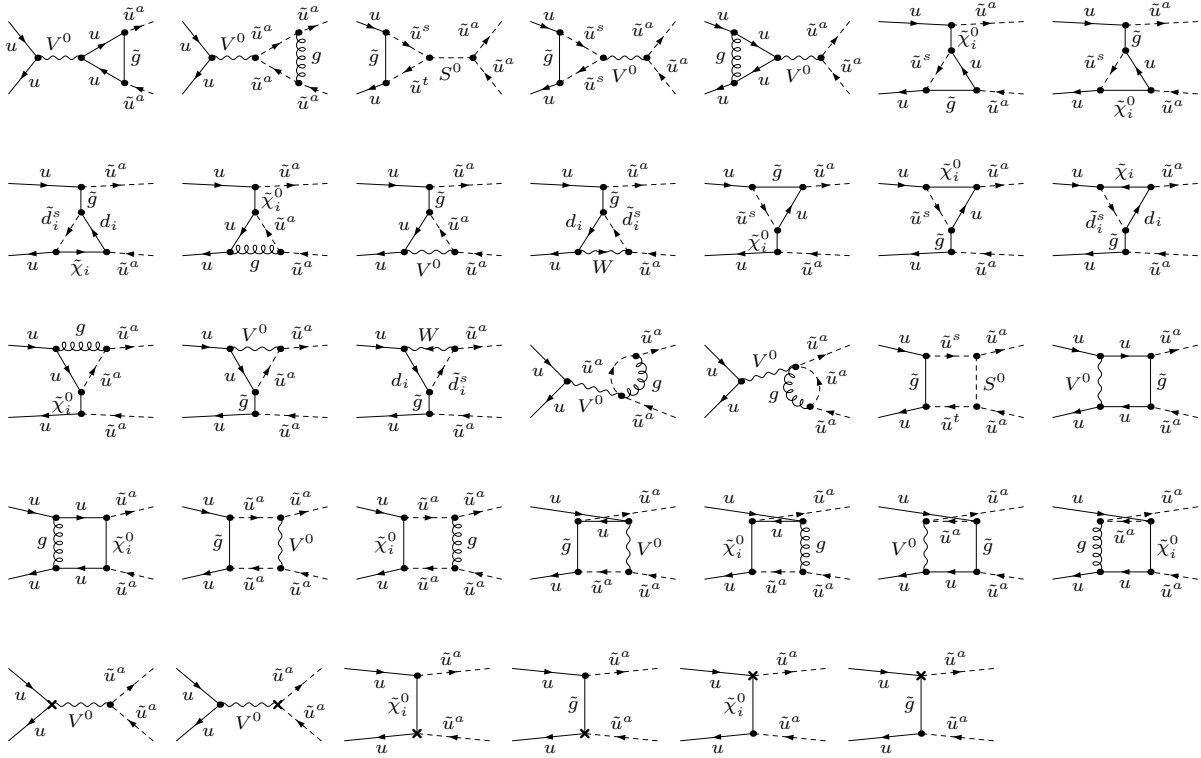}
\caption{One-loop EW diagrams that enter only in the case of the process  
$u  \overline{u} \to \tilde{u}^a\tilde{u}^{a*}$. 
The diagrams in the last row contain
the counter terms listed in Appendix~\ref{S:AppCTT}. 
The renormalization constants in the quark--squark--gluino counter term have to be 
evaluated at $\mathcal{O}(\alpha)$, the other ones at $\mathcal{O}(\alpha_s)$.}
\label{Fig:Q22weak2}
\end{figure}
%
%%%%% QUINTA FIGURA virtual q qbar QCD
%
\begin{figure}[htbp]
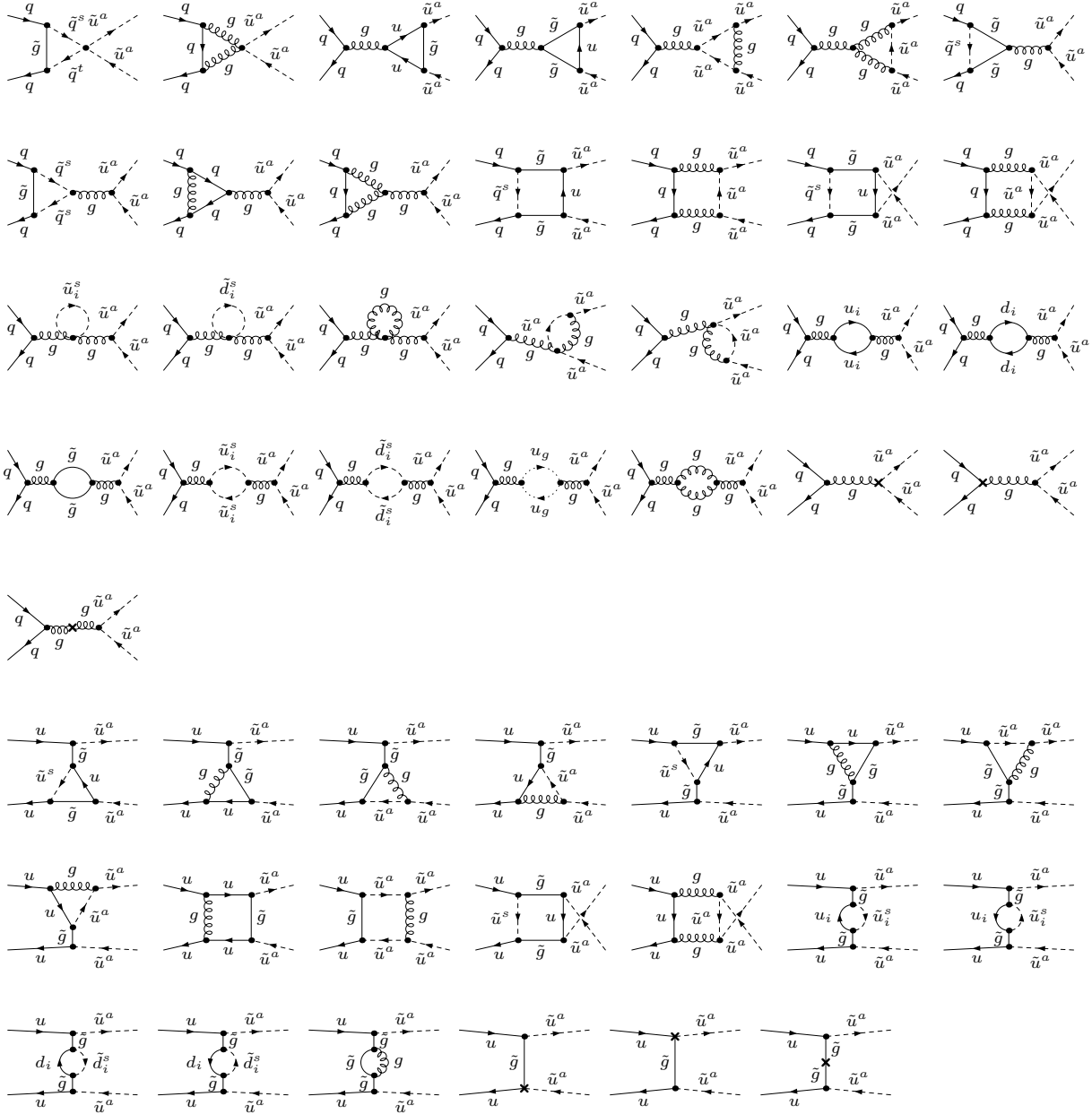

\input{DIAG/Q22strong1.tex}
\input{DIAG/Q22strong2.tex}
\caption{One loop QCD diagrams  for the process 
$q \overline{q} \to  \tilde{u}^a\tilde{u}^{a*}$. 
These diagrams interfere with those of Fig.~\ref{Fig:TREEwkQ} yielding 
$\mathcal{O}(\alpha_s^2\alpha)$ contributions. 
The diagrams containing counter terms can be computed according 
to the Feynman rules listed in Appendix~\ref{S:AppCTT}. 
The renormalization constants appearing in the counter terms  
have to be evaluated at $\mathcal{O}(\alpha_s)$.}
\label{Fig:Q22strong}
\end{figure}
%
%%%%% SESTA FIGURA q qbar + gamma emission
%
\begin{figure}[htbp]
\input{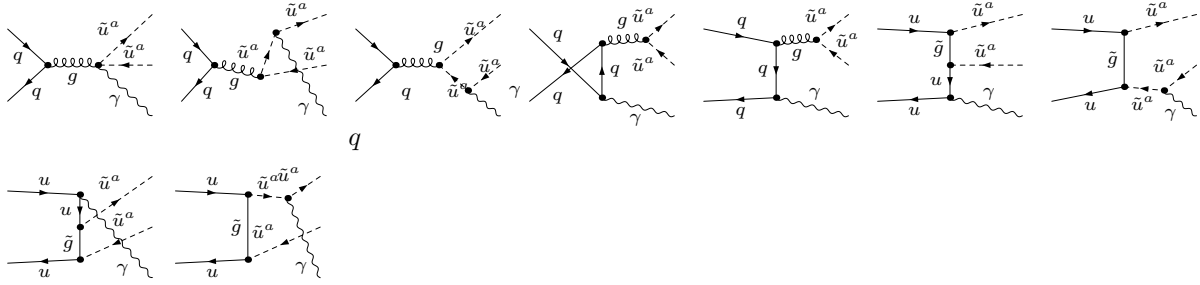}
\caption{$\mathcal{O}(\alpha_s^2 \alpha)$ real photon emission
in $q  \overline{q} \to \tilde{u}^a\tilde{u}^{a*}\gamma$.
The last four diagrams contribute only if $q=u$. }
\label{Fig:2in3gammaQ}
\end{figure}
%
%%%%% SETTIMA FIGURA q qbar + gluon emission
%
\begin{figure}[htbp]
\input{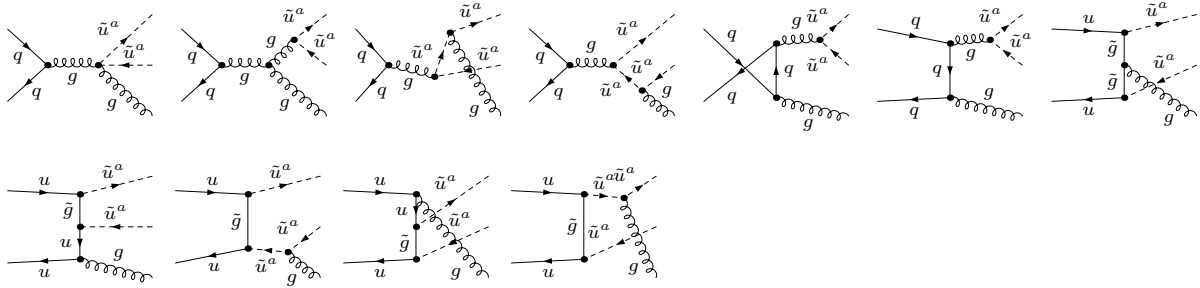}
\begin{center}
(a)
\end{center}
\input{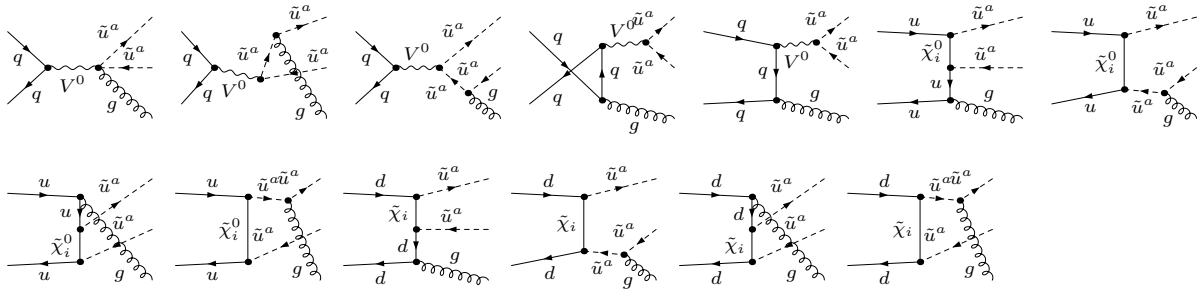}
\begin{center}
(b)
\end{center}
\vskip 0.5cm
\caption{Diagrams for gluon bremsstrahlung from QCD (a) and EW (b) Born  diagrams. 
They  contribute at $\mathcal{O}(\alpha^2_s \alpha)$ through QCD--EW interference }
\label{Fig:Q23strong}
\end{figure}
%
%%%%% OTTAVA FIGURA q gluon fusion

\begin{figure}[htbp]
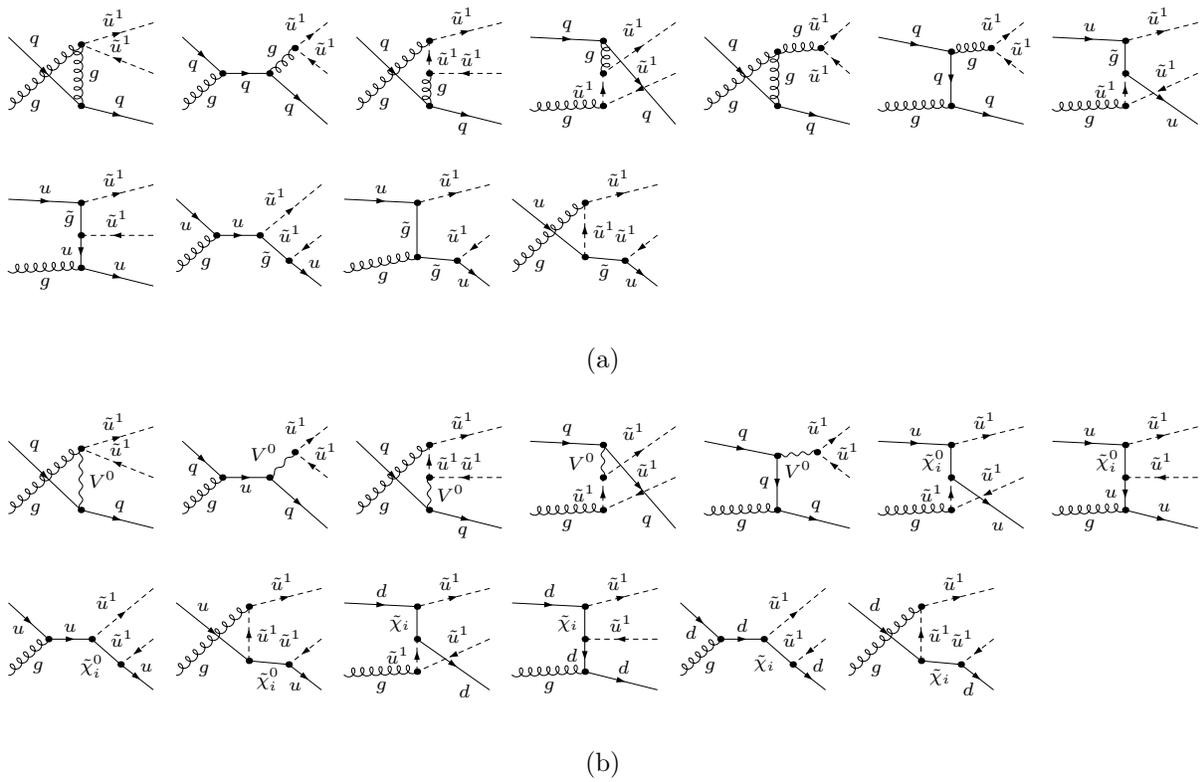

\input{DIAG/QGGIstro.tex}
\begin{center}
(a)
\end{center}
\input{DIAG/QGGIweak.tex}
\begin{center}
(b)
\end{center}
\vskip 0.5cm
\caption{QCD (a) and EW (b) Born  diagrams for quark gluon fusion channels. 
Their interference contributes at $\mathcal{O}(\alpha^2_s \alpha)$.}
\label{Fig:QGfusion}
\end{figure}

\clearpage

%
%
%
%
%%%%% APPENDICE COI CONTROTERMINI  %%%%%%%%%%%%%%%%%%% 
%
\addcontentsline{toc}{section}{Appendix}
\section{Counter terms and renormalization constants}
\label{S:AppCTT}
Here we list the counter terms for renormalization of vertices and propagators 
in the one-loop amplitudes for squark-pair production. For squarks of the 
first two generations, we can neglect $L$-$R$ mixing, and weak eigenstates are also 
mass eigenstates that can be distinguished by their chiralities $a=L,R$. 
The Feynman rules for the counter terms can be expressed in terms of the field
renormalization constants of quarks, squarks, gluons, and  gluinos, 
defined from the relation between bare and renormalized fields,
\ba
\label{EQ:fieldren}  
\Psi^{\mbox{\tiny bare}}_{qa}        = \Psi^{\mbox{\tiny ren}}_{qa}        \left( 1 + \frac{1}{2} \delta Z_{qa}        \right),  &&
\Phi^{\mbox{\tiny bare}}_{\tilde{Q},a} = \Phi^{\mbox{\tiny ren}}_{\tilde{Q},a} \left( 1 + \frac{1}{2} \delta Z_{\tilde{Q},a} \right),\nonumber \\ 
G_\mu^{\mbox{\tiny bare}}                  =  G_\mu^{\mbox{\tiny ren}}                           \left( 1 + \frac{1}{2} \delta Z_{G}         \right),  &&
\Psi^{\mbox{\tiny bare}}_{\tilde{g} } = \Psi^{\mbox{\tiny ren}}_{\tilde{g} } \left( 1 + \frac{1}{2} \delta Z_{\tilde{g} } \right) ,     
\ea
together with the renormalization constants for the strong coupling $g_s$, for the strong Yukawa coupling $\hat{g}_s$, and for the squark masses,
which are defined according to
\be
\label{EQ:renormparameter}
g^{\mbox{\tiny bare}}_s = g^{\mbox{\tiny ren}}_s (1+ \delta Z_{g} ),~~~\hat{g}^{\mbox{\tiny bare}}_s = \hat{g}^{\mbox{\tiny ren}}_s (1+ \delta Z_{\hat{g}} ),~~~
m^{2~\mbox{\tiny bare}}_{\tilde{Q},a} = m^{2~\mbox{\tiny ren}}_{\tilde{Q},a}+ \delta m^{2}_{\tilde{Q},a} \, .
\ee 
The actual expressions of the counterterms that are relevant for our squark-pair production processes are given below.
\begin{itemize}
\item Vertex counter terms involving gauge bosons: \vspace{-0.5 cm}  \\
\vspace{-1 cm}
$\includegraphics[width=3.0cm]{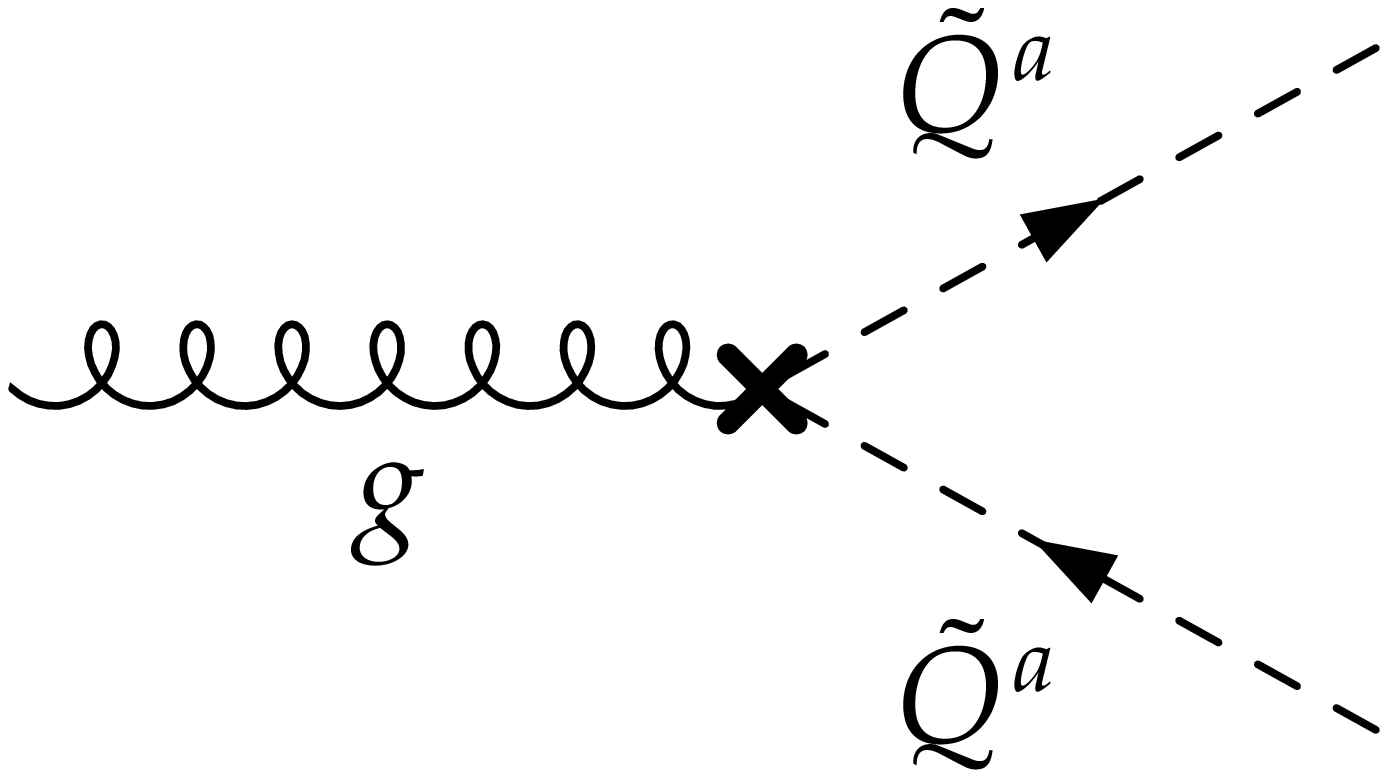}$
     \raisebox{1.65cm}{\hspace{0.8cm} =  \hspace{.8cm}  $-i g_s (\delta Z_{\tilde{Q,a}} + \frac{\delta Z_{G}}{2} + \delta Z_{g}) T^{C}(k+k')_\mu $}  \\ 
\vspace{-1.3 cm}
$\includegraphics[width=3.0cm]{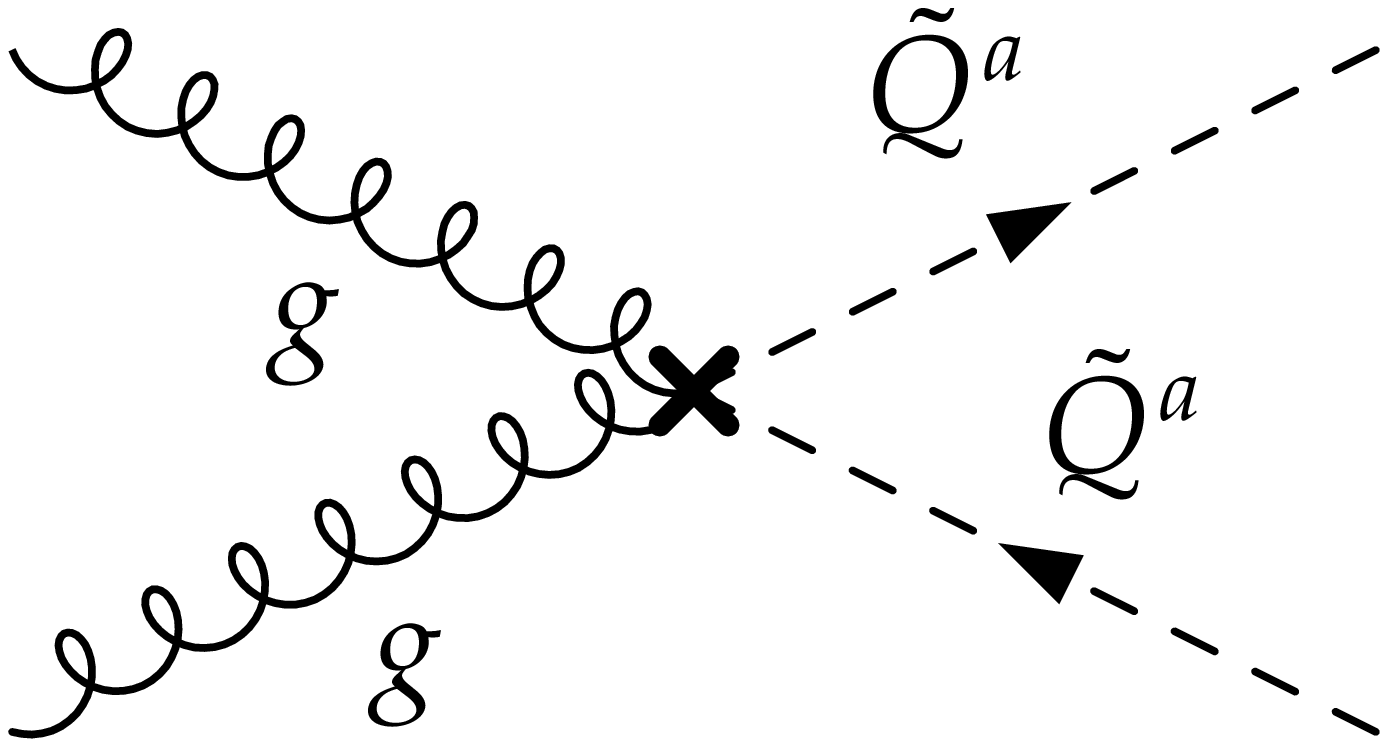}$ 
     \raisebox{1.65cm}{\hspace{.8cm} =  \hspace{.8cm}  $i  g_s^2 \delta Z_{\tilde{Q,a}} ( \frac{1}{3}\delta^{C_1 C_2} +f^{C_1C_2A}T^A ) g_{\mu \nu}  $  }   \\
\vspace{-1.3 cm}
$\includegraphics[width=3.0cm]{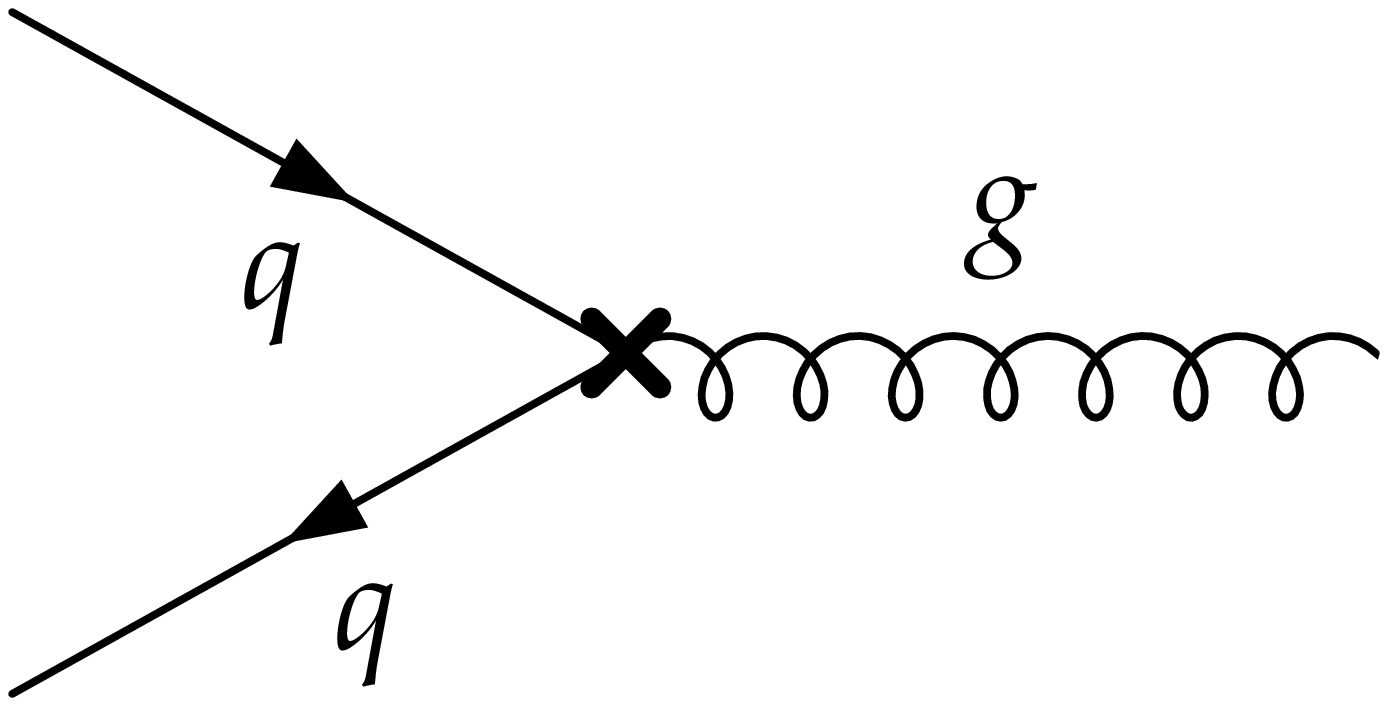}$   
     \raisebox{1.65cm}{\hspace{.8cm} =  \hspace{.8 cm} $-i g_s [  (\frac{\delta Z_{G}}{2} + \delta Z_{g}  + \delta Z_{qL})\gamma_\mu \omega_-  +   
                                                                (\frac{\delta Z_{G}}{2} + \delta Z_{g}  + \delta Z_{qR})\gamma_\mu \omega_+  ]T^C $} \\
\vspace{-1.3 cm}
$\includegraphics[width=3.0cm]{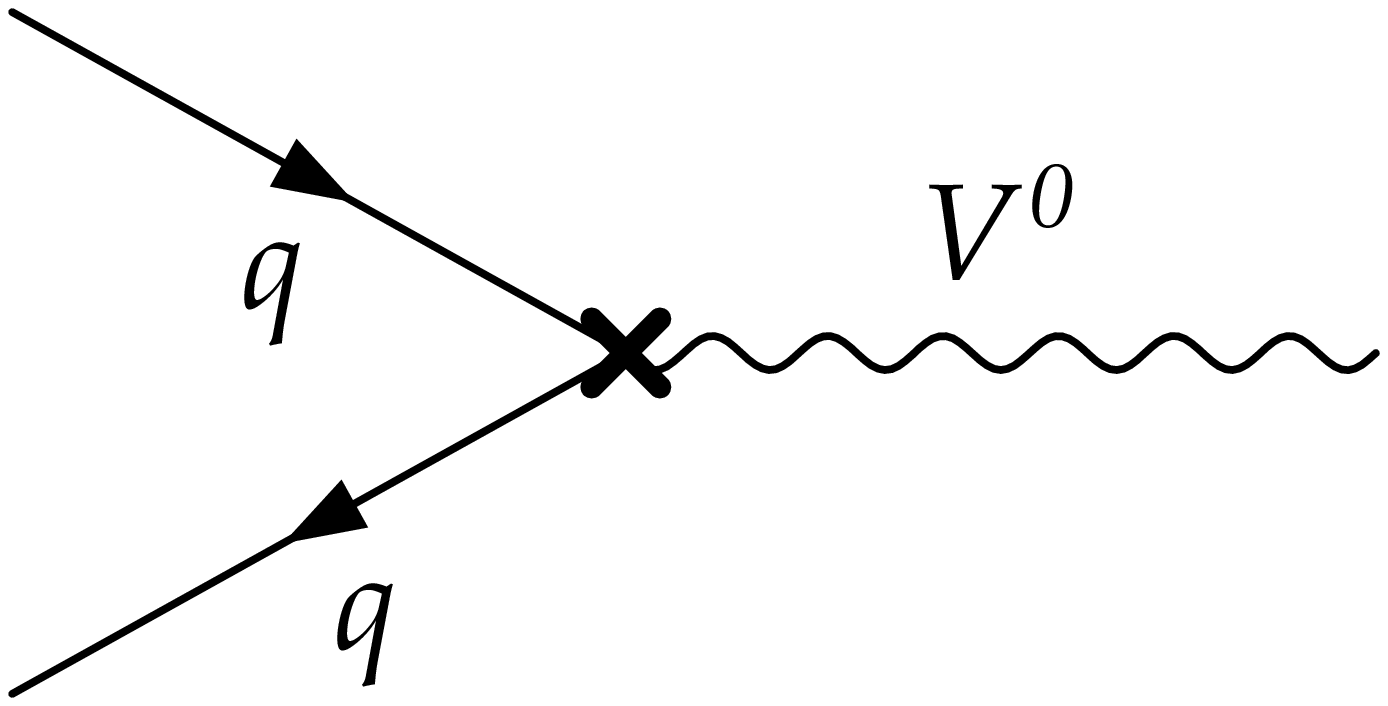}$ 
     \raisebox{1.65cm}{\hspace{.8cm} = \hspace{.8cm}   $-i e [ C_-^V(q) \delta Z_{qL} \gamma_\mu \omega_- + C_+^V(q) \delta Z_{qR} \gamma_\mu \omega_+] $ 
                       \hspace{1.2cm} $V^0=\gamma, Z$}    \\
\vspace{-0.5 cm}
$\includegraphics[width=3.0cm]{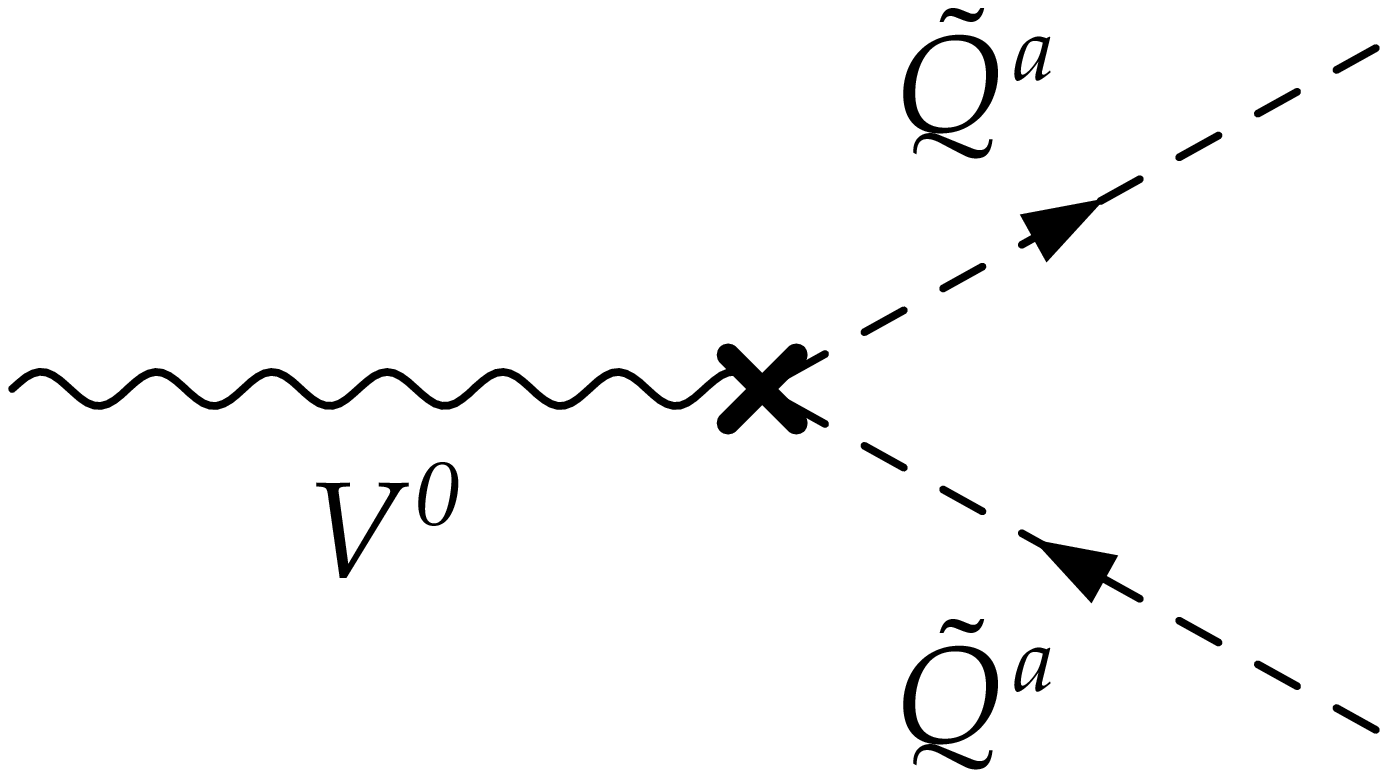}$ 
     \raisebox{1.65cm}{\hspace{.8cm} = \hspace{.8cm}  $-i e [ C_-^V(\tilde{Q})  \delta_{aL}  + C_+^V(\tilde{Q}) \delta_{aR} ] \delta Z_{\tilde{Q,a}}(k+k')_\mu $ 
                       \hspace{1.1cm} $V^0=\gamma, Z$}    \\
$k$ and $k'$ are the momenta of the squark and the antisquark, and they are fixed according to the arrow. $T^C$ are the color matrices and
$f^{ABC}$ the structure constants of the color group. We omit the color indices of fermions and sfermions. 
Moreover, we define
\be
C_{\pm}^\gamma(q) = e_{q},~~~C_-^Z(q) = \frac{1}{c_W s_W} \left(I^3_q - e_q s^2_W \right),~~~C_+^Z(q) = - \frac{s_W}{c_W}e_q \, ,
\ee
where $s_W$ and $c_W$ are sine and  cosine of the electroweak mixing angle $\theta_W$.
\item  Self energy counter terms:  \vspace{-1.0 cm}   \\
\vspace{-1.5 cm}
$\includegraphics[width=3.0cm]{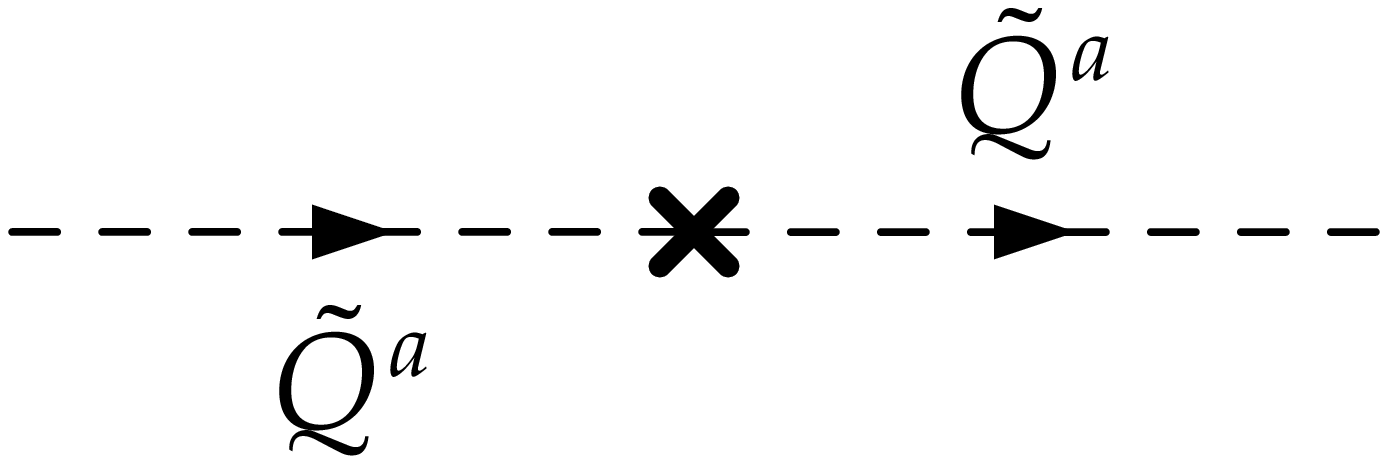}$   
     \raisebox{1.65cm}{\hspace{1cm} =  \hspace{1cm}  $i [(p^2-m^2_{\tilde{Q,a}}) \delta Z_{\tilde{Q,a}} - \delta m^2_{\tilde{Q,a}}   ]$}  \\
\vspace{-1.5 cm}
$\includegraphics[width=3.0cm]{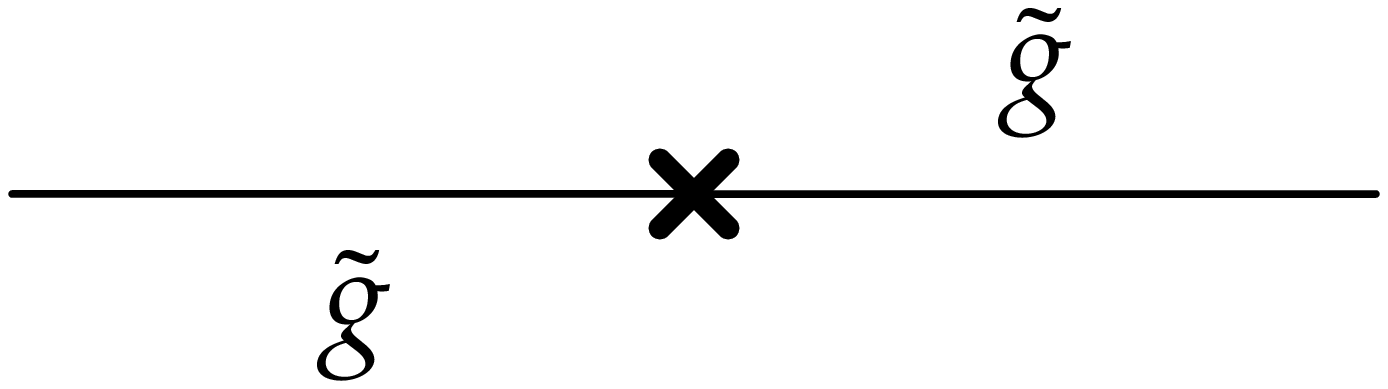}$ 
     \raisebox{1.65cm}{\hspace{1cm} =  \hspace{1cm}  $i [ (\sla{p}- m_{\tilde{g}} ) \delta Z_{\tilde{g}} -\delta m_{\tilde{g}} ]$}   \\ 
\vspace{-1.0 cm}
$\includegraphics[width=3.0cm]{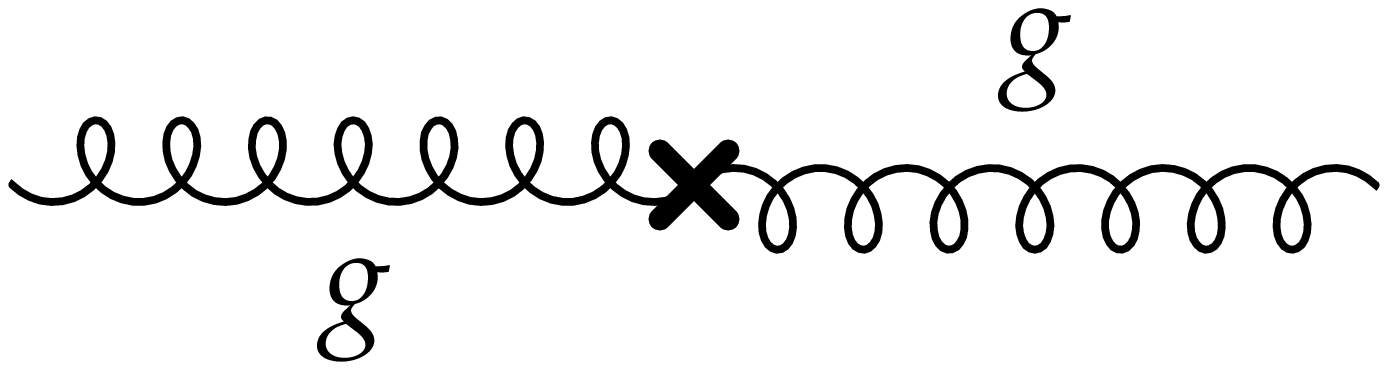}$ 
     \raisebox{1.65cm}{\hspace{1cm} =  \hspace{1cm}  $i (p_\mu p_\nu - g_{\mu \nu} p^2 )\delta Z_G$}   \\ 
\item Vertex counter terms involving gauginos:  \vspace{-0.5 cm} \\
\vspace{-1.3 cm}
$\includegraphics[width=3.0cm]{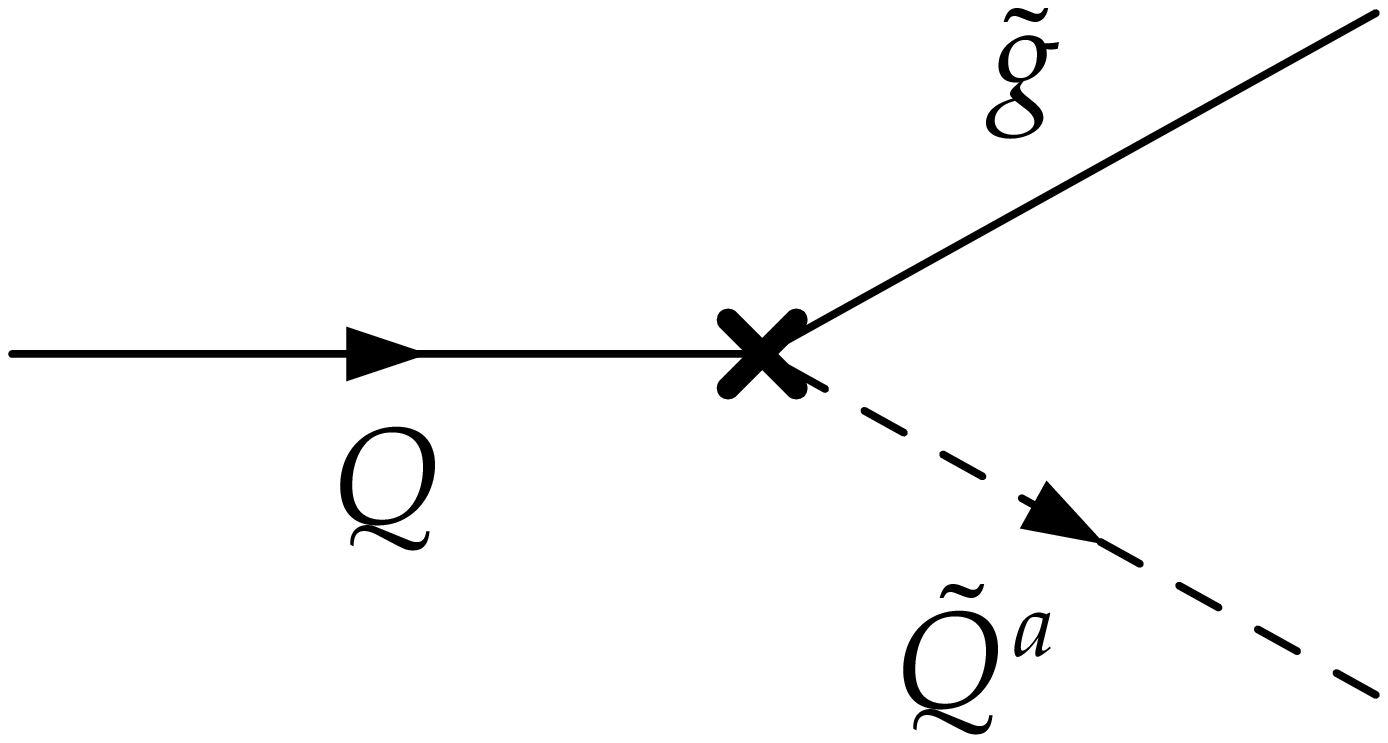}$ 
      \raisebox{1.65cm}{\hspace{.6 cm} =  \hspace{.6 cm} $- i \frac{g_s}{\sqrt{2}} 
                                          [~  (~ \delta Z_{\tilde{Q,a}} + 2 \delta Z_{\hat{g}} +  \delta Z_{\tilde{g}} + \delta Z_{QL} ~)~\delta_{aL}~\omega_-  - $}   \\
\vspace{-2.3 cm}
      \raisebox{1.65cm}{\hspace{5.9 cm}      $(~ \delta Z_{\tilde{Q,a}} + 2 \delta Z_{\hat{g}} +  \delta Z_{\tilde{g}} + \delta Z_{QR} ~)~\delta_{aR}~\omega_+  ~]~T^C$}   \\ 
\vspace{-1.3 cm}
$\includegraphics[width=3.0cm]{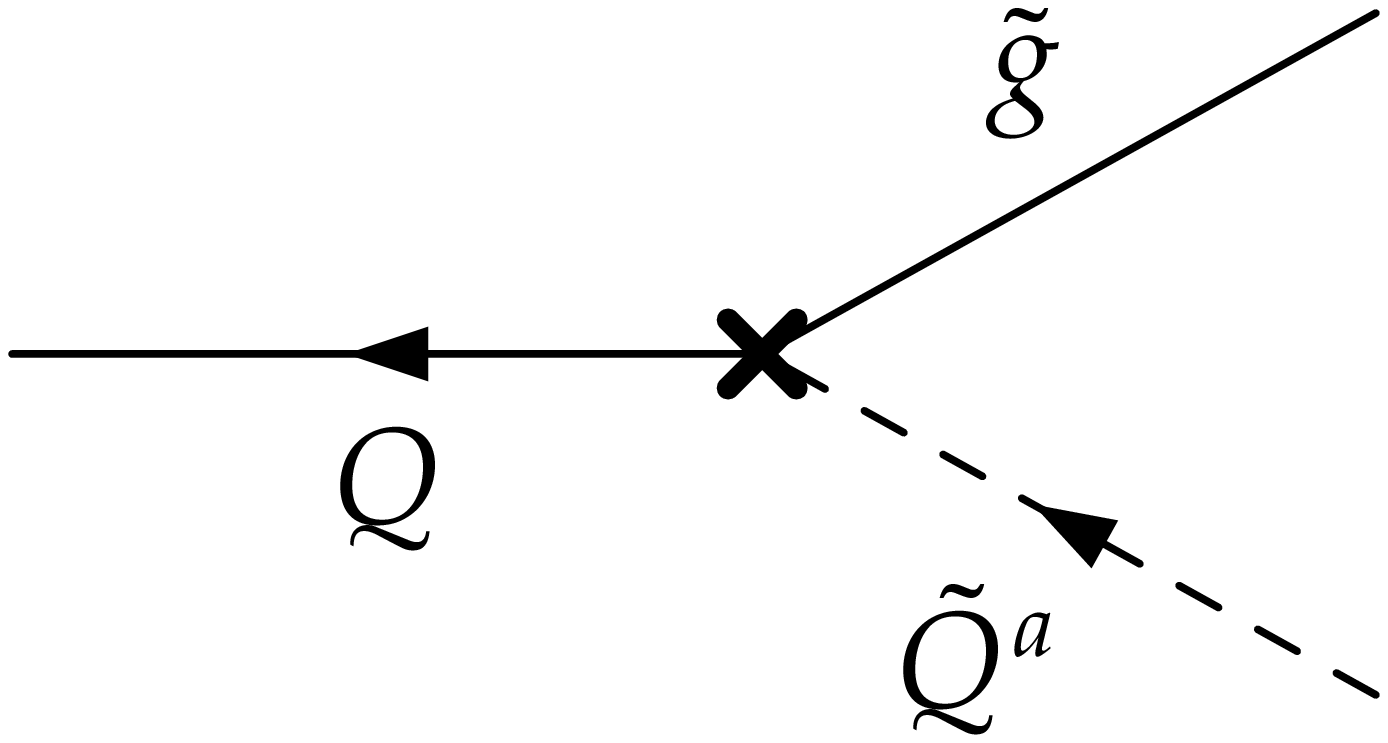}$ 
      \raisebox{1.65cm}{\hspace{.6 cm} =  \hspace{.6 cm} $i \frac{g_s}{\sqrt{2}} 
                                          [~  (~ \delta Z_{\tilde{Q,a}} + 2 \delta Z_{\hat{g}} +  \delta Z_{\tilde{g}} + \delta Z_{QR} ~)~\delta_{aR}~\omega_-  - $}   \\
\vspace{-2.3 cm}
      \raisebox{1.65cm}{\hspace{5.9 cm}      $(~ \delta Z_{\tilde{Q,a}} + 2 \delta Z_{\hat{g}} +  \delta Z_{\tilde{g}} + \delta Z_{QL} ~)~\delta_{aL}~\omega_+  ~]~T^C$}   \\ 
\vspace{-1.3 cm}
$\includegraphics[width=3.0cm]{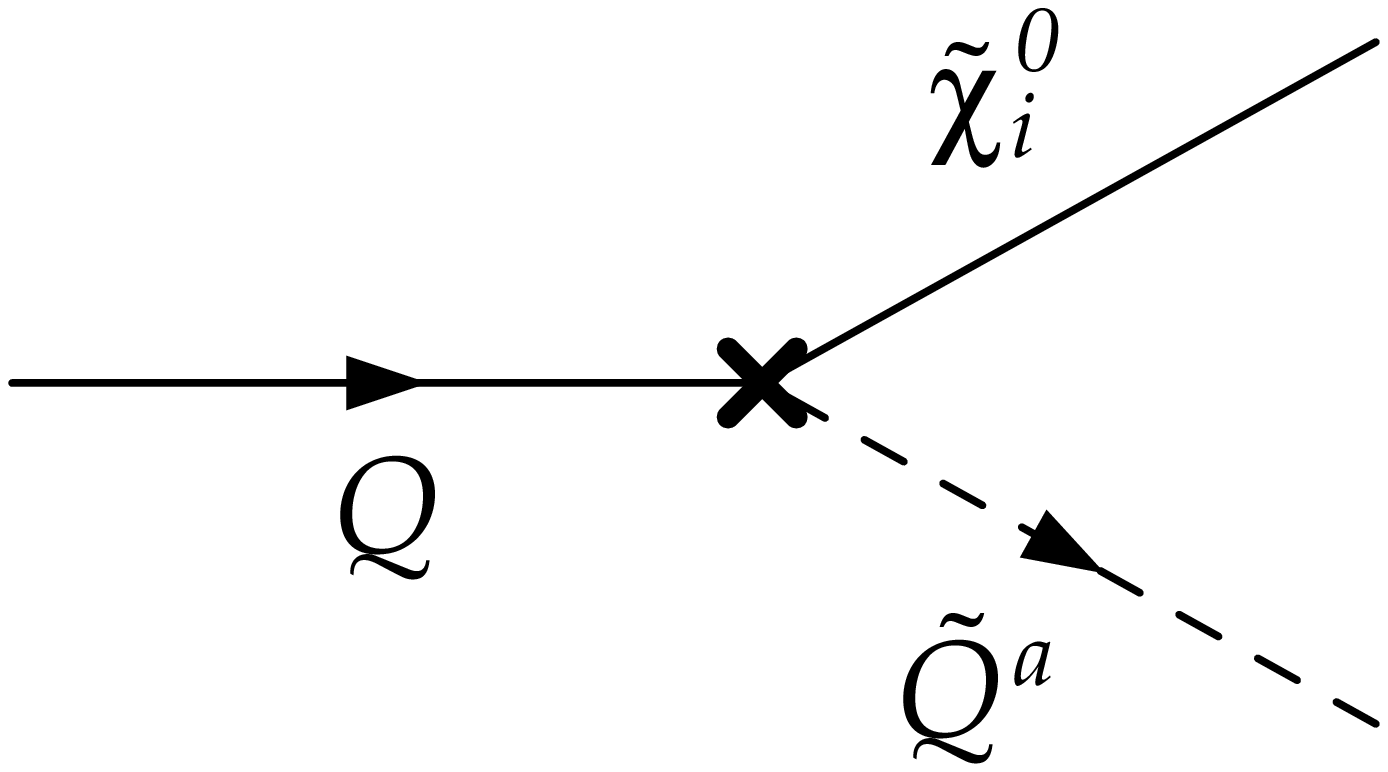}$ 
      \raisebox{1.65cm}{\hspace{.6 cm} = \hspace{.6 cm} $ i e [A_-(Q)~(\delta Z_{\tilde{Q},a} + \delta Z_{QL} )\delta_{aL}~\omega_- 
                                                           + A_+(Q)~(\delta Z_{\tilde{Q},a} + \delta Z_{QR} )\delta_{aR} ~\omega_+  ]$}   \\ 
\vspace{-1.3 cm}
$\includegraphics[width=3.0cm]{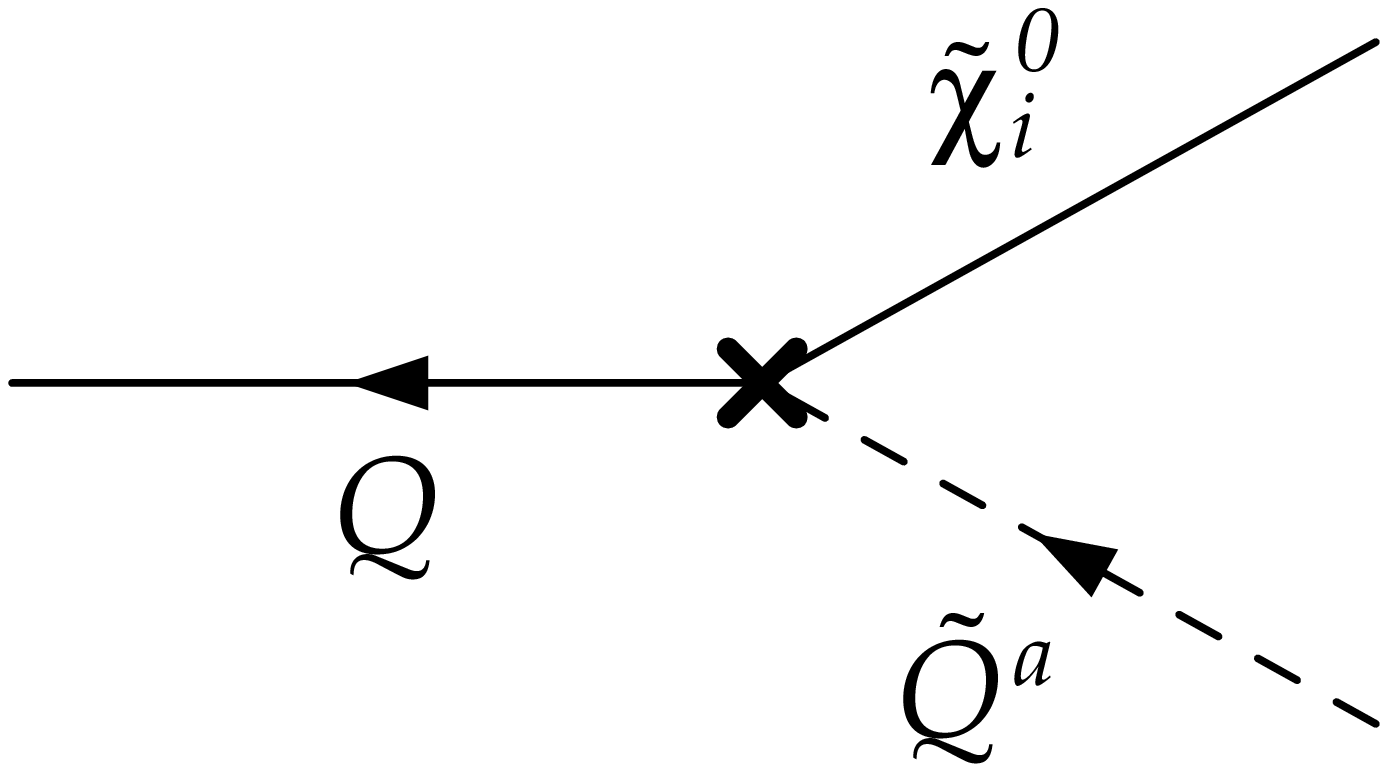}$ 
      \raisebox{1.65cm}{\hspace{.6 cm} = \hspace{.6 cm} $ i e [A_+^*(Q)~(\delta Z_{\tilde{Q},a} + \delta Z_{QR} )\delta_{aR}~\omega_- 
                                                            + A_-^*(Q)~(\delta Z_{\tilde{Q},a} + \delta Z_{QL} )\delta_{aL}~\omega_+  ]$}    \\
\vspace{-1.3 cm}
$\includegraphics[width=3.0cm]{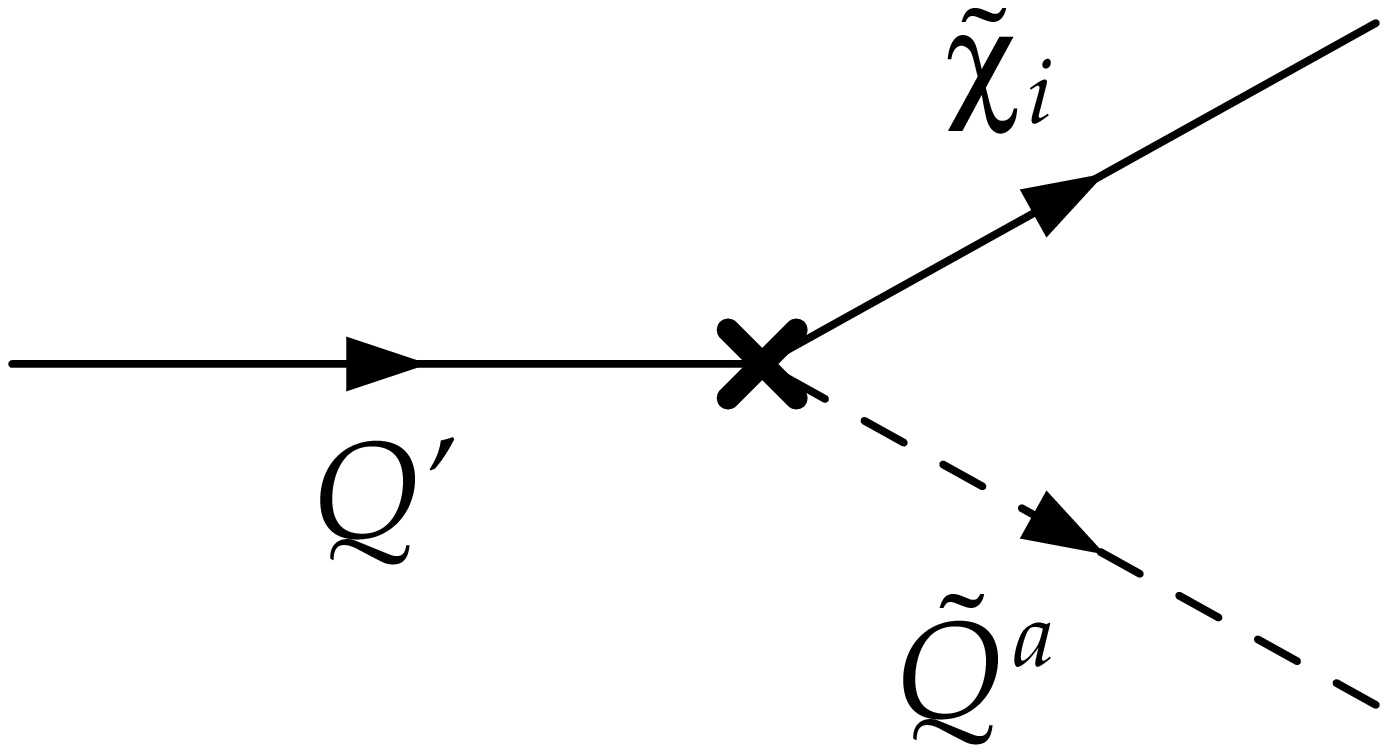}$  
      \raisebox{1.65cm}{\hspace{.6 cm} = \hspace{.6 cm} $ -i e \frac{B(Q')}{2 s_W}(\delta  Z_{\tilde{Q},a} + \delta Z_{Q'L})\delta_{aL}~\omega_- $}   \\ 
\vspace{-0.5 cm}
$\includegraphics[width=3.0cm]{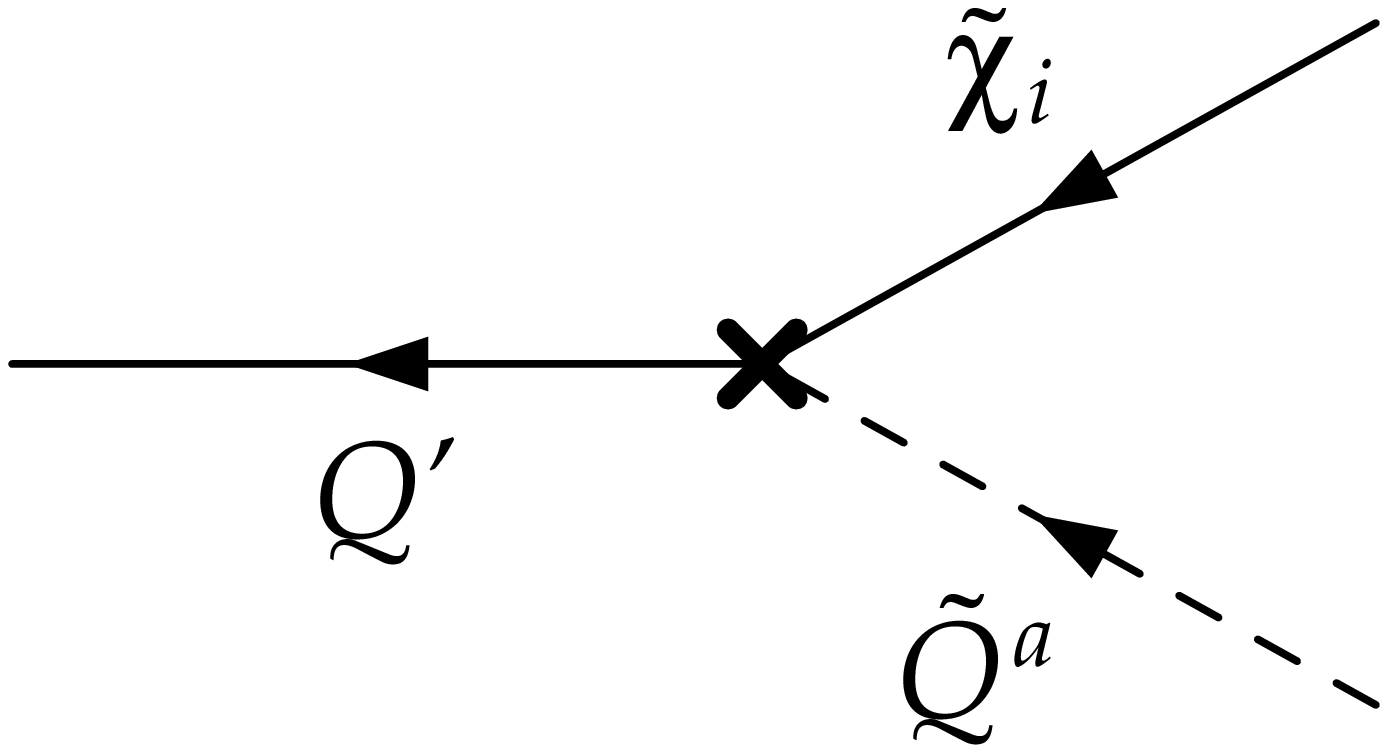}$     
      \raisebox{1.65cm}{\hspace{.6 cm} = \hspace{.6 cm} $ -i e  \frac{B^*(Q')}{2 s_W}(\delta  Z_{\tilde{Q},a} + \delta Z_{Q'L})\delta_{aL}~\omega_+  $}    \\
The Feynman rules involving Majorana particles follow the prescription of Ref.~\cite{Denner:1992me}; 
in particular the fermion flow is fixed according to the arrow depicted in the quark line.
As usual, $Q'$ denotes the SU(2) partner of $Q$. 
The vertices involving neutralinos contain the quantities
\be
A_+(Q) = \frac{1}{\sqrt{2}} \frac{e_Q N_{i1}}{c_W},~~~~A_-(Q)= - \frac{1}{\sqrt{2}}\left( \frac{1}{6}\frac{N_{i1}^*}{c_W} + I_Q\frac{N_{i2}^*}{s_W} \right) ,
\ee
where $N_{ij}$ is the mixing matrix of the neutralinos.  
$B(Q')$ can be expressed in terms of the mixing matrices $U$ and $V$ of the chargino sector: 
$\; B(Q') = U_{i1}^*\;$ $[B(Q')=V_{i1}^*]\;$ for up [down] type quarks.
\end{itemize}

\noindent
The renormalization constants of the squark sector are fixed by on-shell conditions (see also Ref.~\cite{RzehakHollik,RzehakHollik2}),
\ba
\delta Z_{\tilde{Q},a} = -\mathfrak{Re} \left\{ \frac{\partial\Sigma_{\tilde{Q},a}(p^2)}{\partial p^2} \right\}_{\big|{p^2=m^2_{\tilde{Q},a}}}, &&    
\delta m^2_{\tilde{Q},a} =  \mathfrak{Re} \left\{ \Sigma_{\tilde{Q},a}(m^2_{\tilde{Q},a})\right\}, \nonumber \\
\delta Z_{\tilde{Q}',a}= -\mathfrak{Re} \left\{ \frac{\partial\Sigma_{\tilde{Q}',a}(p^2)}{\partial p^2} \right\}_{\big|{p^2=m^2_{\tilde{Q}',a}}},&&
\delta m^2_{\tilde{Q}',R} =  \mathfrak{Re} \left\{ \Sigma_{\tilde{Q}',R}(m^2_{\tilde{Q}',R})\right\} , 
\ea
where $(\tilde{Q}, \tilde{Q}')$ is either of the two $SU(2)$ doublets $(\tilde{u},\tilde{d}),(\tilde{c}, \tilde{s})$, and 
 $\Sigma_{\tilde{Q},a}$ is the self energy of the squark $\tilde{Q^a}$. Due to SU(2) invariance the mass counter term of the
left-handed down-type squark is a dependent quantity,
\be
\delta m^2_{\tilde{Q}',L} = \delta m^2_{\tilde{Q},L} - c_{2 \beta}\; \delta M_W^2  + 4 M_W^2c_\beta^3 s_\beta\; \delta t_\beta \, ,
\ee
(where $c_\theta = \cos\theta$, $s_\theta = \sin\theta$ etc.\ for abbreviation).  
The counter term $\delta t_\beta$  for $\tan \beta$ 
is fixed in the $\overline{\mbox{DR}}$ scheme and can be written
%{\it e.g.}~
in the following way~\cite{Dabelstein,DRbar},
\be
\label{EQ:tanbetaCT}
\delta t_\beta = \frac{1}{2 M_Z c^2_\beta} \mathfrak{Re}\left \{\Sigma^{\mbox{\tiny div}}_{A^0 Z}(m^2_{A^0}) \right \} \, ,
\ee 
where $\Sigma^{\mbox{\tiny div}}$ denotes the divergent part of the $A^0 Z$
self energy in dimensional reduction. 
As pointed out in~\cite{TanBetaRen}, this process-independent condition is also gauge invariant.
Furthermore, the $W$ mass counter term appears in (\ref{EQ:tanbetaCT}), in the on-shell scheme given by
\be
\delta M_W^2 =  \mathfrak{Re} \left\{ \Sigma^T_W(M^2_W)  \right\}  ,                 
\ee 
where $\Sigma^T_W$ is the transverse part of the $W$ self energy. \\

\noindent
The field renormalization constants of the quarks are obtained via on-shell
conditons as follows~\cite{DennerHab},
\be
\delta Z_{qa} = - \mathfrak{Re} \left\{ \Sigma_{qa}(m^2_q)\right\} - 
           m_q^2 \mathfrak{Re}  \left\{ \frac{\partial}{\partial p^2} \left( \Sigma_{qL}(p^2)+ \Sigma_{qR}(p^2) +2\Sigma_{qS}(p^2)    
                 \right) \right\}_{\big|{p^2=m^2_{q}}}      ~~~(a=L,R)
\ee 
with the scalar coefficients in the 
Lorentz decomposition of the self energy,
\be
\Sigma_q(p^2)= \sla{p}\omega_- \Sigma_{qL}(p^2) + \sla{p}\omega_+ \Sigma_{qR}(p^2) + m_q \Sigma_{qS}(p^2).
\ee

\noindent 
Also in the gluino sector we determine the renormalization constants by on-shell conditions,
\ba
\delta m_{\tilde{g}} &=& \frac{1}{2}  \mathfrak{Re}  \left \{ m_{\tilde{g}}\; (\Sigma_{\tilde{g}L}(m^2_{\tilde{g}})+ \Sigma_{\tilde{g}R}(m^2_{\tilde{g}})+
                                                                       2 \Sigma_{\tilde{g}S}(m^2_{\tilde{g}})  ) \right \}  \\
\delta Z_{\tilde{g}} &=& - \mathfrak{Re}  \left \{\Sigma_{\tilde{g}L}(m^2_{\tilde{g}})\right\} - 
           m_{\tilde{g}}^2\, \mathfrak{Re}  \left\{ \frac{\partial}{\partial p^2} \left( \Sigma_{\tilde{g}L}(p^2)+ \Sigma_{\tilde{g}R}(p^2) +2\Sigma_{\tilde{g}S}(p^2)    
                 \right)\right\}_{\big|{p^2=m^2_{\tilde{g}}}} \, . \nonumber
\ea

\bigskip
\noindent
The renormalization of the strong coupling deserves some particular care. As mentioned in Section~\ref{S:loopQuarkCorrection} the strong coupling $g_s$
is renormalized in the $\overline{\mbox{MS}}$ scheme decoupling the heavy particles (top, gluino and squarks) from its runnning. 
Accordingly, the renormalization constant for $g_s$ in (\ref{EQ:renormparameter}) is given by~\cite{Beenakker1996}
\be
\delta Z_{g} = - \frac{\alpha_s}{4 \pi} \left[  \frac{3}{2} \Delta + \ln \left( \frac{m^2_{\tilde{g}} }{\mu^2} \right) 
                         +   \sum_{\tilde{Q},a} \frac{1}{12} \ln \left( \frac{m^2_{\tilde{Q},a}}{\mu^2} \right)  
                         +                    \frac{1}{3} \ln \left( \frac{m^2_{t}         }{\mu^2} \right) \right]
\ee
where $\Delta = 2 /\epsilon - \gamma_E + \ln(4 \pi)$. 
The treatment of UV divergences in dimensional regularization violates supersymmetry at the one-loop level,
introducing a mismatch between the strong Yukawa coupling and $g_s$. 
In order to restore  supersymmetry in physical amplitudes, cancellation of
this extra term is required, which at one-loop order can be achieved by modifying the renormalization constant
for $\hat{g}_s$ to be different from $\delta Z_g$:
\be
\delta Z_{\hat{g}} = \delta Z_{g} + \frac{\alpha_s}{3 \pi}  \, . 
\ee

\bigskip
\noindent
For completeness we quote also
the field renormalization constant of the gluon in Eq.~(\ref{EQ:fieldren}), 
\be
\delta Z_{G} =  2 \delta Z_{g} \, . 
\ee
At $\mathcal{O}(\alpha_s^2\alpha)$
it enters only the one-loop amplitude 
$\mathcal{M}^{1,\mbox{\tiny qcd}}_{q \overline{q}\to \tilde{Q}^a{Q}^{a*}}$, 
but since the gluon only appears in internal lines, $\delta Z_{G}$
is cancelled in the sum of self energy and vertex counter terms.
%
%
%
%
%%%%% APPENDICE COL PHASE SPACE SLICING %%%%%%%%%%%%%%%%%%% 
%
\addcontentsline{toc}{section}{Appendix}
\section{Bremsstrahlung integrals}
\label{S:AppPSS}
Here we list the IR and collinearly singular integrals that appear in the phase space integration 
of the bremsstrahlung processes, with either photons or gluons radiated.
In  the phase space slicing method, cuts are imposed: $\Delta E = 2 \delta_{s} {\sqrt s}$
on the energy of the emitted photon (gluon), and an angle cut $\delta_c$ 
on the angle between the photon/gluon and the radiating quark
via $\cos\theta > 1- \delta_{c}$.
The phase space is thus split into a soft and a collinear region that are
singular and a complementary non-singular region, which is integrated numerically.

%A general discussion about this method can be found in Ref.~\cite{PPSrev}. 

The integration over the soft region can be performed analytically,
regularizing the singularities by small masses for
the photon (gluon) and  the light quarks.
With the help of explicit formulae for the IR integrals \cite{DennerHab, Veltman:Int}, 
one obtains the factorized expressions for the cross section given below.
In the collinear region, the integration can be expressed as a convolution of the lowest-order
cross section and a radiator function.

\subsection*{Process {\boldmath{$gg \to \tilde{Q}^{a}\tilde{Q}^{a*} \gamma$}}  }
This process is affected by IR singularities only. 
Integrated over the soft region, the differential cross section reads as follows,
\be
d \sigma^{2,1}_{gg \to \tilde{Q}^a\tilde{Q}^{a *}\gamma} = 
-\frac{\alpha}{ \pi}e^2_{\tilde{Q}}( \delta_{F}- \delta_{FF})\;   d \sigma^{2,0}_{gg \to \tilde{Q}^a\tilde{Q}^{a *}}
\ee
where
\ba
\delta_{F}  &=& \ln \left( \frac{4 \Delta E^2}{\lambda^2}\right)+ \frac{1}{\beta}\ln \left(\frac{1-\beta}{1+\beta}  \right) \nonumber \\ 
\delta_{FF} &=& \frac{2}{\beta}
             \left( \frac{s-2m^2_{\tilde{Q},a}}{s} \right)
             \left[ \frac{1}{2}\ln \left( \frac{1+\beta}{1-\beta}\right)\ln \left( \frac{4 \Delta E^2}{\lambda^2}\right)   -
                    \mbox{Li}_2\left( \frac{2 \beta}{1+ \beta}  \right) -
                    \frac{1}{4} \ln^2\left(\frac{1+\beta}{1-\beta}  \right) \right] \nonumber. 
\ea
$\lambda$ is the infinitesimal mass regularizing the IR  divergencies, and
$\beta= \sqrt{ 1- (4m^2_{\tilde{Q},a} / s)}$.

\subsection*{Process {\boldmath{$q \overline{q} \to \tilde{Q}^{a}\tilde{Q}^{a*} \gamma$}}  }
The differential cross section integrated over the soft region 
can be expressed in terms of the $\mathcal{O}(\alpha_s^2)$ cross section for
$q \overline{q} \to \tilde{Q}^{a}\tilde{Q}^{a*}$ and a IR-singular factor,
\be
d \sigma^{2,1}_{q \overline{q} \to \tilde{Q}^a\tilde{Q}^{a *}\gamma} = 
-\frac{\alpha}{ \pi}\left[ 
                            e^2_{q}              (\delta_{I}- \delta_{II}) + 
                            e^2_{\tilde{Q}}       (\delta_{F}- \delta_{FF}) +
                            e_{q}e_{\tilde{Q}}  (\delta_{IF} - \delta_{FI}) \right] 
d \sigma^{2,0}_{q \overline{q} \to \tilde{Q}^a\tilde{Q}^{a *}}
\ee
where $e_q = \frac{2}{3}$ if $q=u,c$ and $e_q = - \frac{1}{3}$ otherwise. Furthermore,
\ba
\delta_{I}  &=& \ln \left( \frac{4 \Delta E^2}{\lambda^2}\right)+ \ln \left(\frac{m_q^2}{s}  \right) ,  \\
\delta_{II} &=& \ln \left( \frac{s}{m_q^2} \right) \ln \left( \frac{4 \Delta E^2}{\lambda^2}\right)- \frac{\pi^2}{3} - 
               \frac{1}{2}\ln^2\left( \frac{s}{m_q^2} \right) ,  \nonumber \\
\delta_{IF} &=& \ln \left( \frac{m_q^2 m_{\tilde{Q},a}^2}{(t -m_{\tilde{Q},a}^2)^2} \right)\ln \left( \frac{4 \Delta E^2}{\lambda^2}\right) +
                \frac{1}{2} \left[ \ln^2\left( \frac{m_q^2}{s} \right) - \ln^2\left( \frac{1 -\beta}{1+\beta} \right) \right] + \frac{\pi^2}{3} +
                2 \mbox{Li}_2\left(1 +\frac{st}{(m^2_{\tilde{Q},a}-t)^2} \right)  \nonumber \\
            &-& 2 \left[  \mbox{Li}_2\left( 1+ \frac{(1-\beta)st}{2 m^2_{\tilde{Q},a}(m^2_{\tilde{Q},a}-t)} \right) +  
                          \mbox{Li}_2\left( 1+ \frac{(1+\beta)st}{2 m^2_{\tilde{Q},a}(m^2_{\tilde{Q},a}-t)} \right)  \right]; \nonumber 
\ea
$\delta_{FI}$ can be obtained from $\delta_{IF}$ by  the substitution $t \to u$.\\

The differential cross section integrated over the collinear region
can be written in terms of a convolution integral,
\be
\label{Eq:CollinearGamma}
 d\sigma^{2,1}_{q\overline{q}\to \tilde{Q}^{a}\tilde{Q}^{a*}\gamma}(s)=
\frac{\alpha e_q^2}{2 \pi} \int_{x_0}^{1- \delta_s}dz~\left\{ \left[ \ln\left( \frac{s \delta_c}{2 m_q^2}  \right) -1 \right]P_{qq}(z) + (1-z)     \right \} 
d\sigma^{2,0}_{q\overline{q}\to \tilde{Q}^a\tilde{Q}^{a*}}(zs),
\ee
with $x_0 = (4 m^2_{\tilde{Q},a})/ s$ and the quark splitting function $P_{qq}(z)$ from Eq.~(\ref{EQ:splittingfunction}).

\subsection*{Process {\boldmath{$q \overline{q} \to \tilde{Q}^{a}\tilde{Q}^{a*} g$}}  }
The singularities affecting this radiative process are Abelian-like, similar to the case of photon radiation, and thus 
can be treated by  mass regularization as well.
The differential cross section integrated over the soft region 
can also be expressed in terms of $\mathcal{O}(\alpha_s\alpha)$ contributions
to the cross section for $q \overline{q} \to \tilde{Q}^{a}\tilde{Q}^{a*}$,
but only together with a rearrangement of the color structure. The emission of a gluon as a colored particle
leads to color correlations in the eikonal current, which can be taken into account following  
the  prescription of Ref.~\cite{DipoleCatani}, yielding the result
\ba
d \sigma^{2,1}_{q \overline{q} \to \tilde{Q}^{a}\tilde{Q}^{a*}g} &=& -\frac{\alpha_s}{2 \pi} \Big\{ 
                                                                       \Big [ C_F \Big(2 \delta_I + 2\delta_F \Big) 
                                                                               +2 \Big ( C_F +\frac{1}{N} \Big) \delta_{FI} 
                                                                               + \frac{1}{N} \Big (\delta_{II}+ \delta_{FF}-\delta_{IF} \Big) \Big ] 
                                                                               d \sigma^{1,1}_{q \overline{q} \to \tilde{Q}^{a}\tilde{Q}^{a*}}  \nonumber \\
                                                                    &-&    \Big [  \delta_{FI}- \delta_{IF}  \Big ]    
                                                                               d \bar{\sigma}^{1,1}_{q \overline{q} \to \tilde{Q}^{a}\tilde{Q}^{a*}} - 
                                                                        \Big [  \delta_{FI}+ \delta_{II} + \delta_{FF} \Big ]    
                                                                               d \tilde{\sigma}^{1,1}_{q \overline{q} \to \tilde{Q}^{a}\tilde{Q}^{a*}} \Big \} ,
\ea
with $C_F = \frac{4}{3}$ and $N = 3$. In order to specify the color-modified ``cross sections''
 $d \bar{\sigma}$ and $d \tilde{\sigma}$, we first separate the tree-level amplitudes
for $q \overline{q} \to \tilde{Q}^{a}\tilde{Q}^{a*}$ into color factors and reduced matrix elements,
according to the $s$- and $t$-channel diagrams in Fig.~\ref{Fig:TREE}:
\ba
\mathcal{M}^{0,\mbox{\tiny qcd}~[c_1,c_2,c_3,c_4]}_{q\overline{q}\to \tilde{Q}^a\tilde{Q}^{a*}}
&=& \sum_{C} \left(
T^C_{c_2c_1}T^C_{c_3c_4} \mathcal{M}^{0,\mbox{\tiny qcd\, (s)}}_{q\overline{q}\to \tilde{Q}^a\tilde{Q}^{a*}} +
    T^C_{c_3c_1}T^C_{c_2c_4} \mathcal{M}^{0,\mbox{\tiny qcd\, (t)}}_{q\overline{q}\to \tilde{Q}^a\tilde{Q}^{a*}} 
\right ) ,  \nonumber \\
\mathcal{M}^{0,\mbox{\tiny ew}~[c_1,c_2,c_3,c_4]}_{q\overline{q}\to \tilde{Q}^a\tilde{Q}^{a*}}
&=& \delta_{c_1c_2}\delta_{c_3c_4} \mathcal{M}^{0,\mbox{\tiny ew \, (s)}}_{q\overline{q}\to \tilde{Q}^a\tilde{Q}^{a*}} +
    \delta_{c_1c_3}\delta_{c_2c_4} \mathcal{M}^{0,\mbox{\tiny ew \, (t)}}_{q\overline{q}\to \tilde{Q}^a\tilde{Q}^{a*}}  
\ea
where $T^C$ are the color matrices in the fundamental representation.
With this notation we can write for the color-rearranged contributions,
\ba
d \bar{\sigma}^{1,1}_{q \overline{q} \to \tilde{Q}^{a}\tilde{Q}^{a*}} &=& 
\frac{dt}{16 \pi s^2} \frac{1}{N^2}~2\, \mathfrak{Re} \left \{ 
\left ( \mathcal{M}^{0,\mbox{\tiny qcd}~[c_1,c_2,c_1,c_3]}_{q\overline{q}\to \tilde{Q}^a\tilde{Q}^{a *}} \right )^*
\mathcal{M}^{0, \mbox{\tiny ew}~[c_4,c_2,c_4,c_3]}_{q\overline{q}\to \tilde{Q}^a\tilde{Q}^{a*}}  \right\} , \nonumber \\
d \tilde{\sigma}^{1,1}_{q \overline{q} \to \tilde{Q}^{a}\tilde{Q}^{a*}} &=& 
\frac{dt}{16 \pi s^2} \frac{1}{N^2} ~2\, \mathfrak{Re} \left \{ 
\left ( \mathcal{M}^{0,\mbox{\tiny qcd}~[c_1,c_1,c_2,c_3]}_{q\overline{q}\to \tilde{Q}^a\tilde{Q}^{a *}} \right )^*
\mathcal{M}^{0, \mbox{\tiny ew}~[c_4,c_4,c_2,c_3]}_{q\overline{q}\to \tilde{Q}^a\tilde{Q}^{a*}}  \right\} ,
\ea 
where color summation has to be performed over each pair of equal indices. 
On top, average over the initial helicities is assumed. Owing to the particular color structure, 
$d \tilde{\sigma}$ is different from zero only if $q=Q$. \\

The differential cross section integrated over the collinear region
can be written in terms of a convolution integral
similar to Eq.~(\ref{Eq:CollinearGamma}),
\be
\label{Eq:CollinearGluon}
 d\sigma^{2,1}_{q\overline{q}\to \tilde{Q}^{a}\tilde{Q}^{a*}g}(s)=
\frac{\alpha_s C_F}{2 \pi} \int_{x_0}^{1- \delta_s}dz~\left\{ \left[ \ln\left( \frac{s \delta_c}{2 m_q^2}  \right) -1 \right]P_{qq}(z) + (1-z)     \right \} 
d\sigma^{1,1}_{q\overline{q}\to \tilde{Q}^a\tilde{Q}^{a*}}(zs) \, ,
\ee
with $d\sigma^{1,1}_{q\overline{q}\to \tilde{Q}^a\tilde{Q}^{a*}}$  instead of
$d\sigma^{2,0}_{q\overline{q}\to \tilde{Q}^a\tilde{Q}^{a*}}$.

\subsection*{Processes {\boldmath{$q g \to \tilde{Q}^{a}\tilde{Q}^{a*} q$}} and
{\boldmath{$\overline{q} g \to \tilde{Q}^{a}\tilde{Q}^{a*} \overline{q}$}}    }
These processes exhibit singularities when the final (anti-)quark is
emmitted off the gluon in the collinear region. In that region the
differential cross section can be written, in analogy to~\cite{Tobi}, 
as follows,
\ba
 d\sigma^{2,1}_{qg\to \tilde{Q}^{a}\tilde{Q}^{a*}q}(s) &=&
\frac{\alpha_s T_F}{2 \pi} \int_{x_0}^{1}dz~\left\{\ln\left( \frac{s
      (1-z)^2\delta_c}{2 m_q^2}\right)P_{qg}(z) 
+ 2z(1-z)     \right \} 
d\sigma^{1,1}_{q\overline{q}\to \tilde{Q}^a\tilde{Q}^{a*}}(zs) \\
d\sigma^{2,1}_{\overline{q}g\to \tilde{Q}^{a}\tilde{Q}^{a*}\overline{q}}(s) &=&
\frac{\alpha_s T_F}{2 \pi} \int_{x_0}^{1}dz~\left\{\ln\left( \frac{s
      (1-z)^2\delta_c}{2 m_{\overline{q}}^2}\right)P_{qg}(z) 
+ 2z(1-z)     \right \} 
d\sigma^{1,1}_{q\overline{q}\to \tilde{Q}^a\tilde{Q}^{a*}}(zs) \nonumber 
\label{Eq:CollinearQuark}
\ea
with the splitting function $P_{qg}$ from Eq.~(\ref{EQ:splittingfunction}).

\newpage
\bibliographystyle{JHEP}
\bibliography{ref}

\newpage 

%%%%%%%%%%%%%%%%%%%%%%%%%%%%%%%%%%%%%%%%%%%%%%%%
%%%%% STUDY OF DIFFERENT  IR REGULARIZATIONS %%%
%%%%%%%%%%%%%%%%%%%%%%%%%%%%%%%%%%%%%%%%%%%%%%%%

\begin{figure}
\centering
\epsfig{file=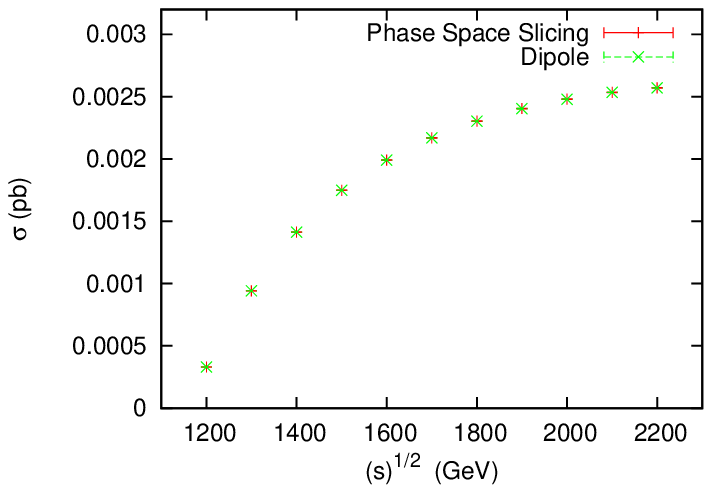, width= 7.3cm}
\epsfig{file=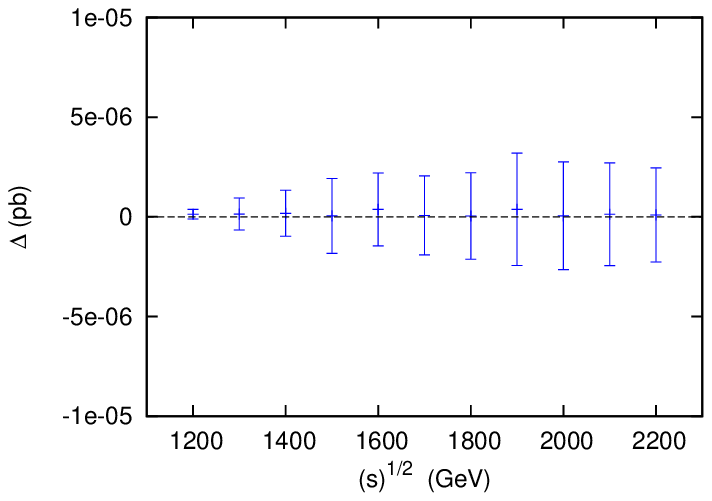, width= 7.3cm}
\epsfig{file=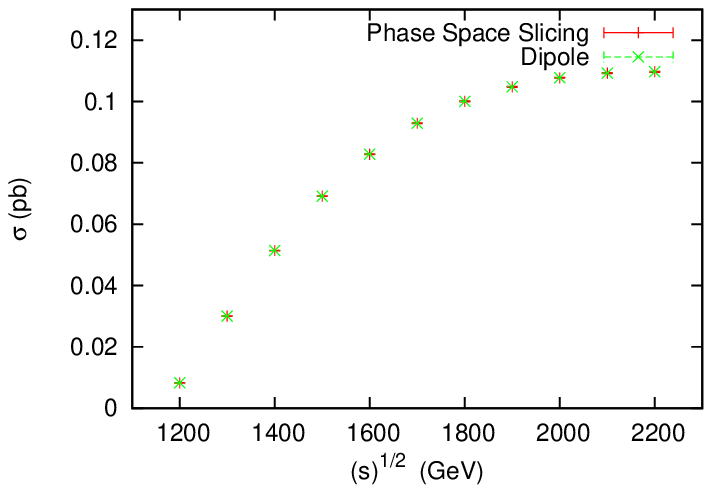, width= 7.3cm}
\epsfig{file=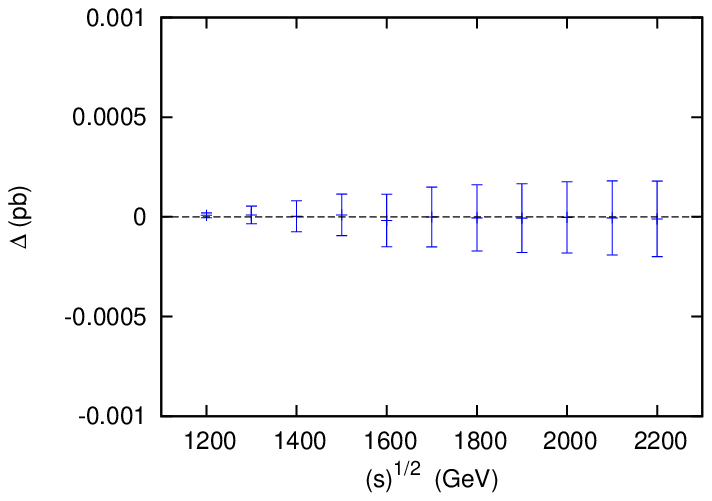, width= 7.3cm}
\epsfig{file=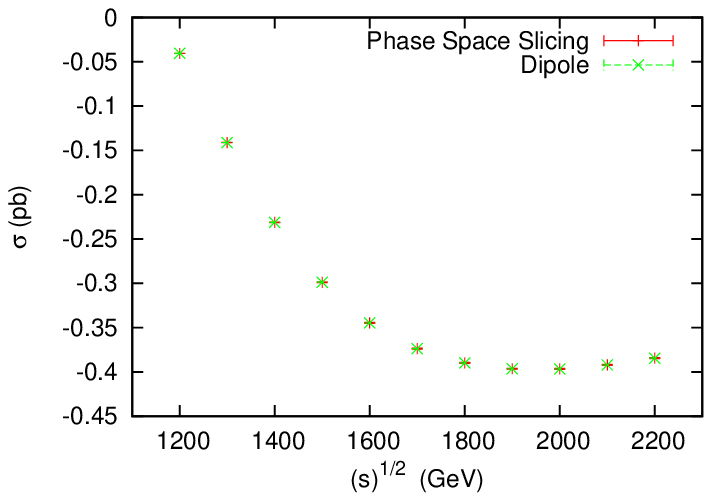, width= 7.3cm}
\epsfig{file=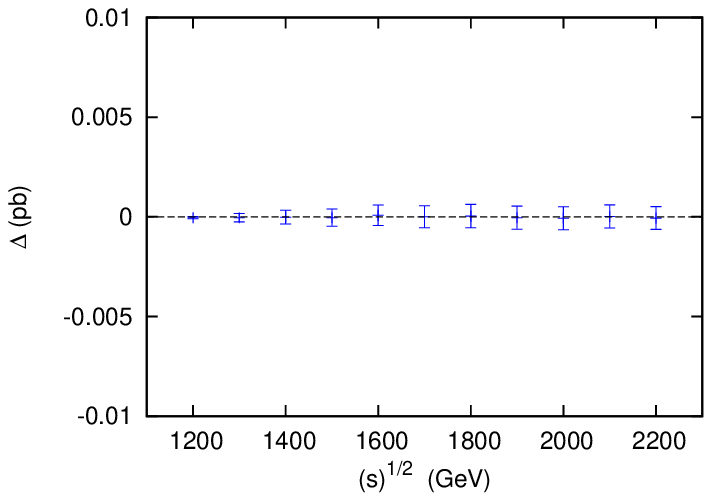, width= 7.3cm}
\epsfig{file=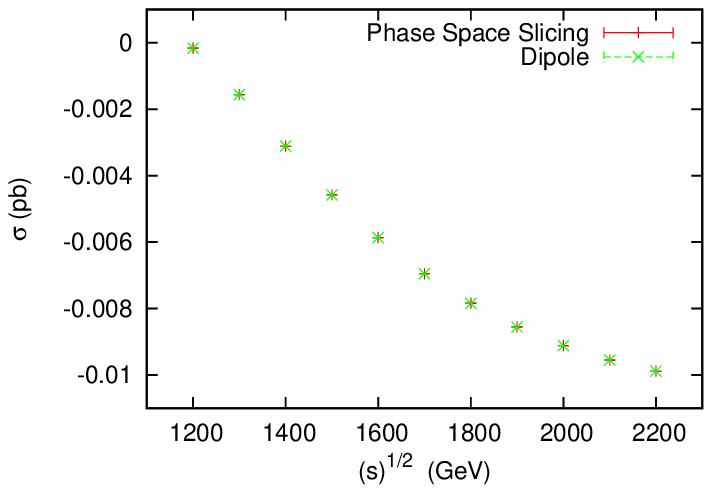, width= 7.3cm}
\epsfig{file=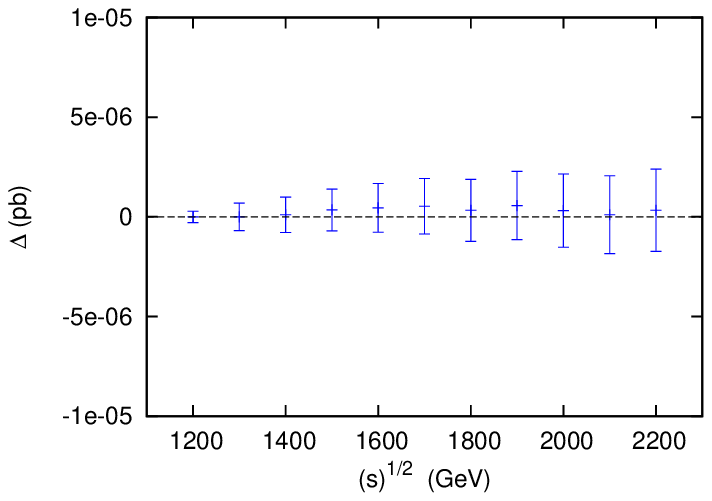, width= 7.3cm}
\caption{Lowest order  partonic cross sections for the process $g g  \to \tilde{u}^L \tilde{u}^{L*}\gamma$ (first panel),
$u \overline{u} \to \tilde{u}^{L} \tilde{u}^{L*}\gamma$ (second  panel), $u \overline{u} \to \tilde{u}^L \tilde{u}^{L*} g$ (third panel)
and $ug \to\tilde{u}^L \tilde{u}^{L*} u$ (fourth panel), computed with the 
two different methods. $\Delta$ is defined as $\Delta =\sigma^{\mbox{\tiny Slicing}} -\sigma^{\mbox{\tiny Dipole}}$. The error bars represent the 
integration uncertainty. 
The SUSY parameters are those of the SPS1a$'$ point~\cite{SPA}.}
\label{Fig:METODI}
\end{figure}

%%%%%%%%%%%%%%%%%%%%%%%%%%%%%%%%%%%%%%%%%%%%%%%%
%%%%% STUDY OF DIFFERENT SQUARKS CASES       %%%
%%%%%%%%%%%%%%%%%%%%%%%%%%%%%%%%%%%%%%%%%%%%%%%%

\begin{figure}
\centering
\underline{$\tilde{u}^{R}$}\\
\epsfig{file= 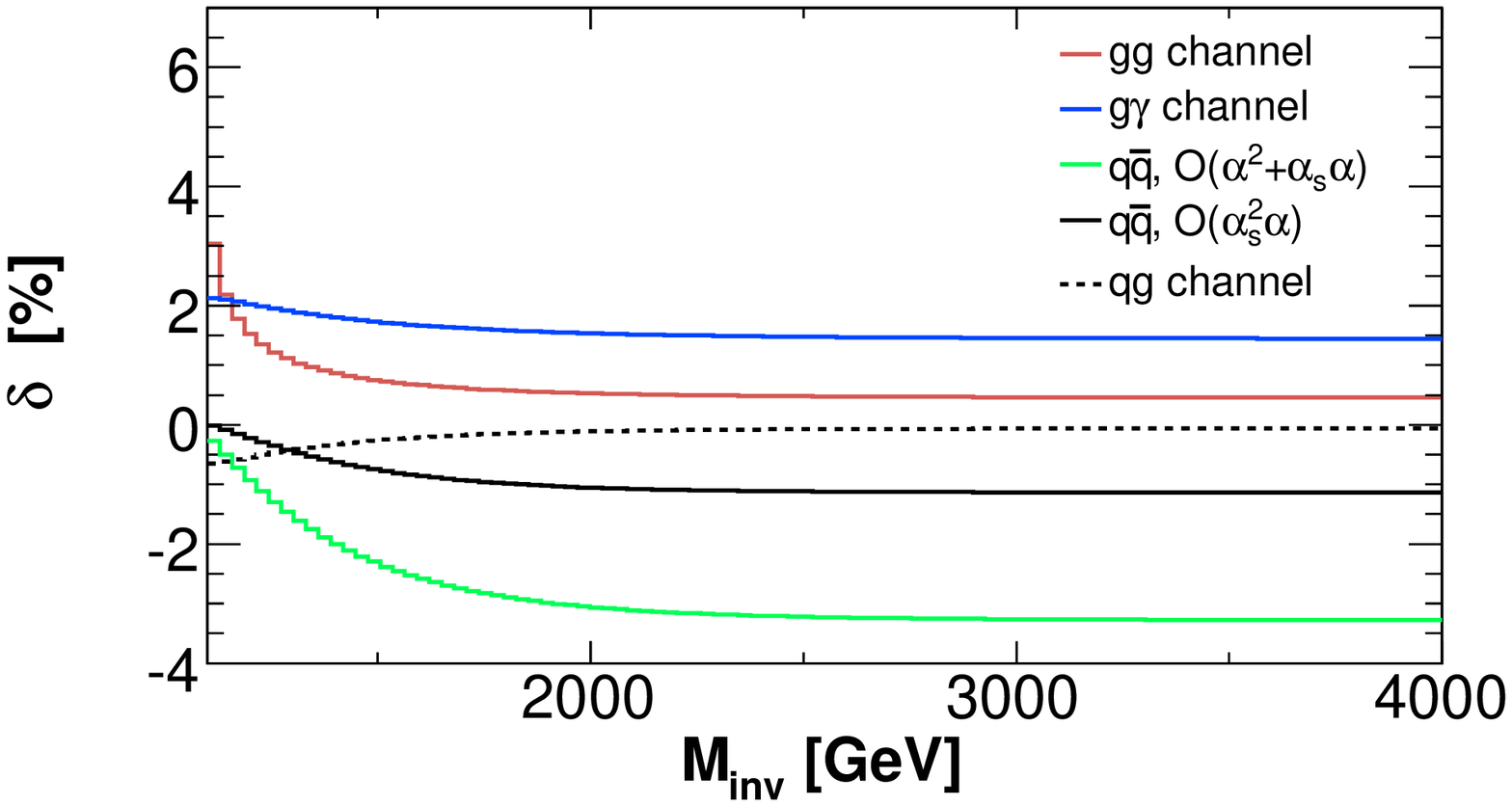, width=7.9cm}
\epsfig{file= 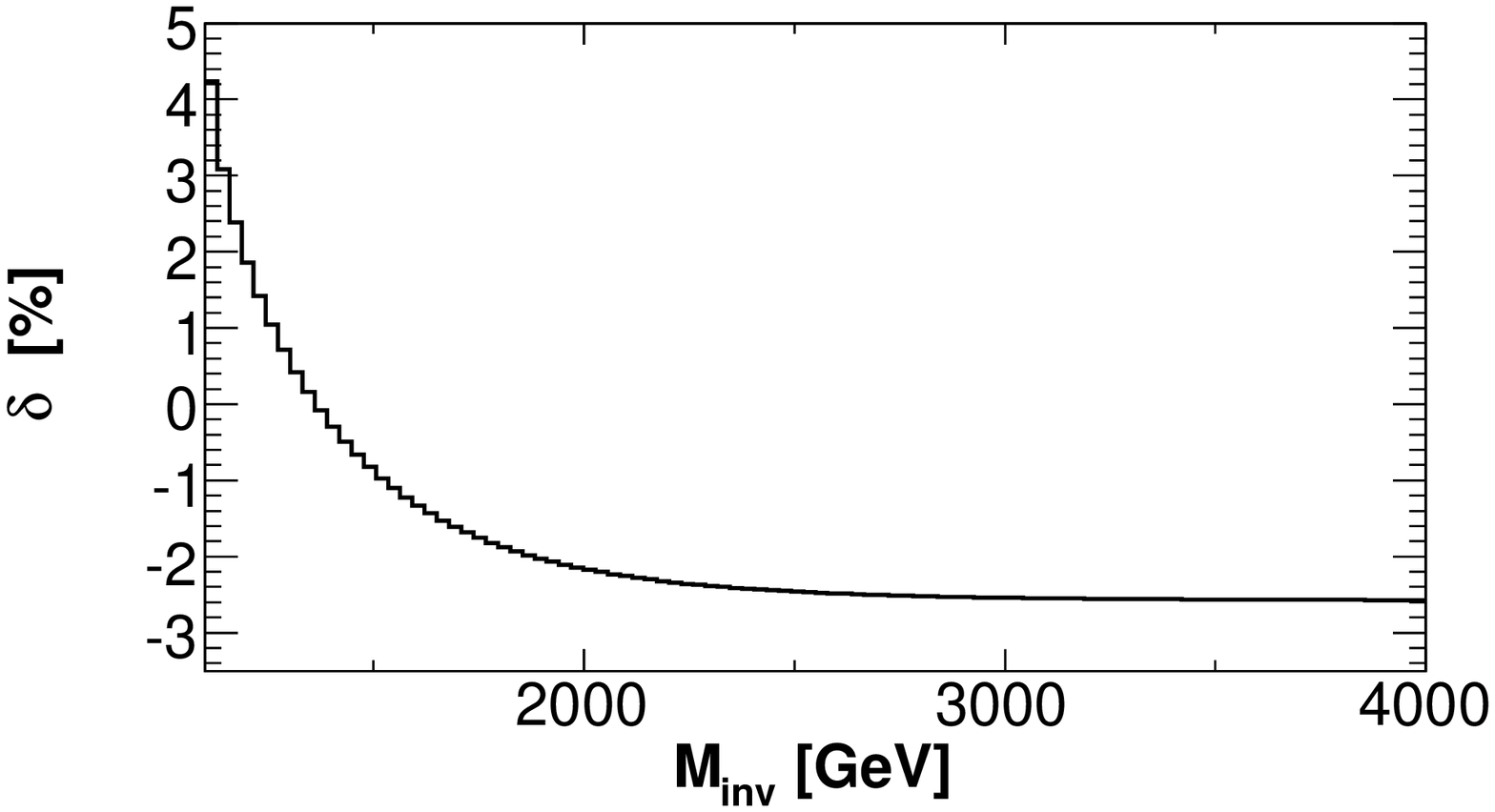, width=7.9cm}
\underline{$\tilde{u}^{L}$}\\
\epsfig{file= 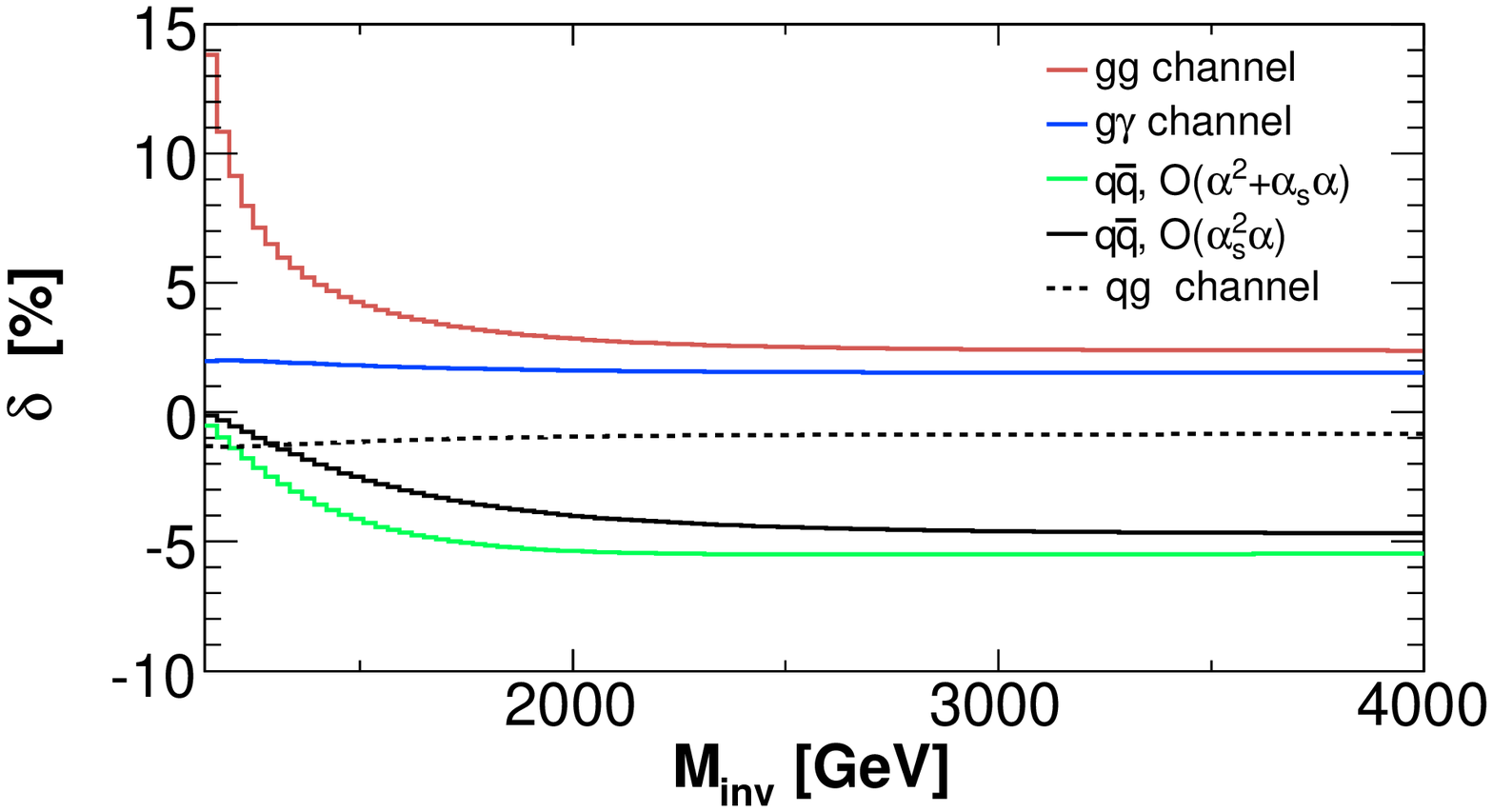, width=7.9cm}
\epsfig{file= 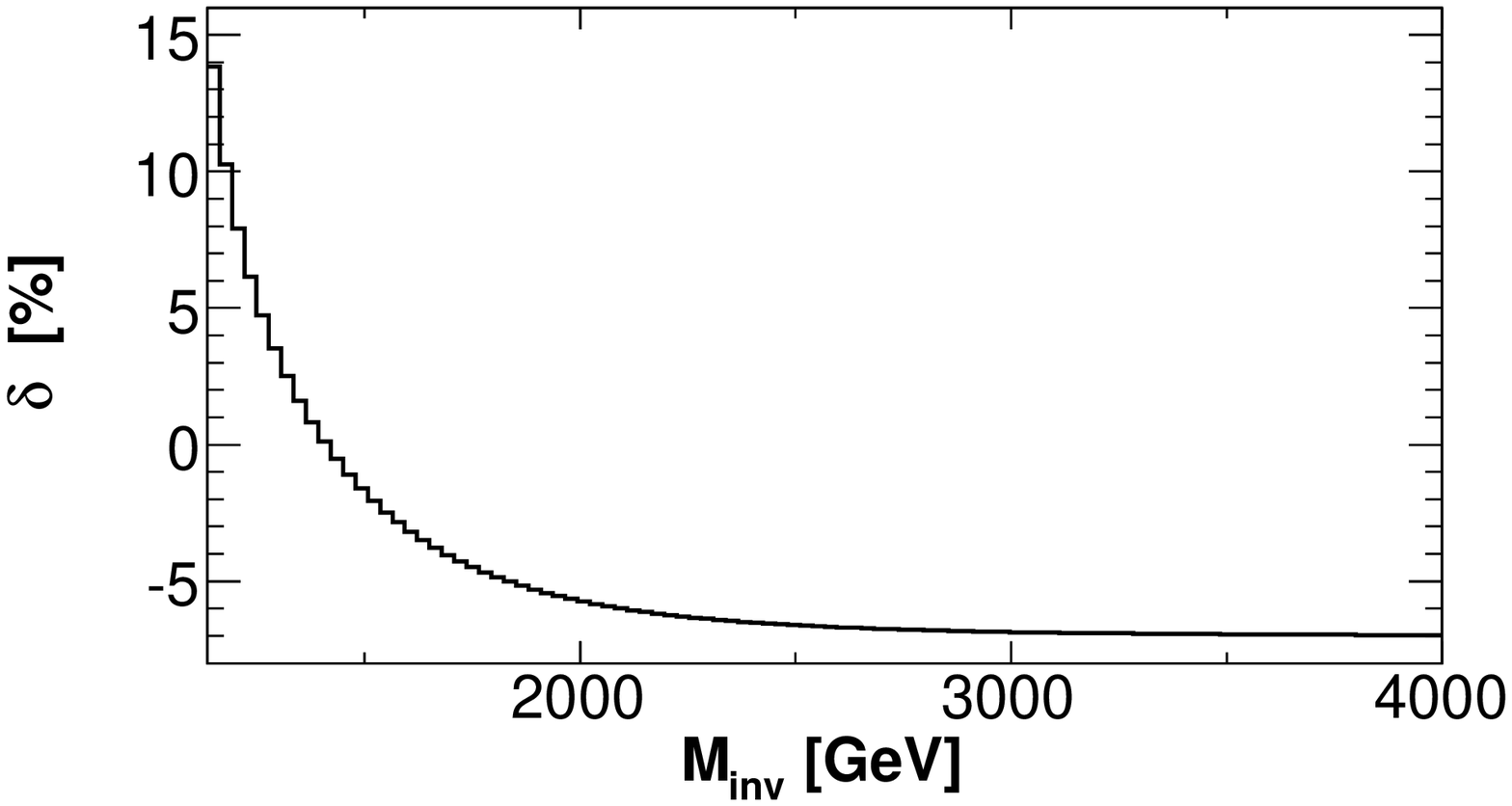, width=7.9cm}
\underline{$\tilde{d}^{L}$}\\
\epsfig{file= 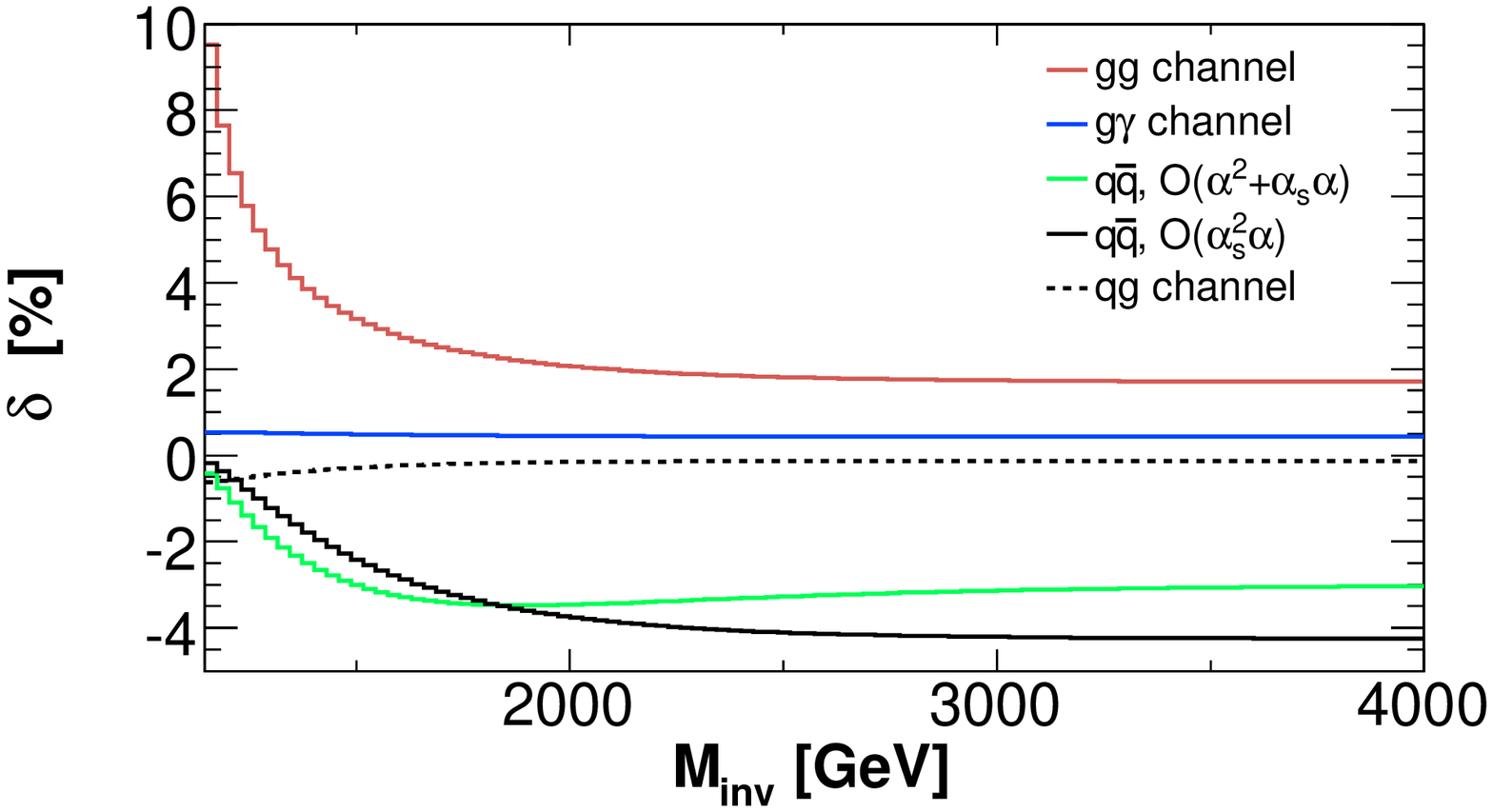, width=7.9cm}
\epsfig{file= 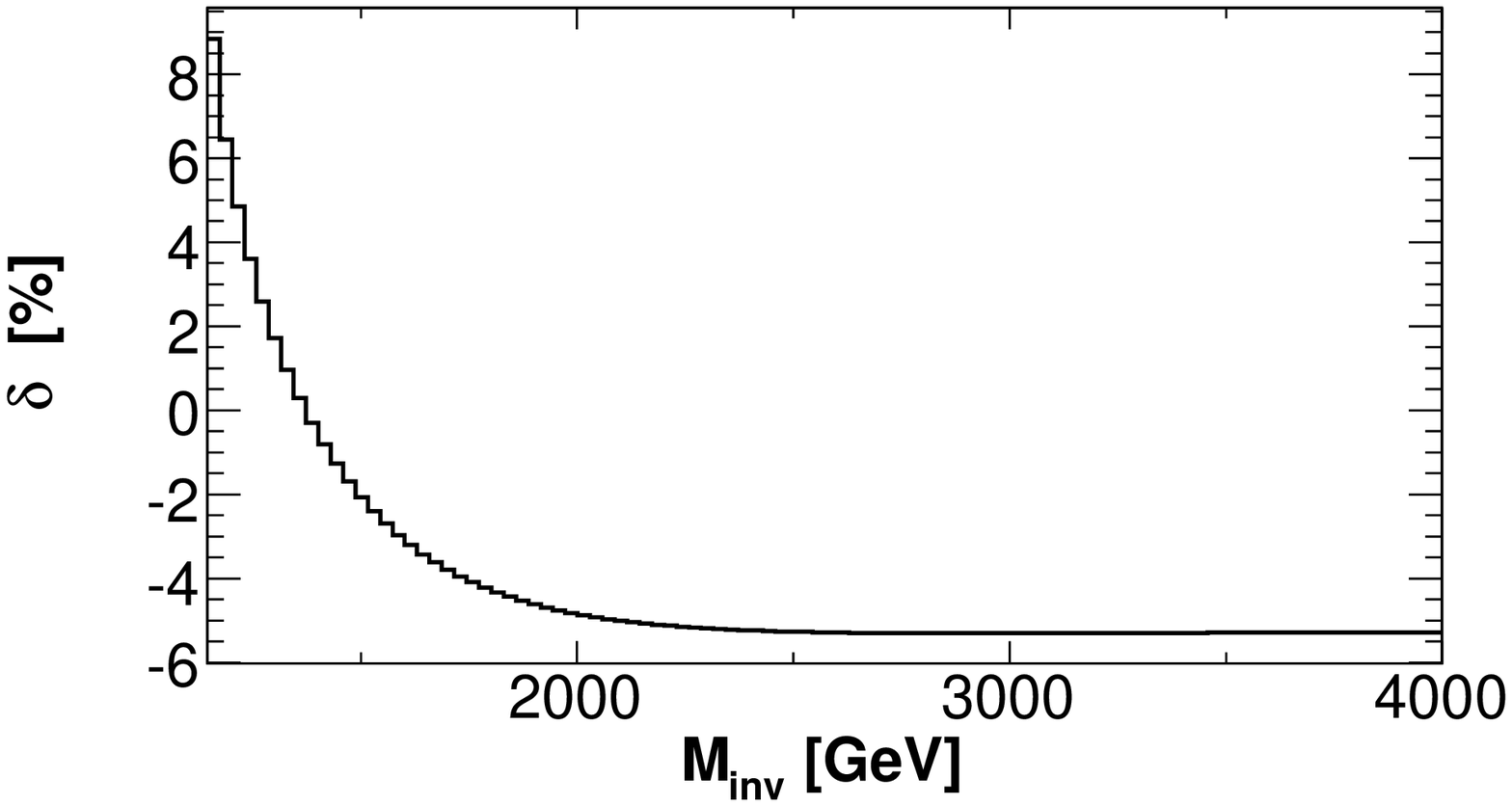, width=7.9cm}
\underline{$\tilde{c}^{L}$}\\
\epsfig{file= 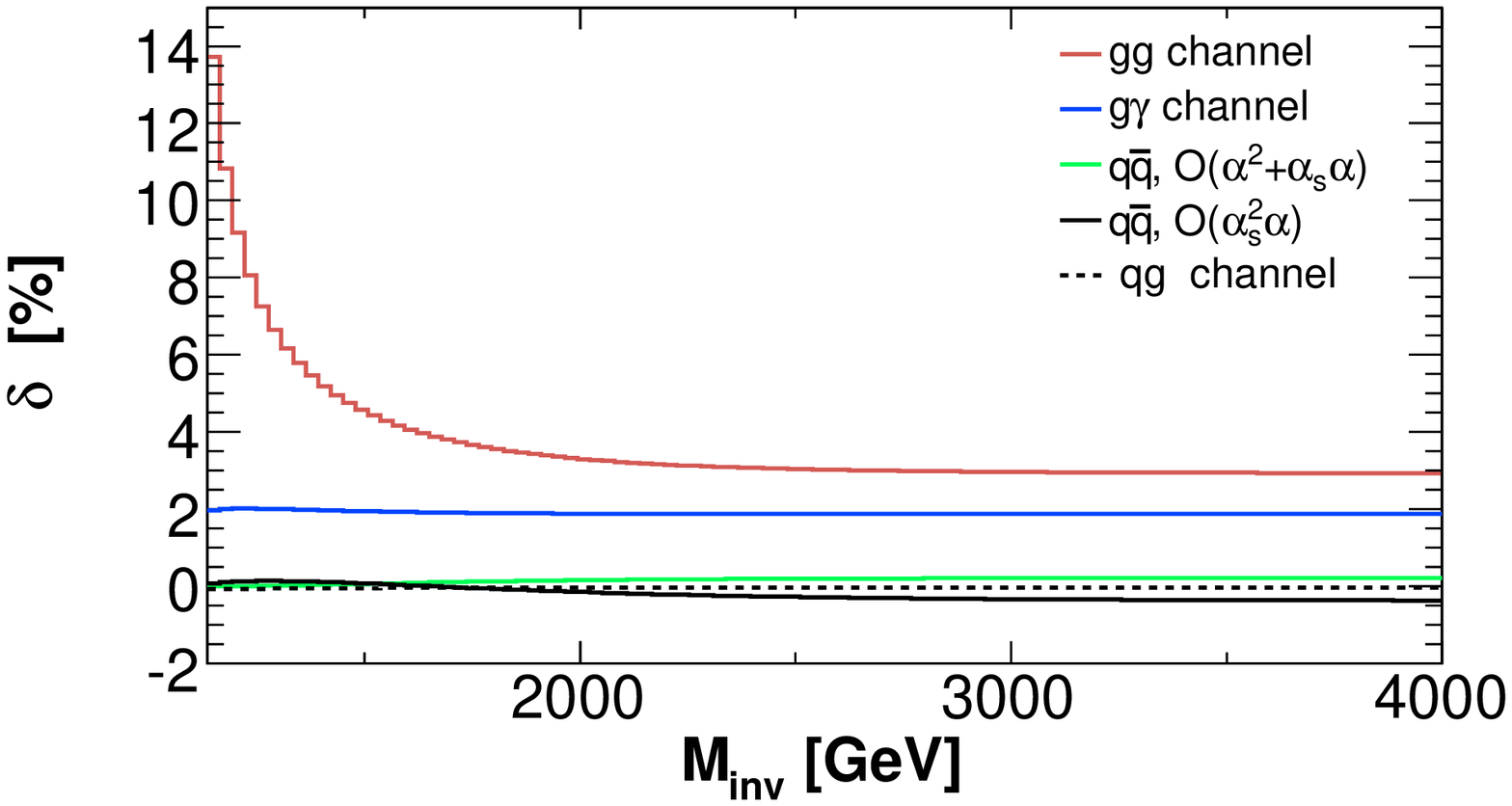, width=7.9cm}
\epsfig{file= 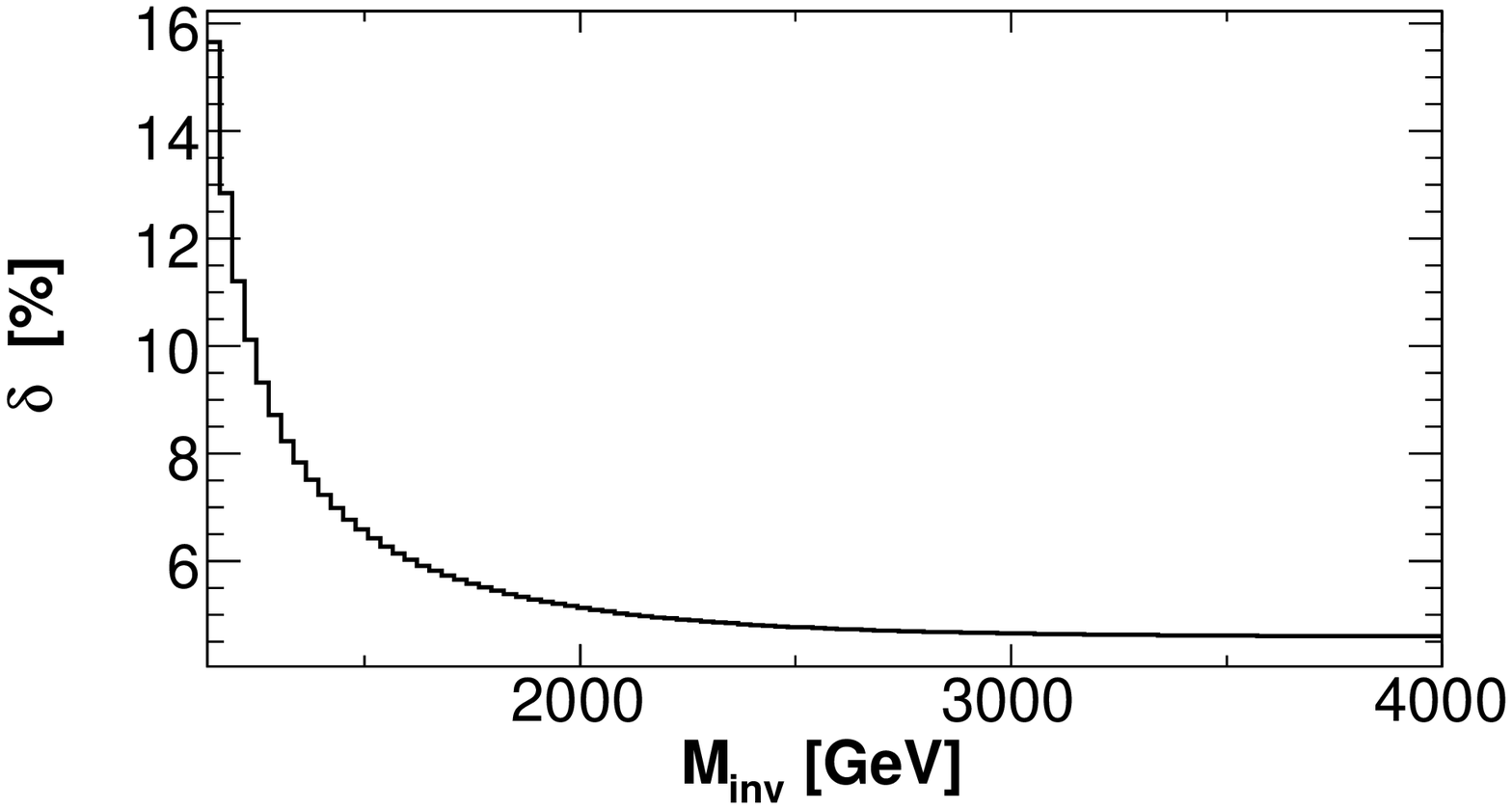, width=7.9cm}
\caption{Cumulative invariant mass distribution for different species of squark pairs,
defined as the cross section 
integrated up to $M_{\mbox{\tiny inv}}$ of the invariant mass of the squark-antisquark pair. 
The left panels show the relative contributions from the various channels,
the right ones show the complete EW contribution.
The SUSY parameter point corresponds to SPS1a$'$.}
\label{Fig:IC_SQ}
\end{figure}

%%%%%%%%%%%%%%%%%%%

\begin{figure}
\centering
\underline{$\tilde{u}^{R}$}\\
\epsfig{file= 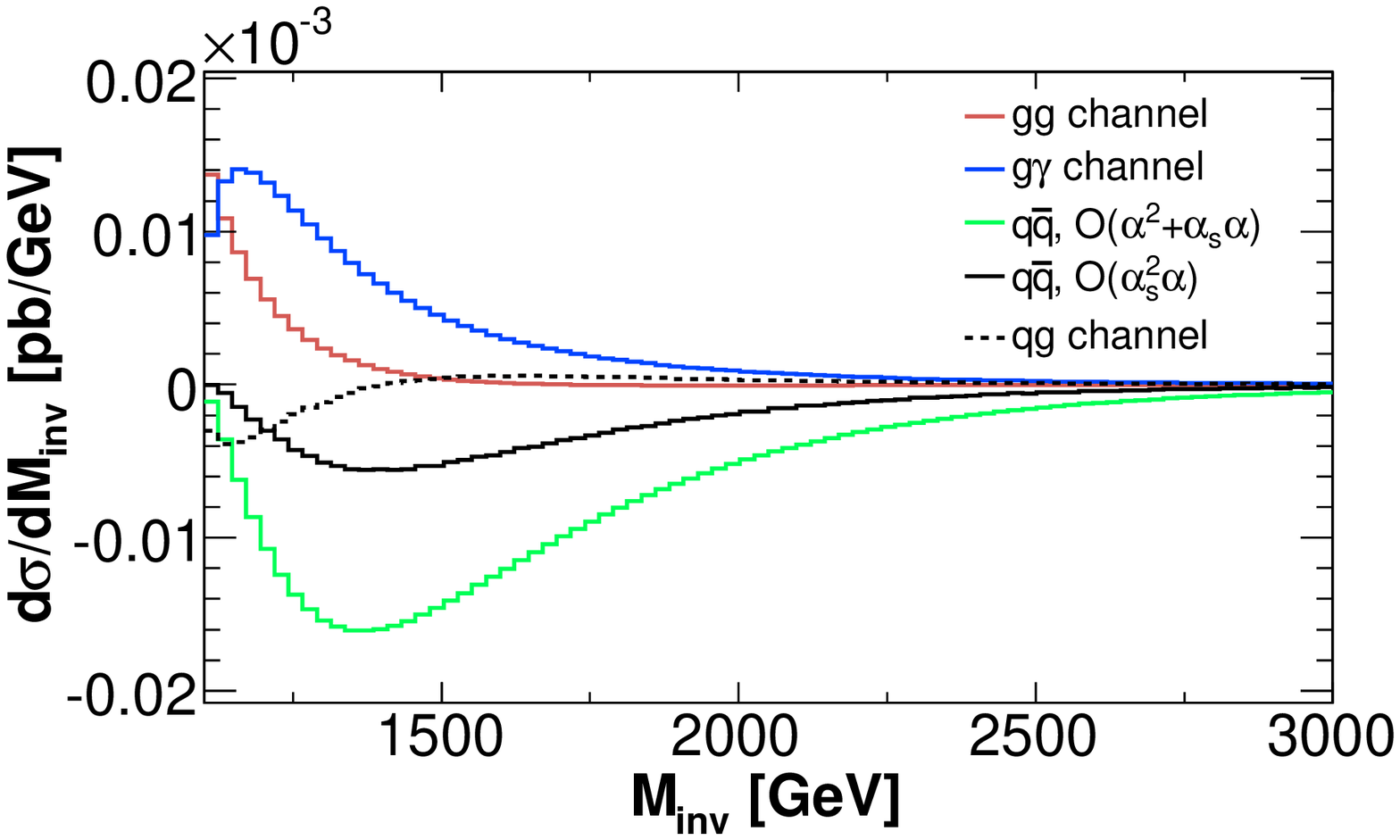, width=7.9cm}
\epsfig{file= 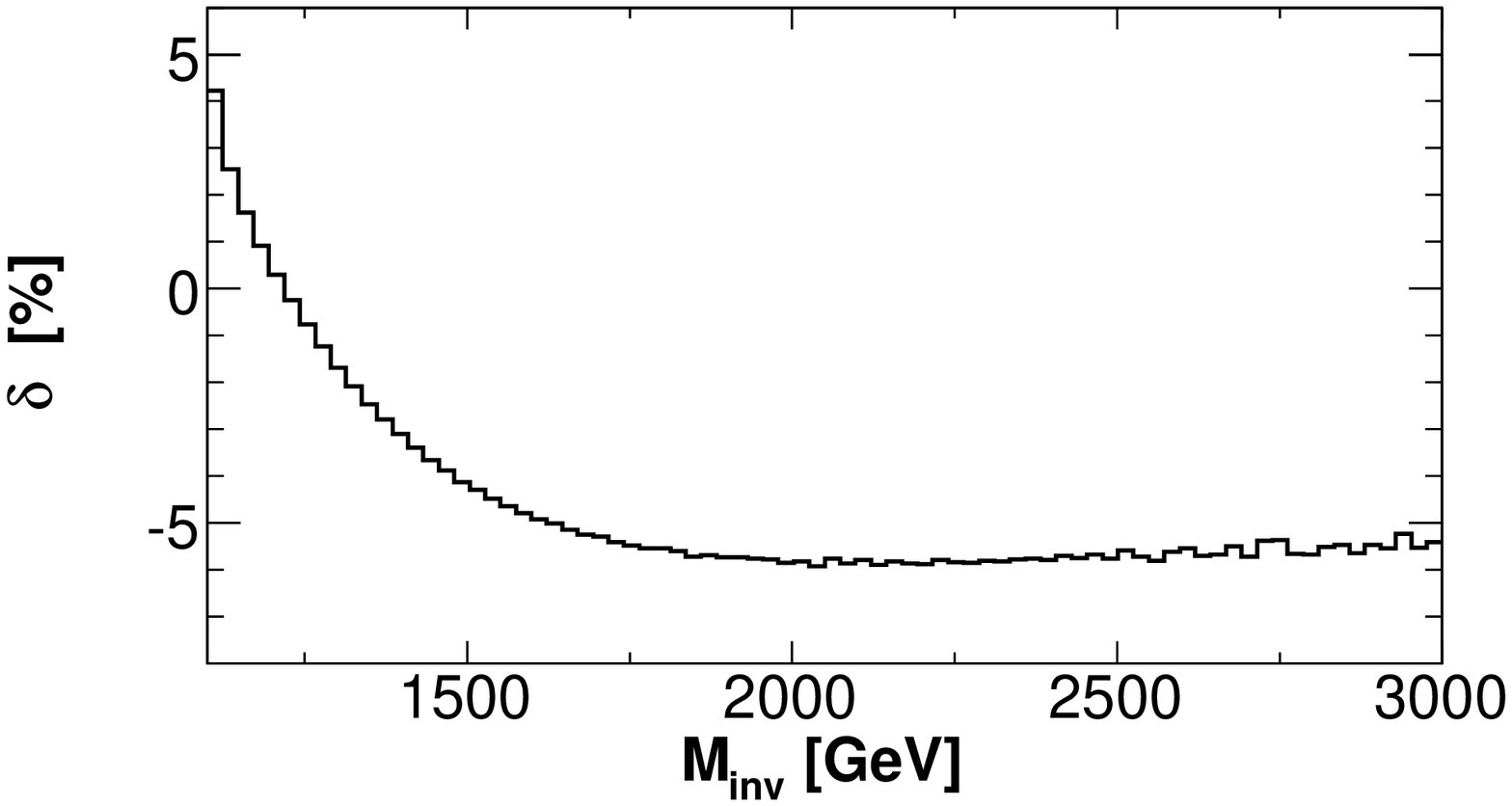, width=7.9cm}
\underline{$\tilde{u}^{L}$}\\
\epsfig{file= 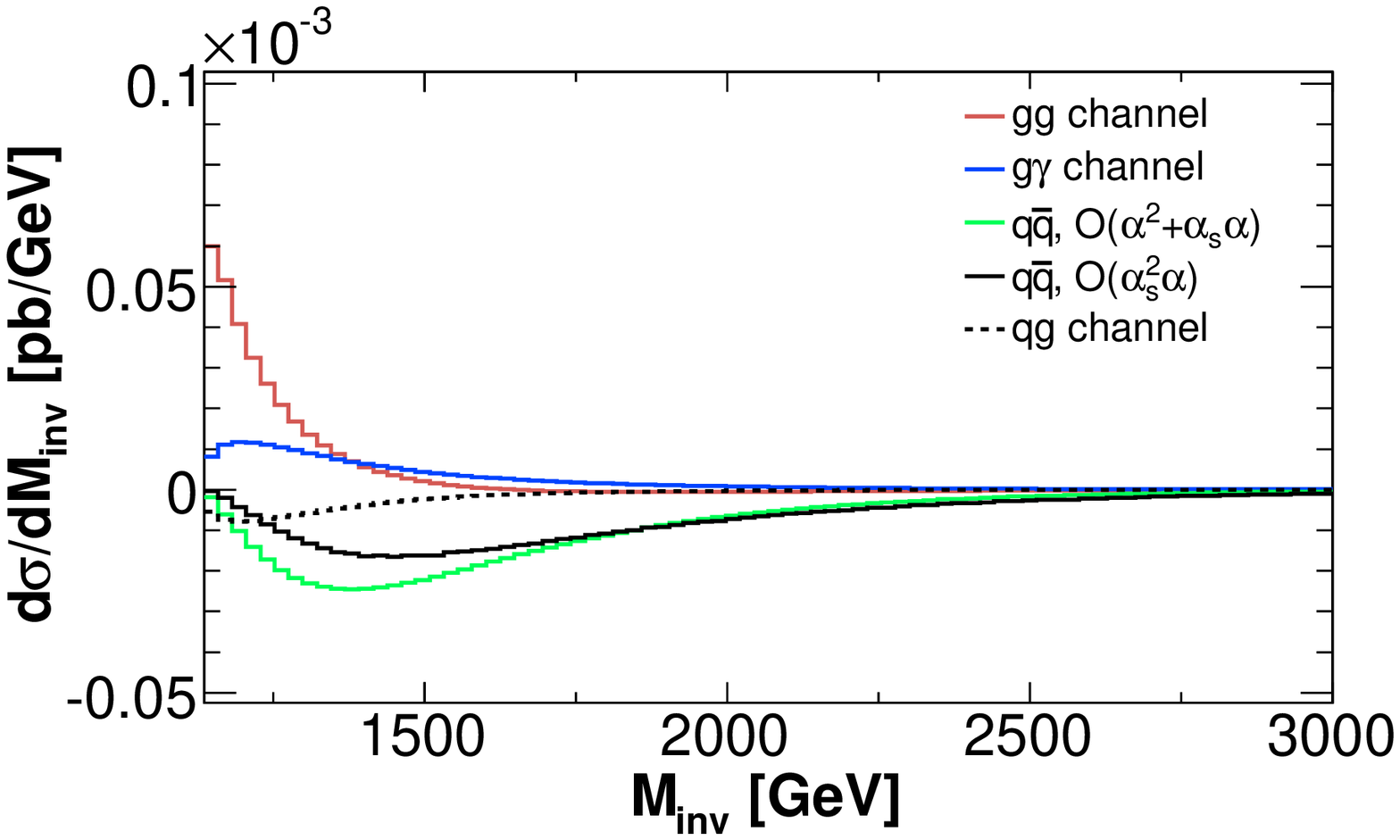, width=7.9cm}
\epsfig{file= 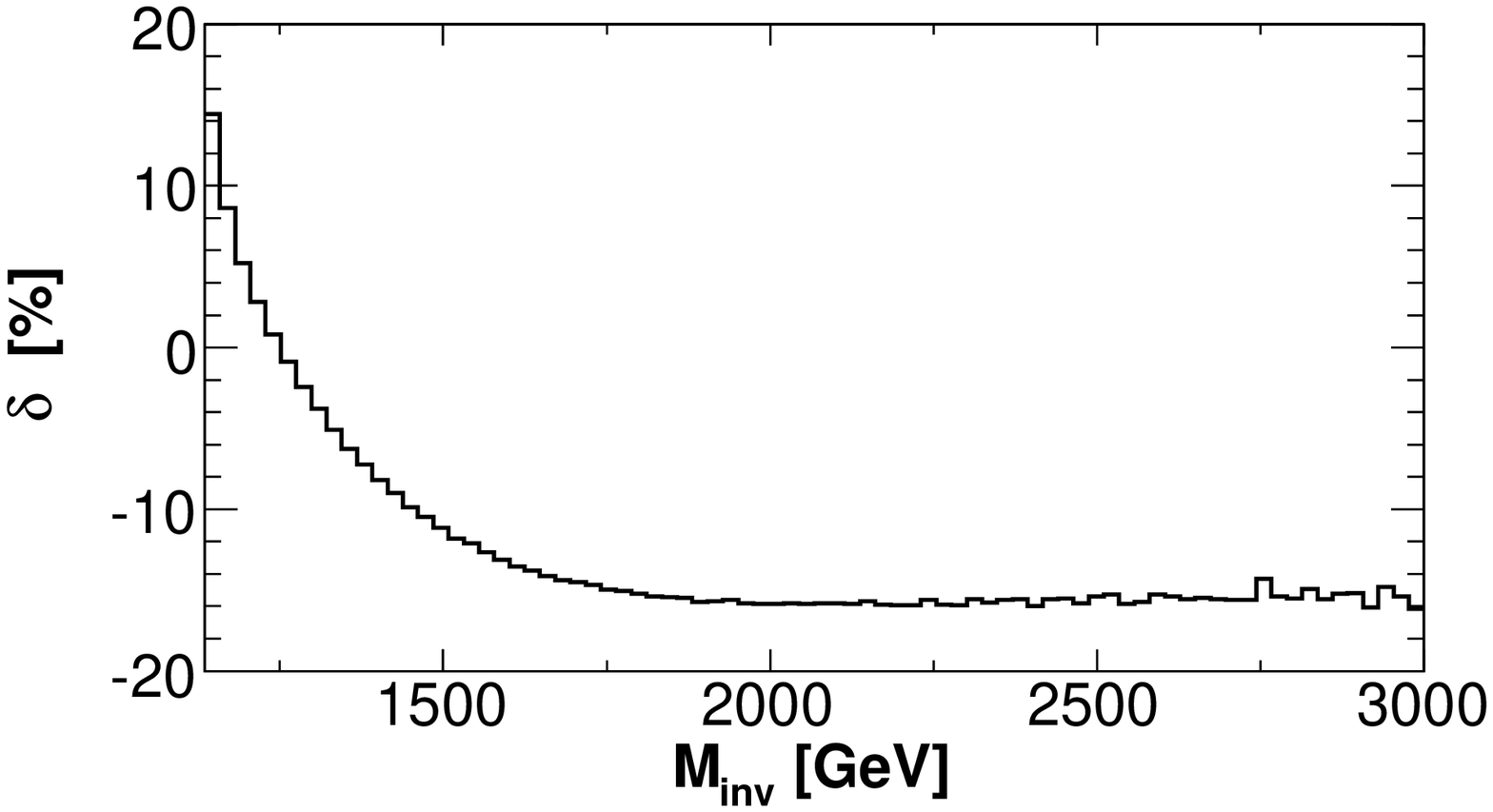, width=7.9cm}
\underline{$\tilde{d}^{L}$}\\
\epsfig{file= 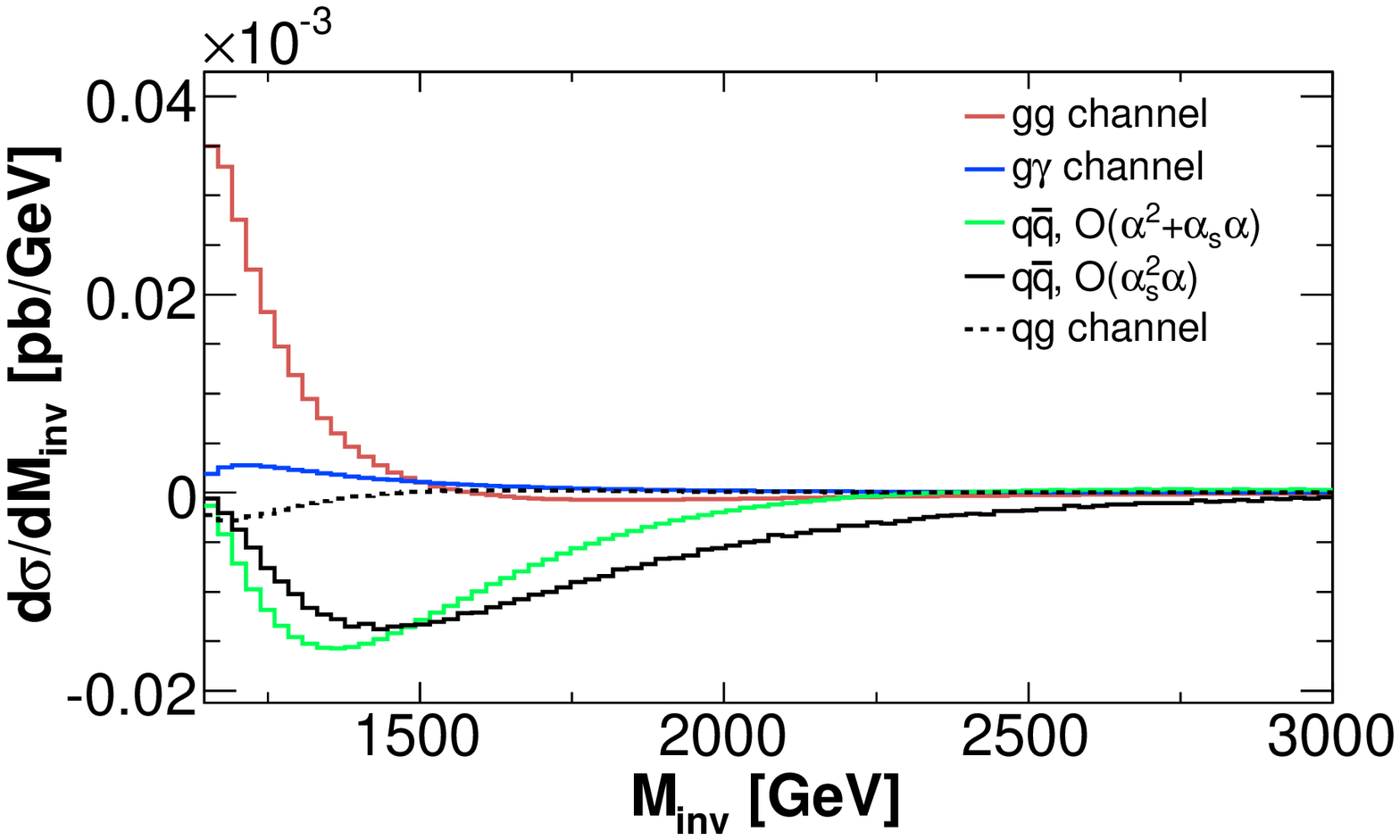, width=7.9cm}
\epsfig{file= 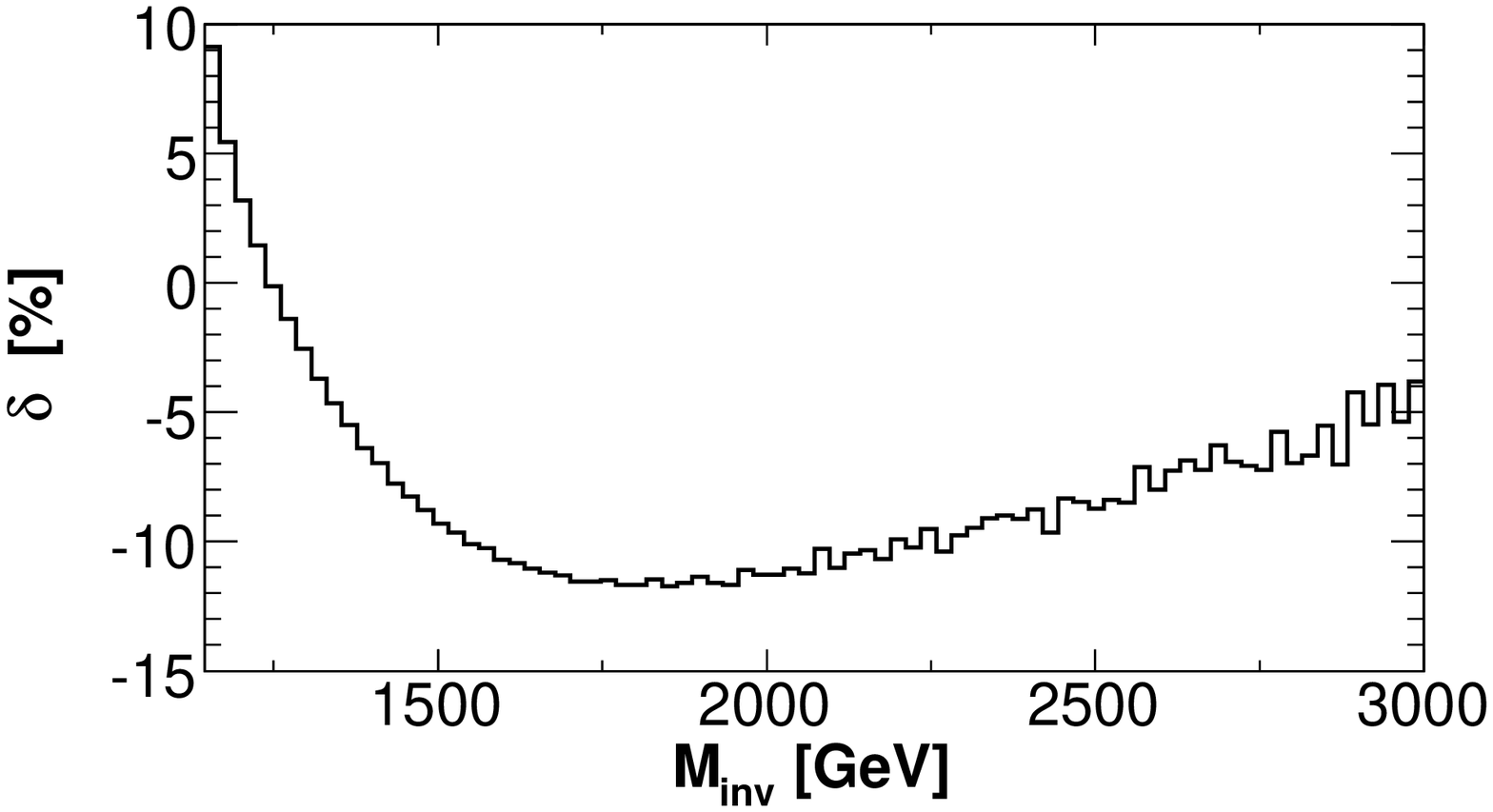, width=7.9cm}
\underline{$\tilde{c}^{L}$}\\
\epsfig{file= 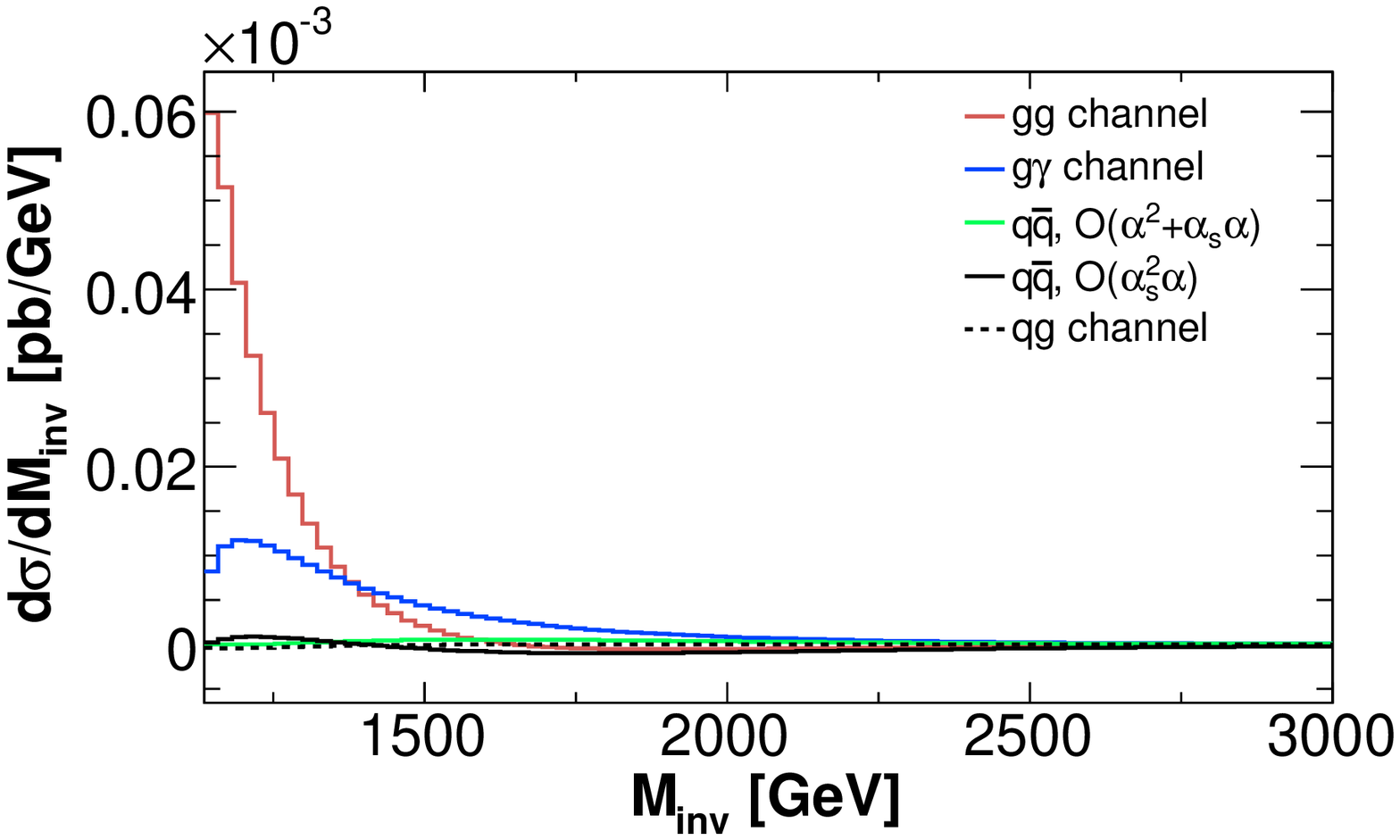, width=7.9cm}
\epsfig{file= 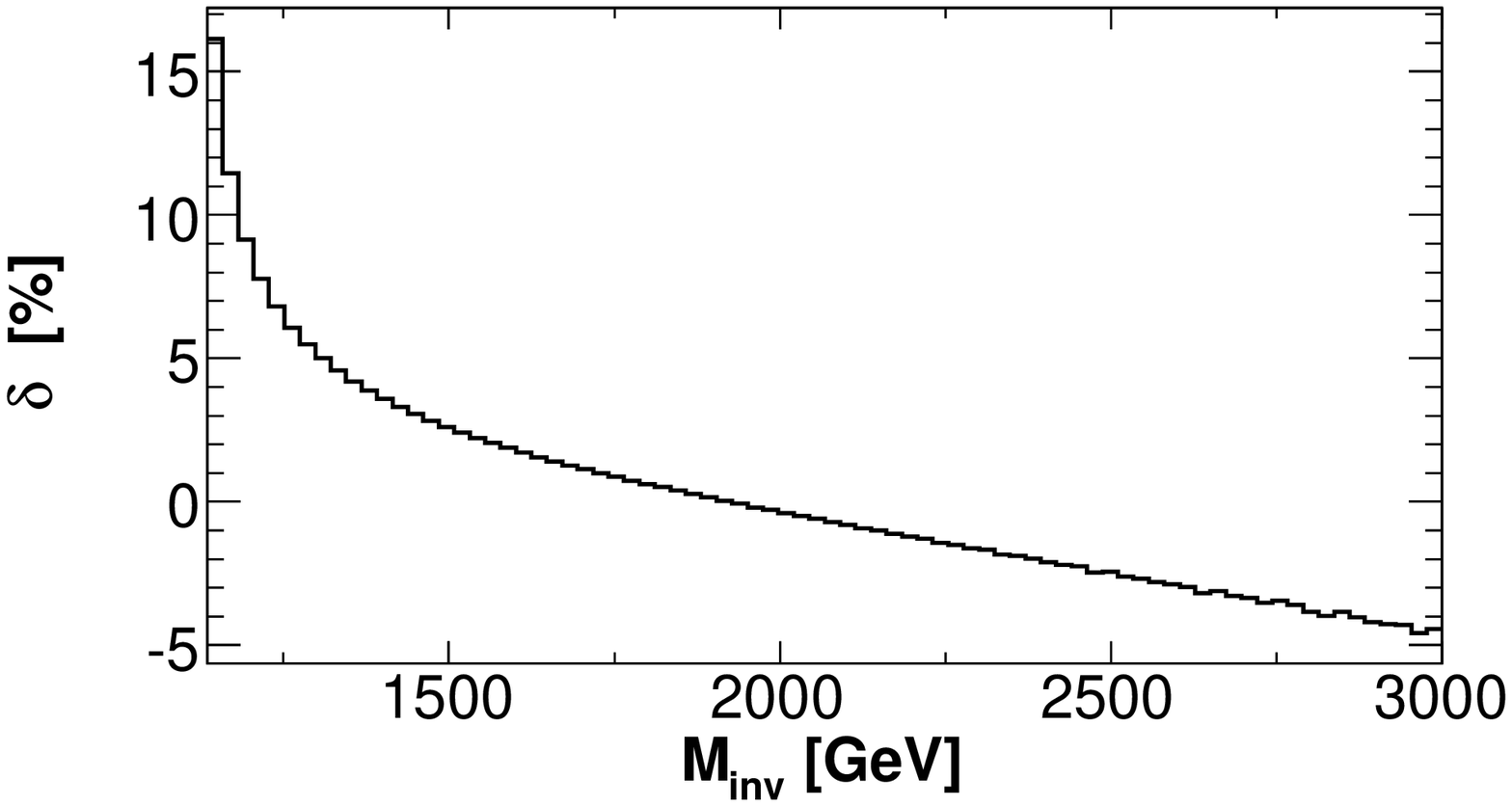, width=7.9cm}
\caption{Invariant mass distribution for different species of squark pairs, 
for the SUSY parameter point corresponding to SPS1a$'$.
The left panels show the relative contributions from the various channels,
the right ones show the complete EW contribution.} 
\label{Fig:IM_SQ}
\end{figure}

%%%%%%%%%%%%%%%%%

\begin{figure}
\centering
\underline{$\tilde{u}^{R}$}\\
\epsfig{file= 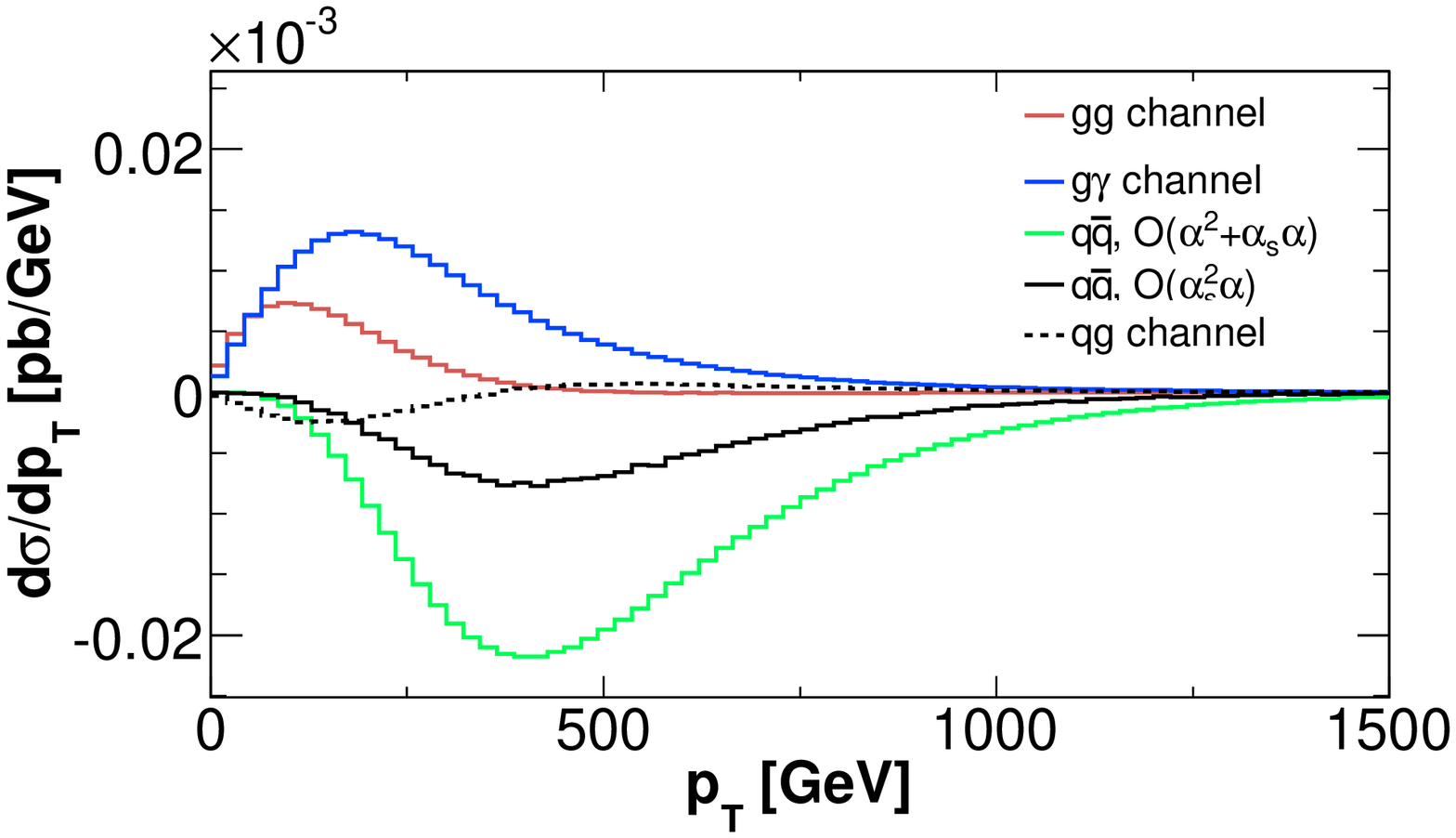, width=7.9cm}
\epsfig{file= 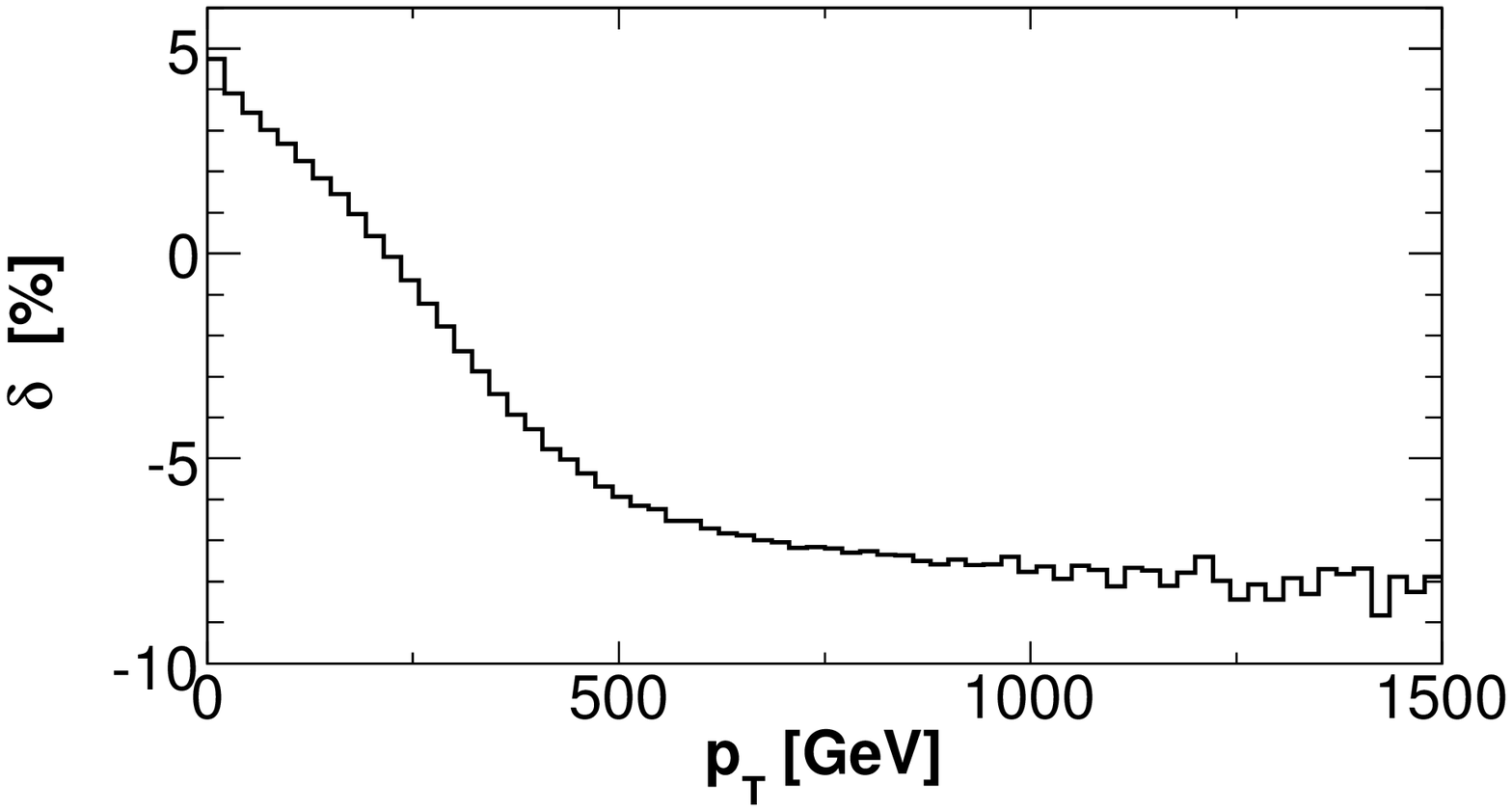, width=7.9cm}
\underline{$\tilde{u}^{L}$}\\
\epsfig{file= 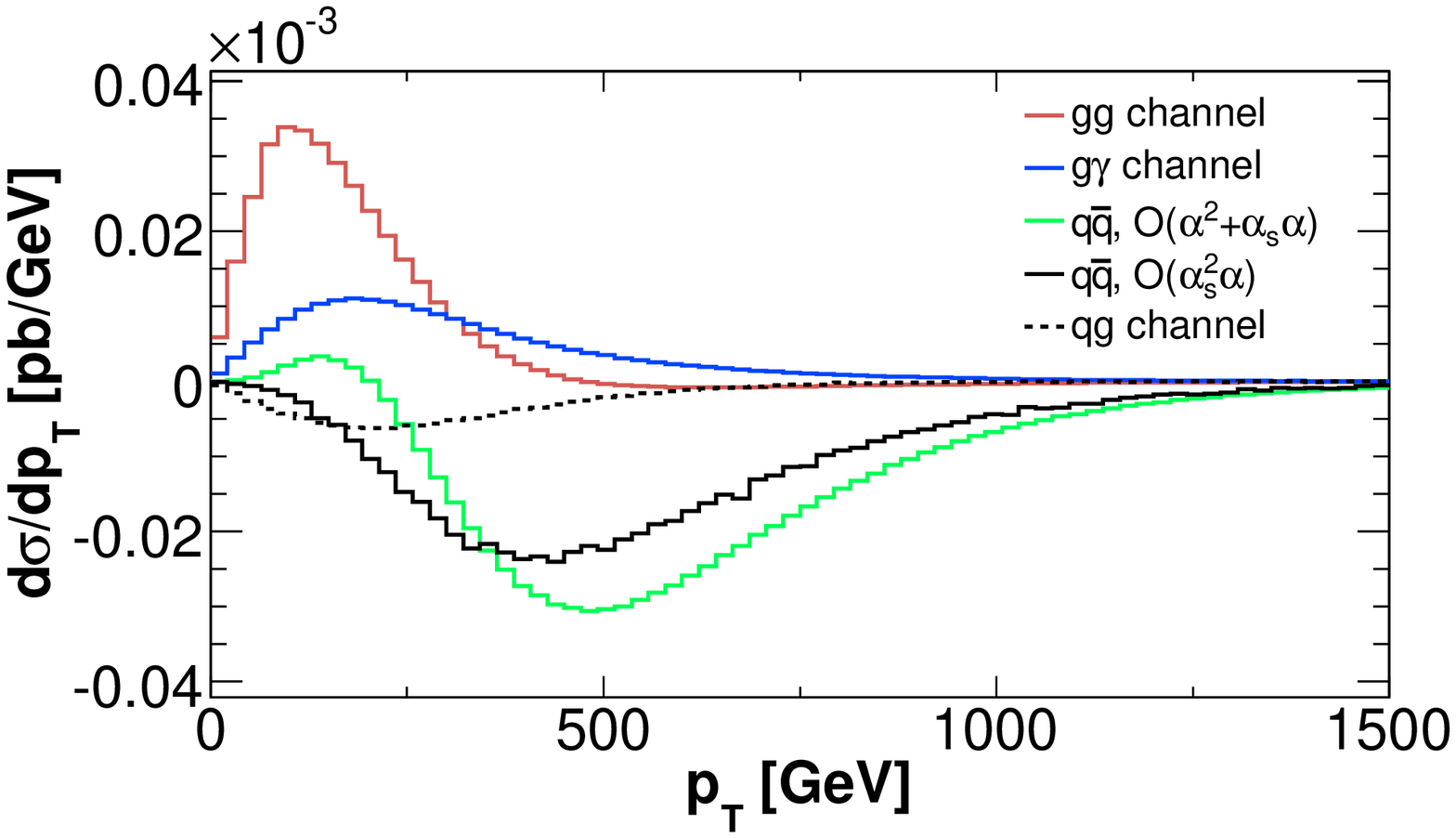, width=7.9cm}
\epsfig{file= 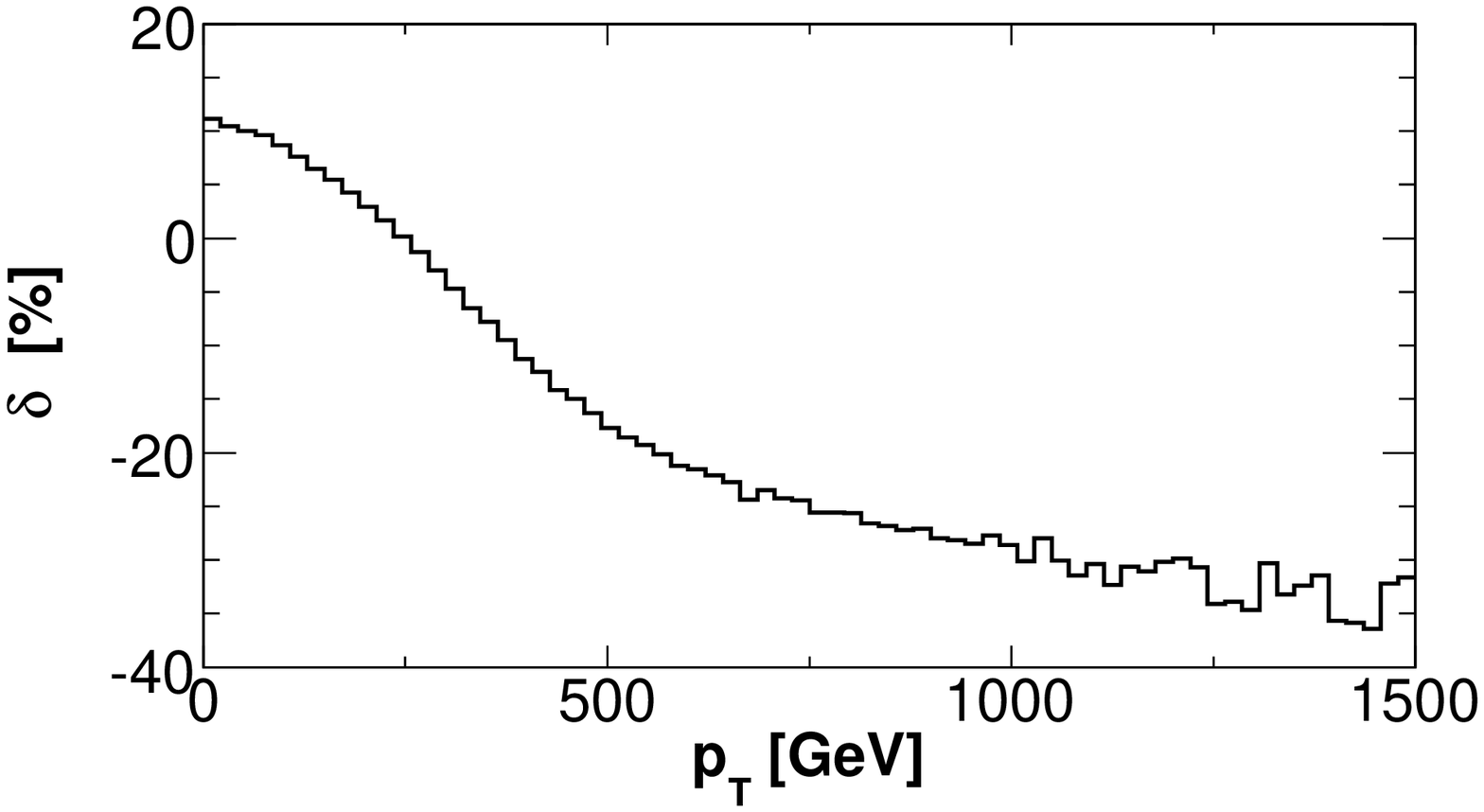, width=7.9cm}
\underline{$\tilde{d}^{L}$}\\
\epsfig{file= 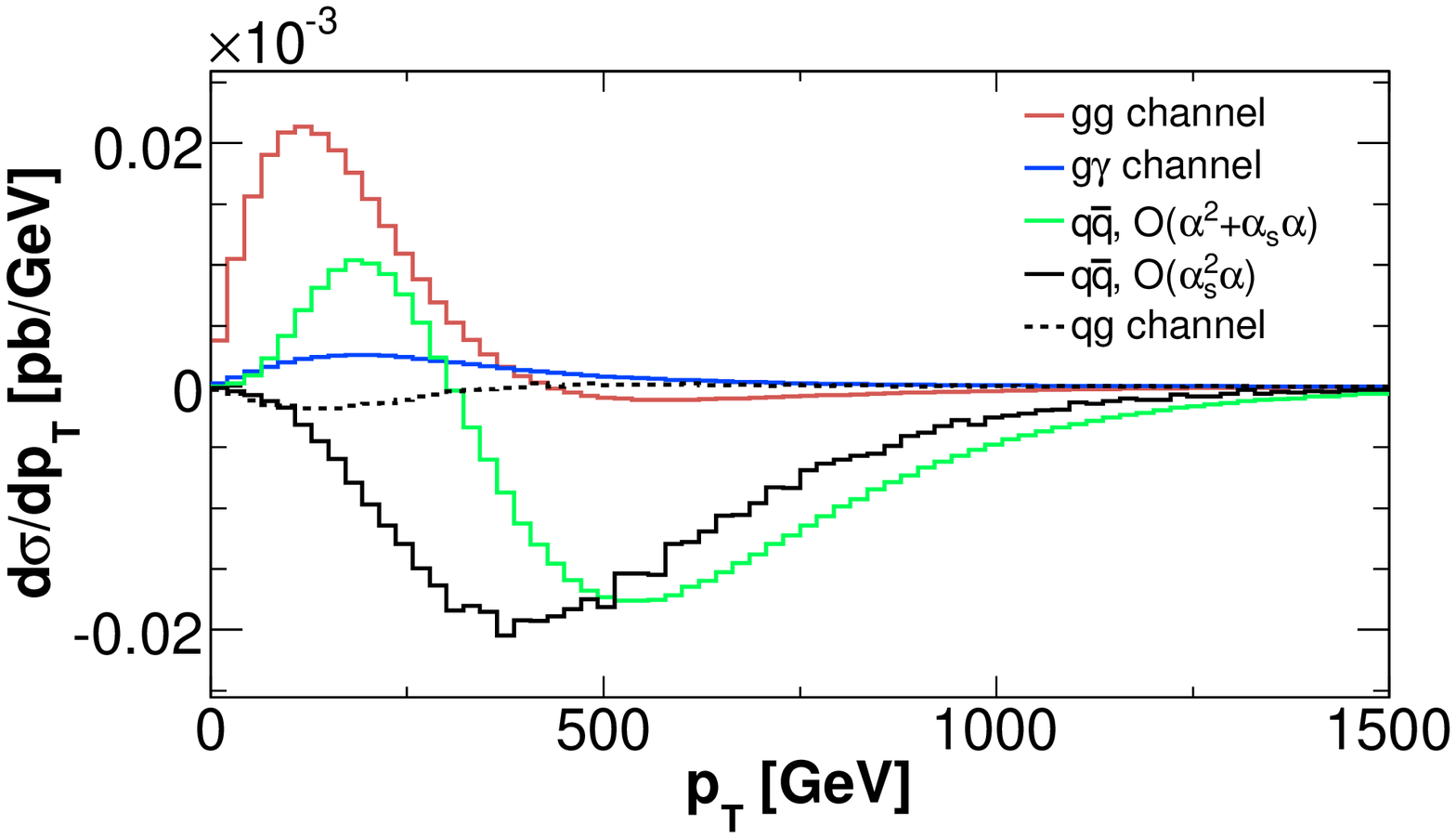, width=7.9cm}
\epsfig{file= 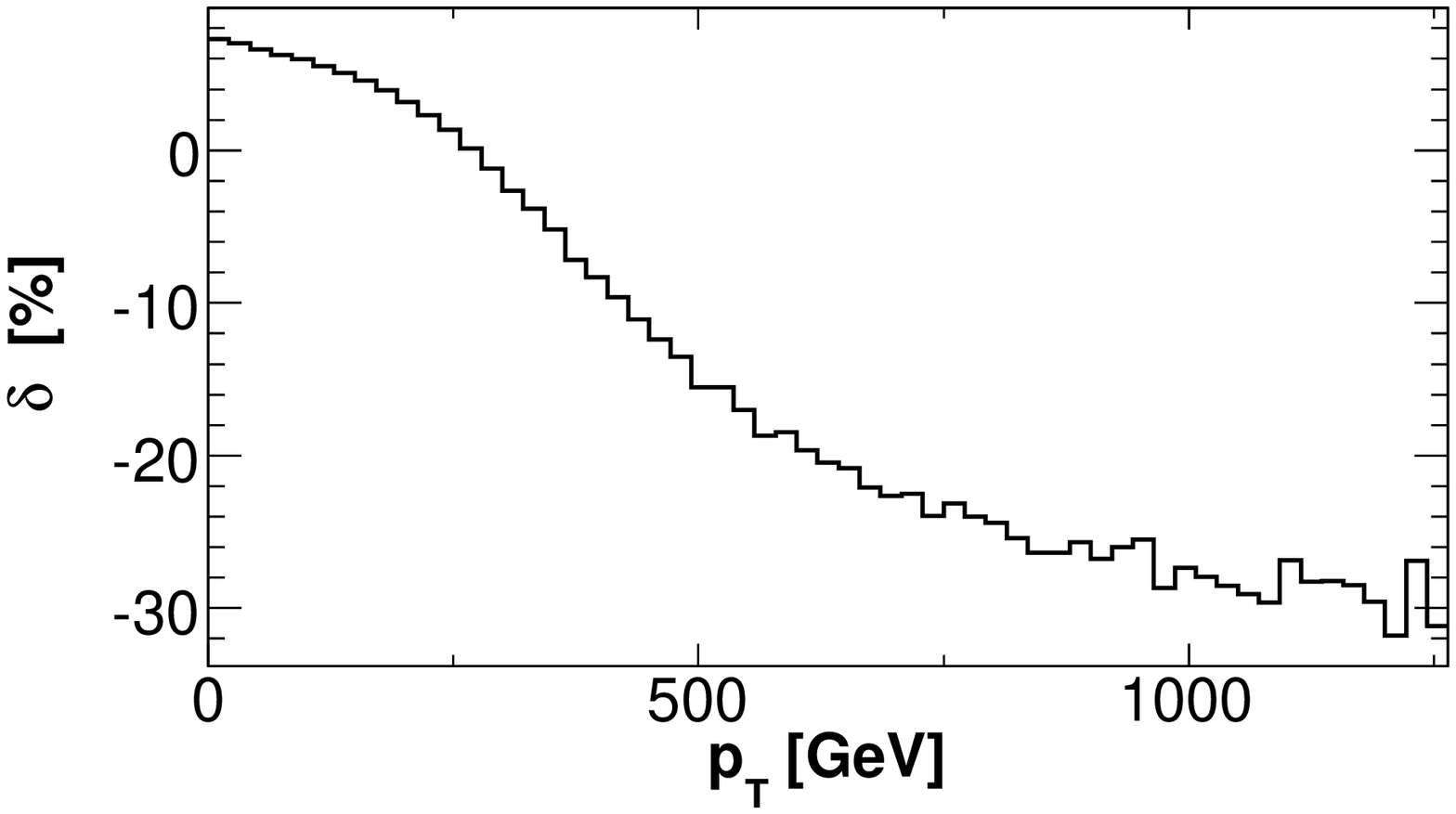, width=7.9cm}
\underline{$\tilde{c}^{L}$}\\
\epsfig{file= 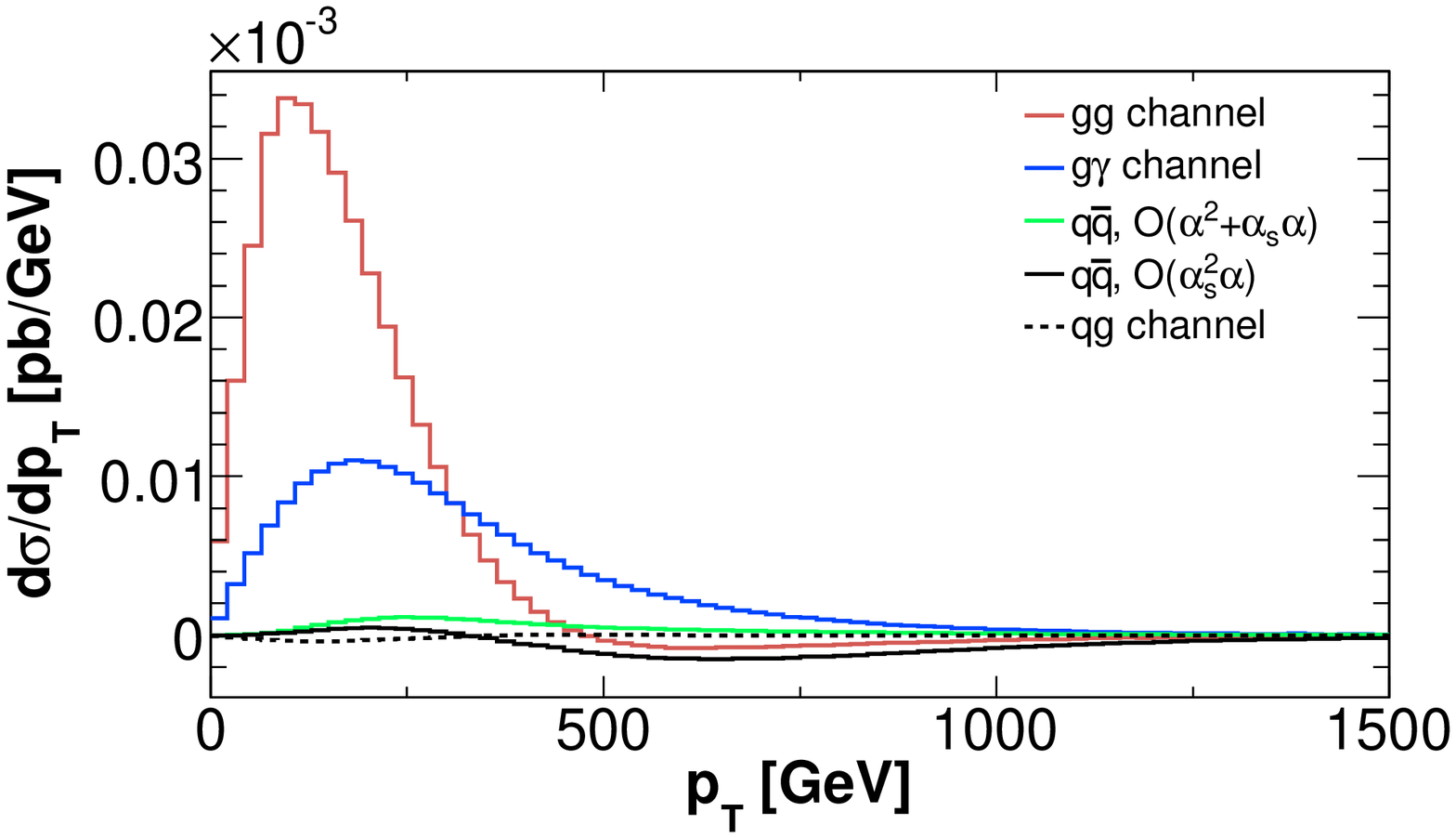, width=7.9cm}
\epsfig{file= 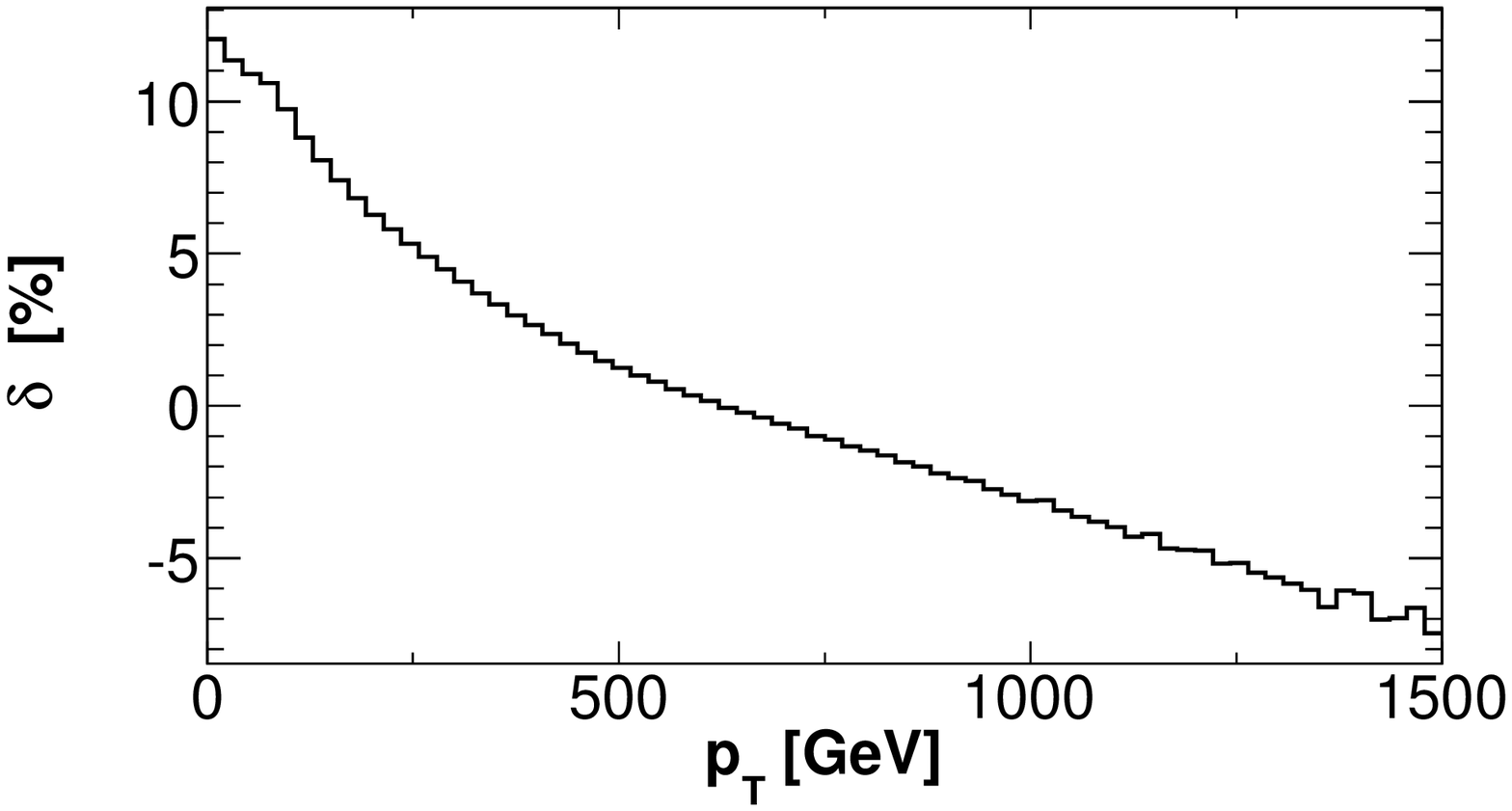, width=7.9cm}
\caption{Transverse momentum distribution for different species of squark pairs.
Notations and input parameters as in \ref{Fig:IM_SQ}.}
\label{Fig:PT_SQ}
\end{figure}

%%%%%%%%%%%%%%%%%%%

%\begin{figure}
%\centering
%
%\underline{$\tilde{u}^{R}$}\\
%\epsfig{file= PLOT/UP1/SPS1a_YY_Ch.eps, width=7.9cm}
%\epsfig{file= PLOT/UP1/SPS1a_YY_Lo.eps, width=7.9cm}
%
%\underline{$\tilde{u}^{L}$}\\
%\epsfig{file= PLOT/UP2/SPS1a_YY_Ch.eps, width=7.9cm}
%\epsfig{file= PLOT/UP2/SPS1a_YY_Lo.eps, width=7.9cm}
%
%\underline{$\tilde{d}^{L}$}\\
%\epsfig{file= PLOT/DO2/SPS1a_YY_Ch.eps, width=7.9cm}
%\epsfig{file= PLOT/DO2/SPS1a_YY_Lo.eps, width=7.9cm}
%
%\underline{$\tilde{c}^{L}$}\\
%\epsfig{file= PLOT/CH2/SPS1a_YY_Ch.eps, width=7.9cm}
%\epsfig{file= PLOT/CH2/SPS1a_YY_Lo.eps, width=7.9cm}
%\caption{Rapidity distribution for different species of squark pairs at LO and NLO 
%(right columns).
%The left panels show the EW contribution from the various channels. 
%The SUSY parameter point corresponds to SPS1a$'$.}
%\label{Fig:YY_SQ}
%\end{figure}

%%%%%%%%%%%%%%%%%%%%%%%%%%%%%%%%%%%%%%%%%%%%%%%%
%%%%% SCAN ON THE MASS OF THE SQUARKS      %%%%%
%%%%%%%%%%%%%%%%%%%%%%%%%%%%%%%%%%%%%%%%%%%%%%%%

\begin{figure}
\centering
\underline{$\tilde{u}^R$}\\
\epsfig{file= 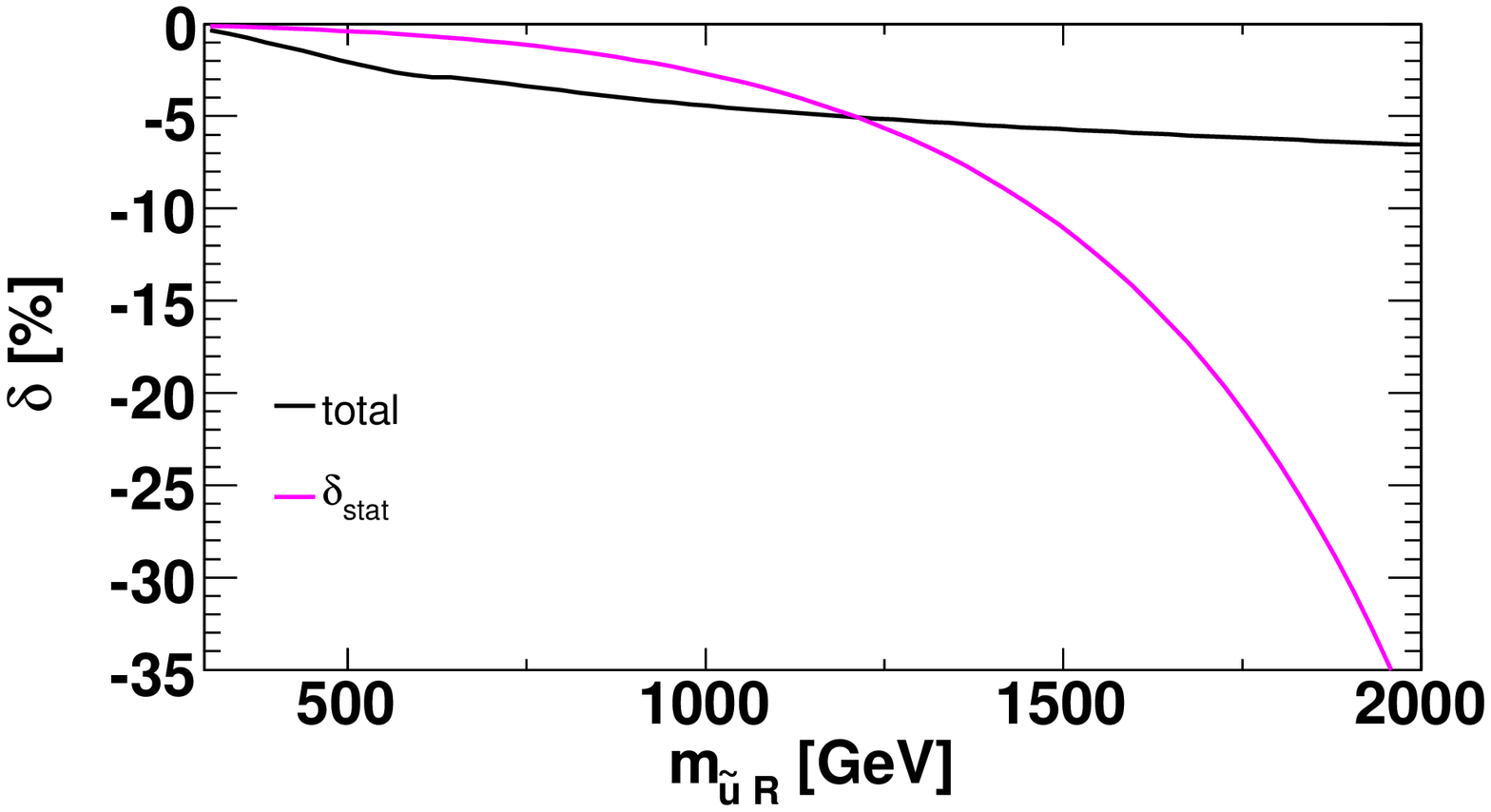, width= 5.2cm}
\epsfig{file= 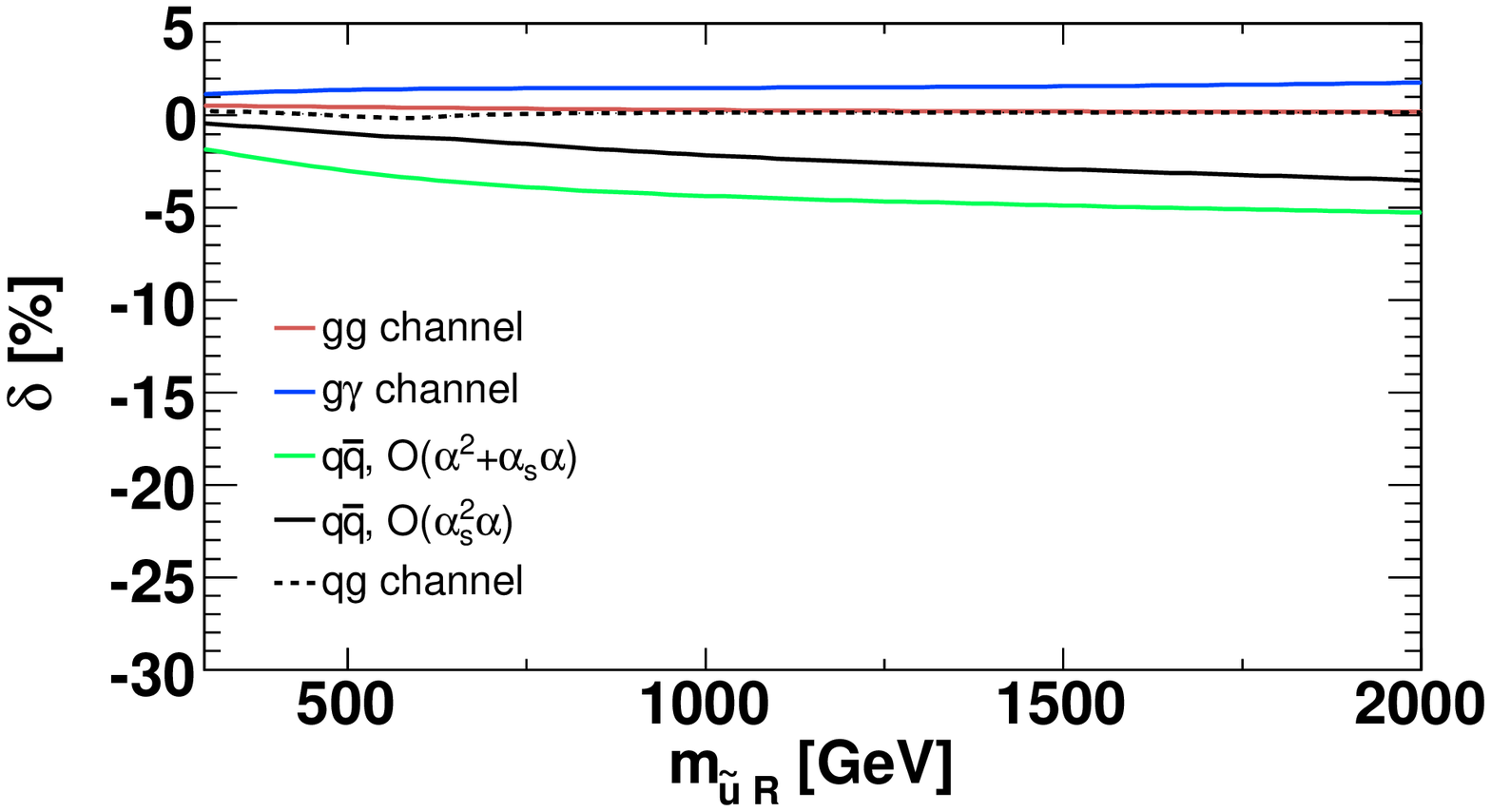, width=5.2cm}
\epsfig{file= 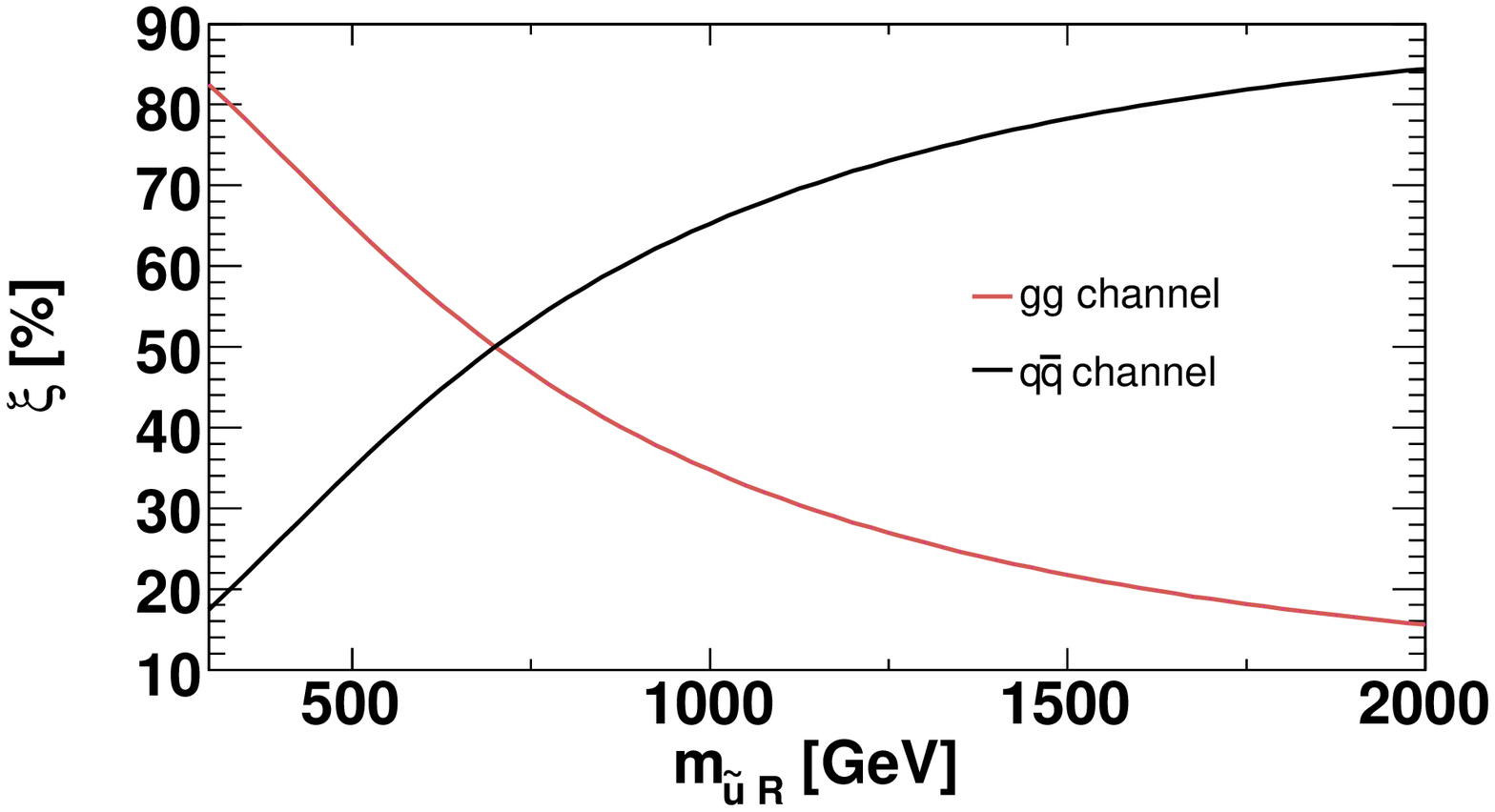, width=5.2cm}
\underline{$\tilde{u}^L$}\\
\epsfig{file= 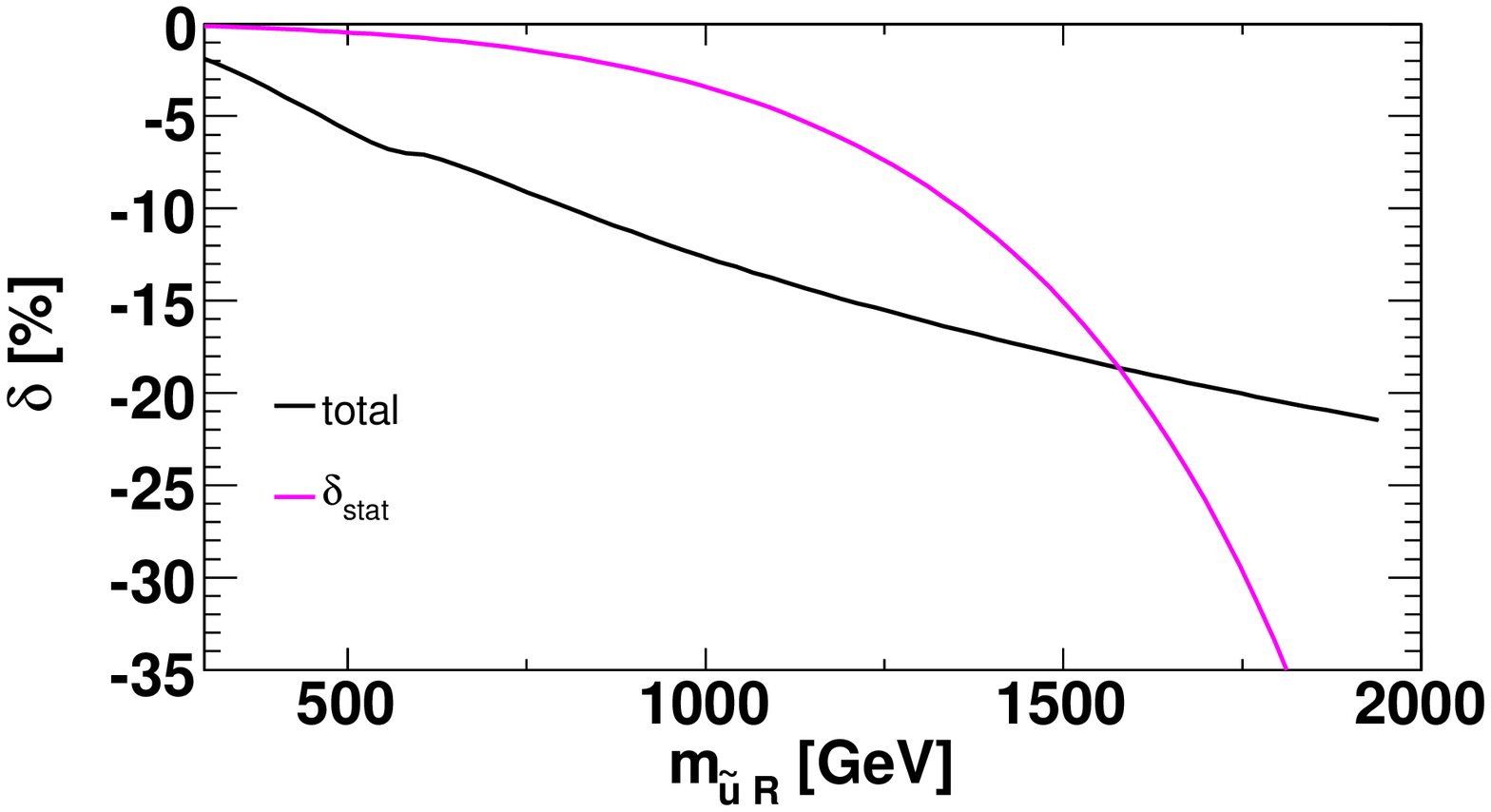, width= 5.2cm}
\epsfig{file= 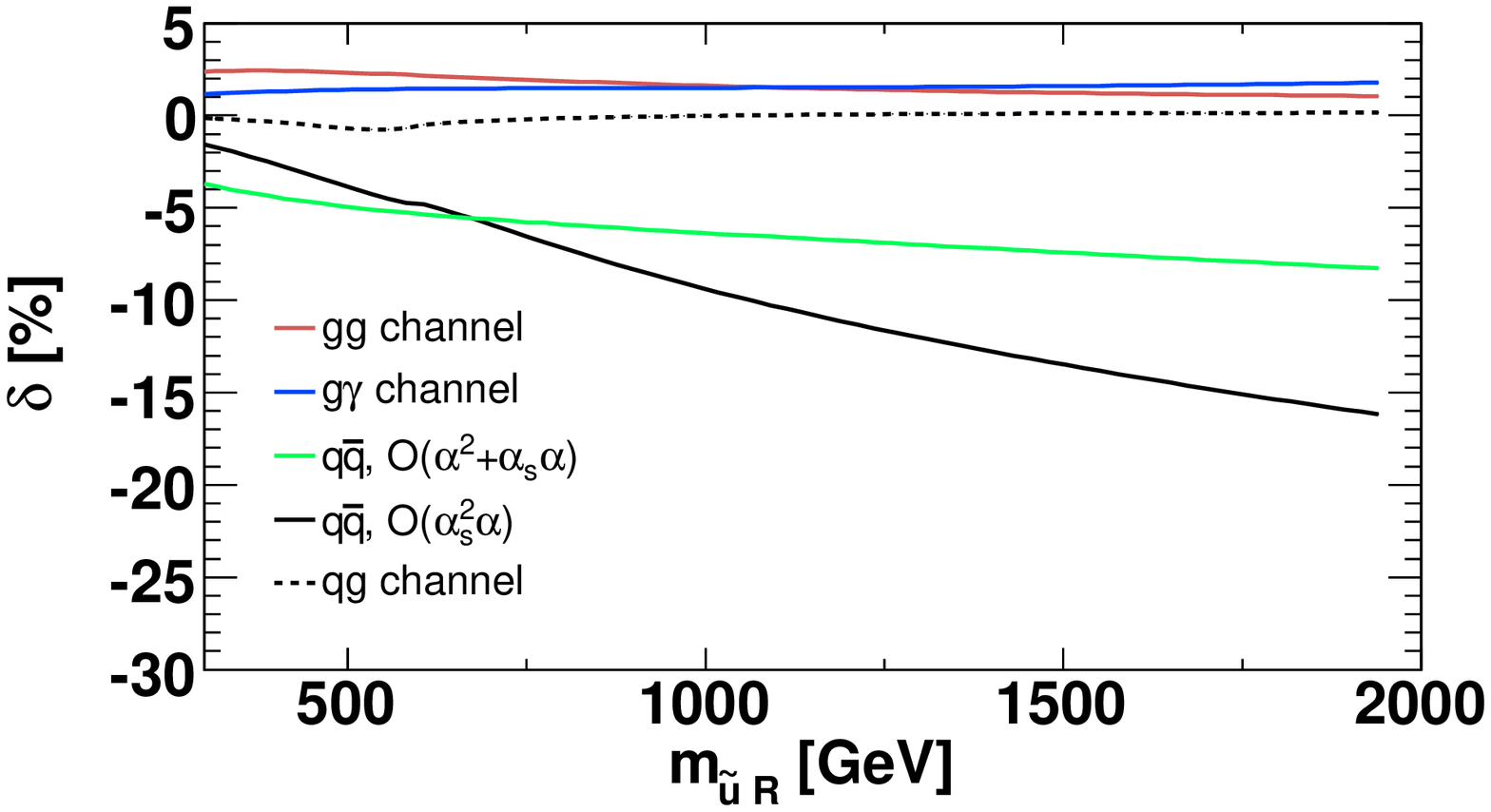, width=5.2cm}
\epsfig{file= 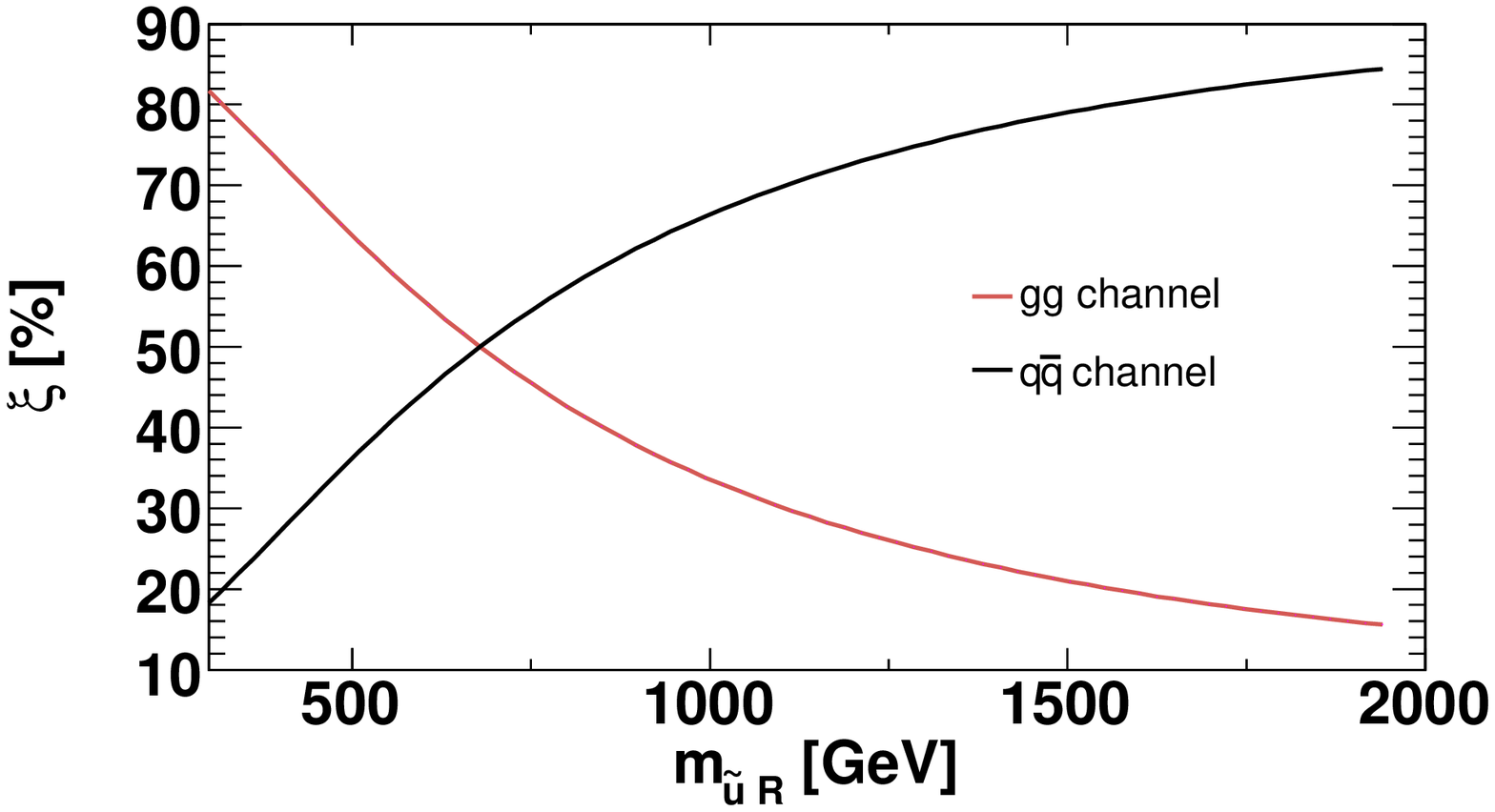, width=5.2cm}
\underline{$\tilde{d}^L$}\\
\epsfig{file= 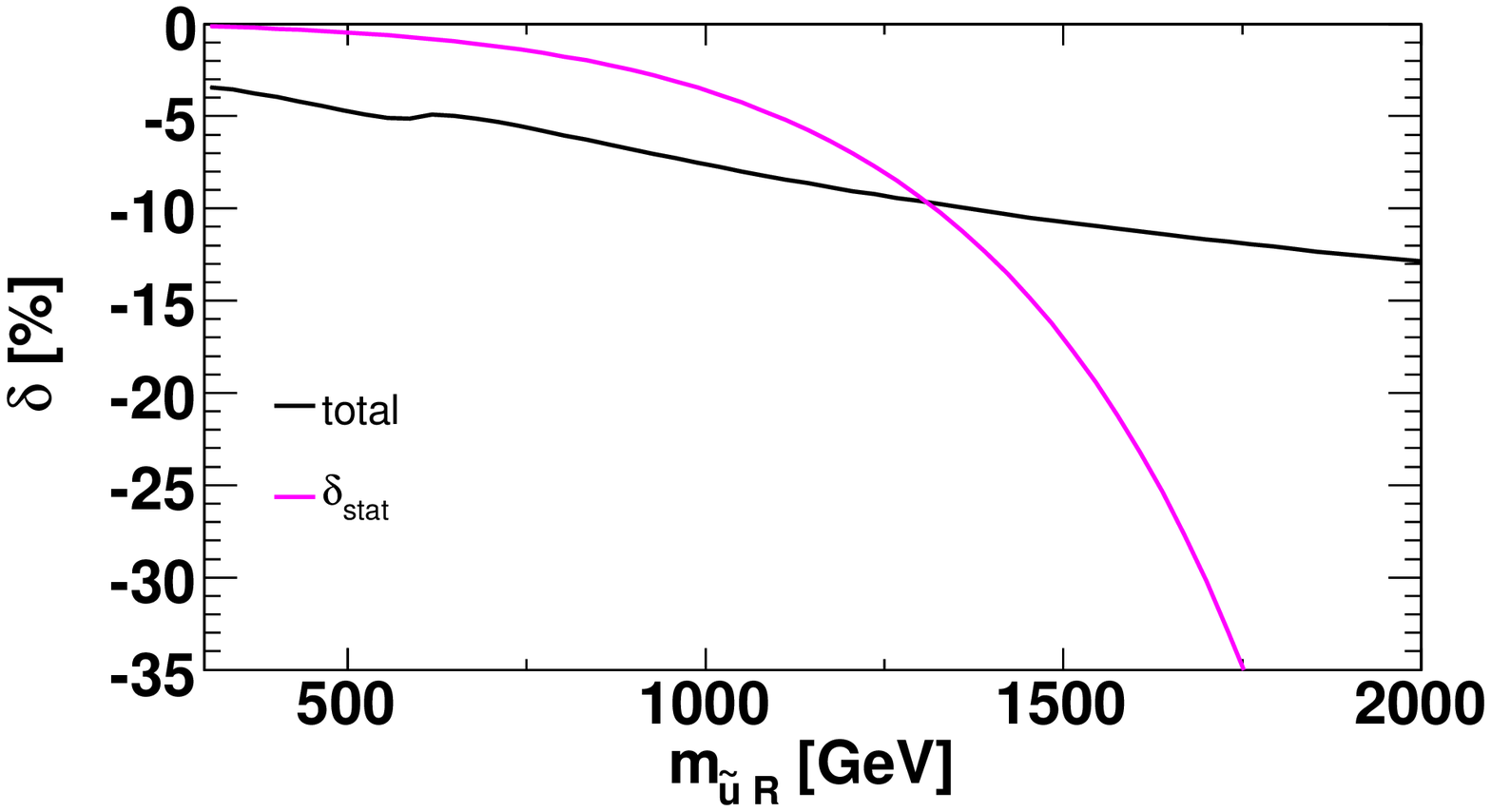, width= 5.2cm}
\epsfig{file= 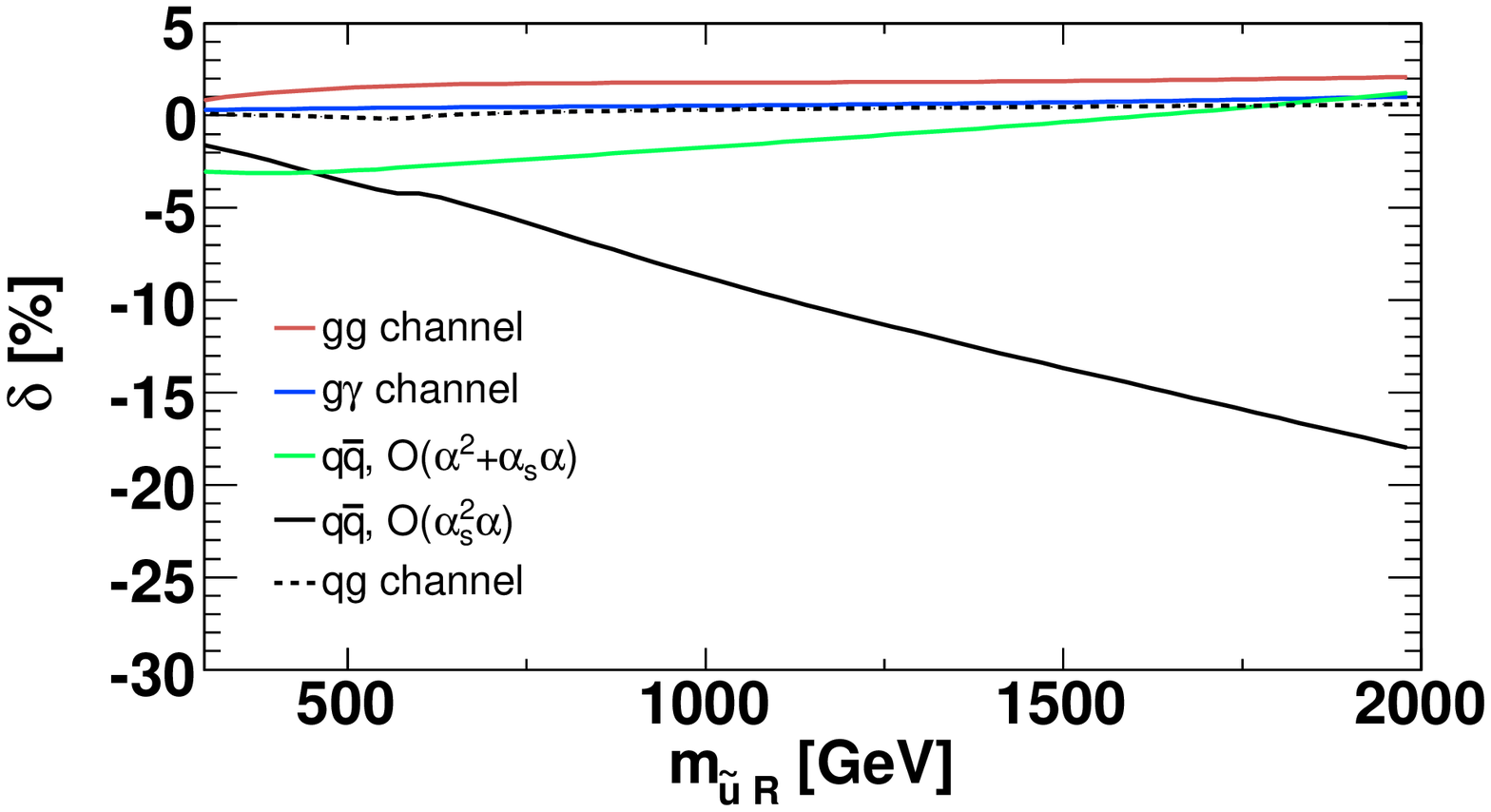, width=5.2cm}
\epsfig{file= 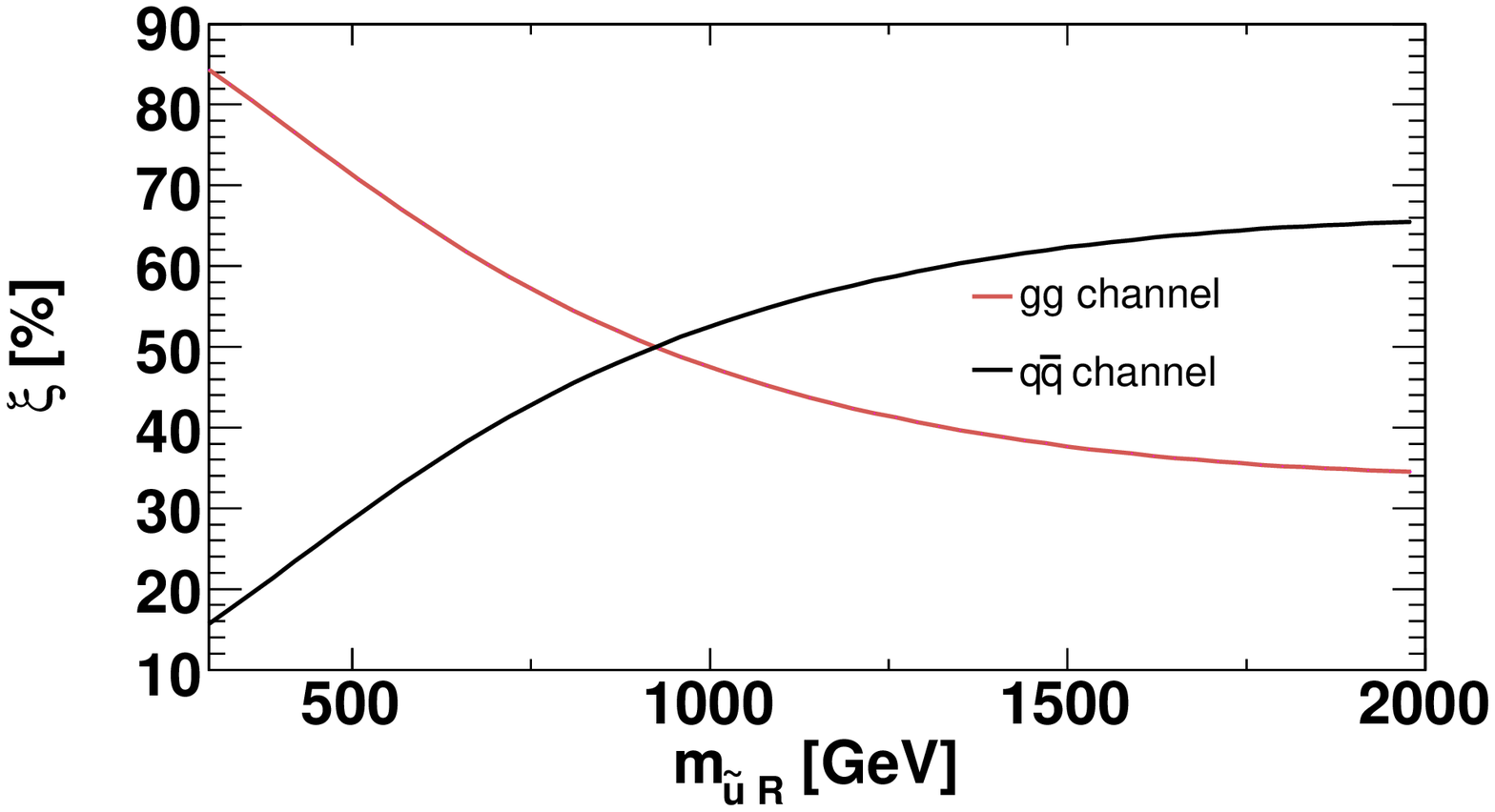, width=5.2cm}
\underline{$\tilde{c}^L$}\\
\epsfig{file= 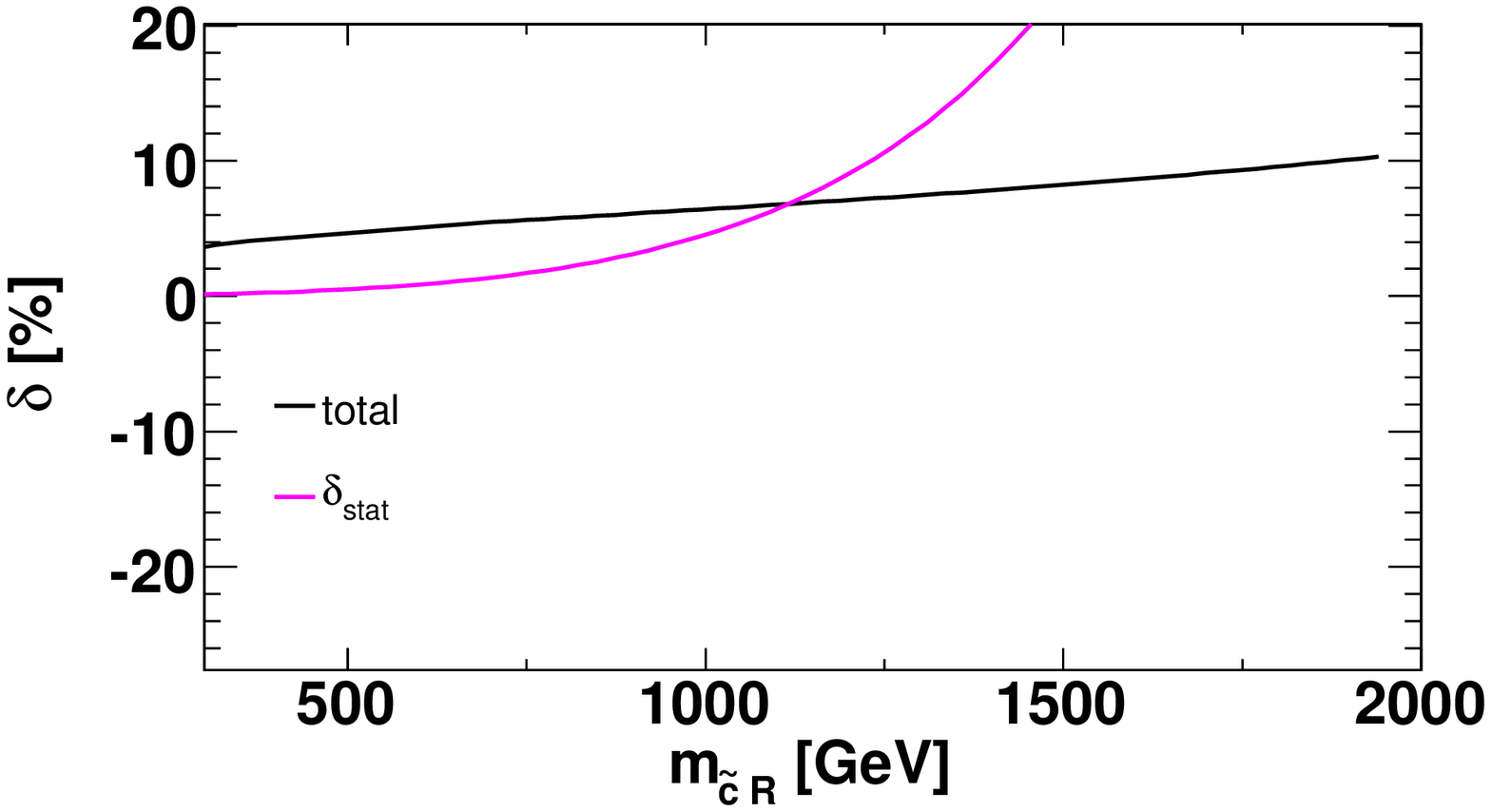, width= 5.2cm}
\epsfig{file= 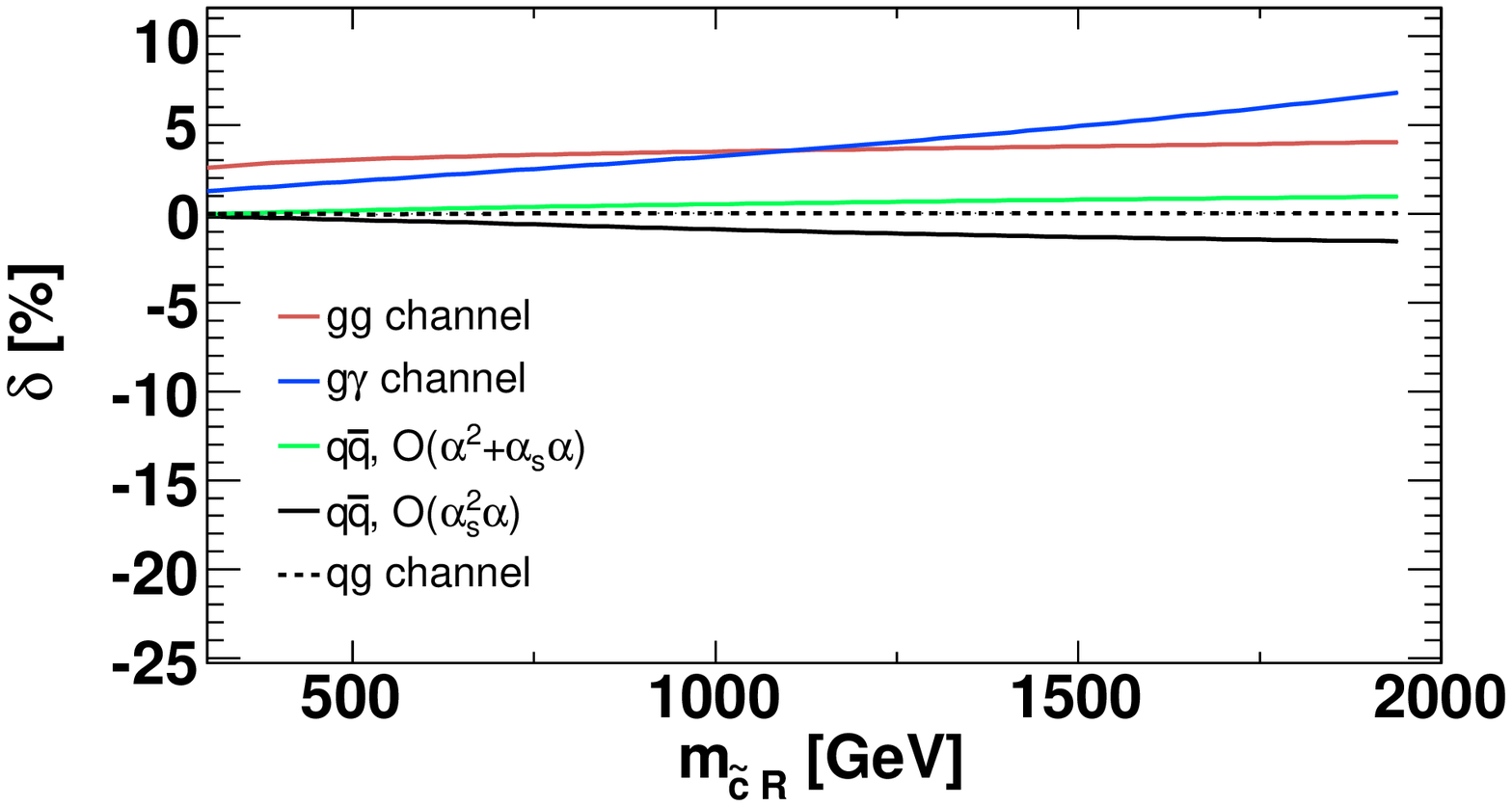, width=5.2cm}
\epsfig{file= 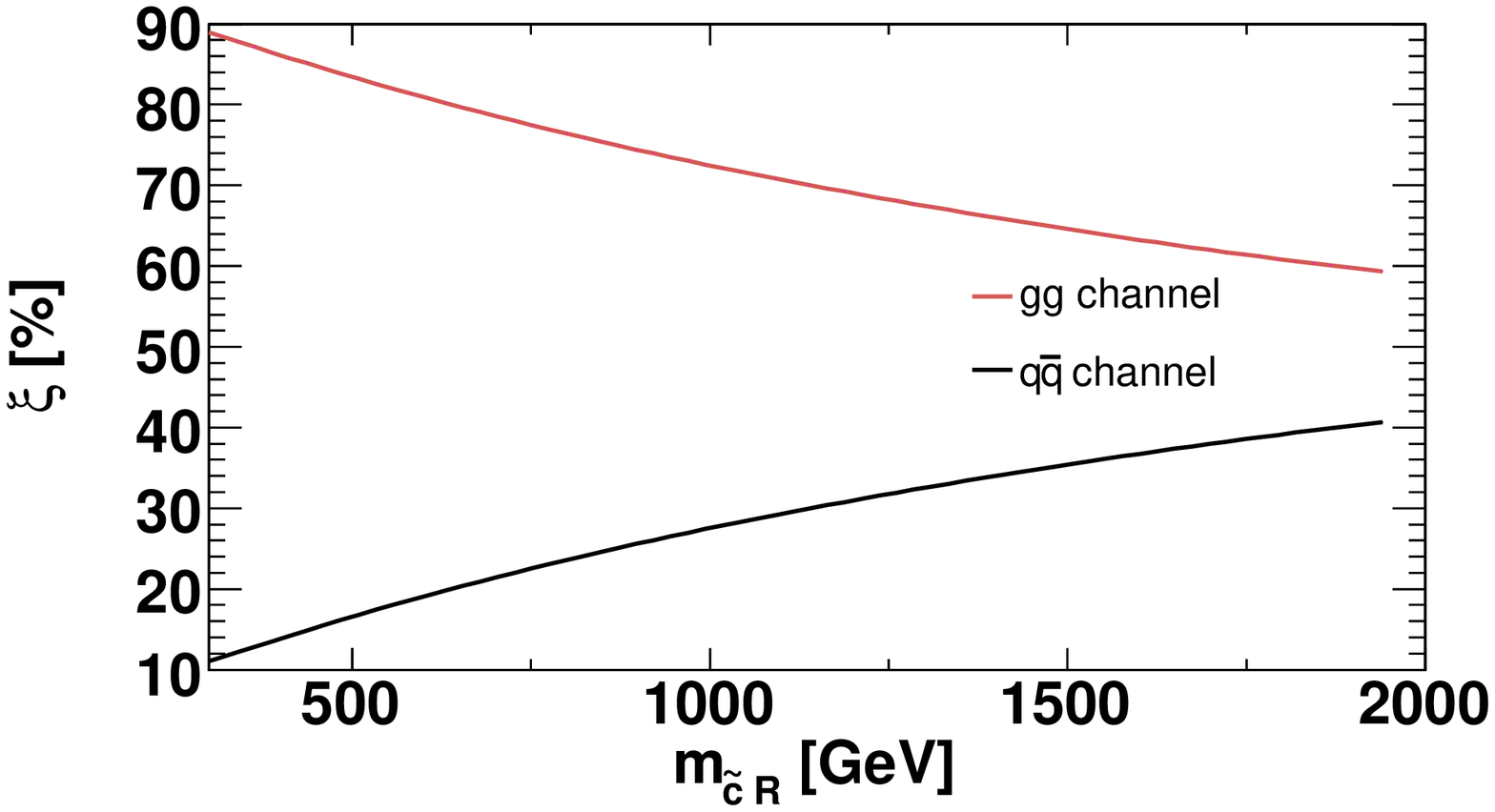, width=5.2cm}
\caption{Squark-mass dependence of the EW contributions.
Total EW contribution (left), individual contributions from the various channels (central).
The panels in the right column show the relative yield of the two channels 
that contribute at LO.}
\label{Fig:Scan2}
\end{figure}

%%%%%%%%%%%%%%%%%%%%%%%%%%%%%%%%%%%%%%%%%%%%%%%%
%%%%% STUDY OF DIFFERENT  BENCHMARK POINTS %%%%%
%%%%%%%%%%%%%%%%%%%%%%%%%%%%%%%%%%%%%%%%%%%%%%%%

\newpage

\begin{figure}
\centering
\underline{SPS1a$'$}\\
\epsfig{file= PLOT/UP2/SPS1a_IC_Ch.eps, width=7.9cm}
\epsfig{file= PLOT/UP2/SPS1a_IC_Per.eps, width=7.9cm}
\underline{SPS5}\\
\epsfig{file= 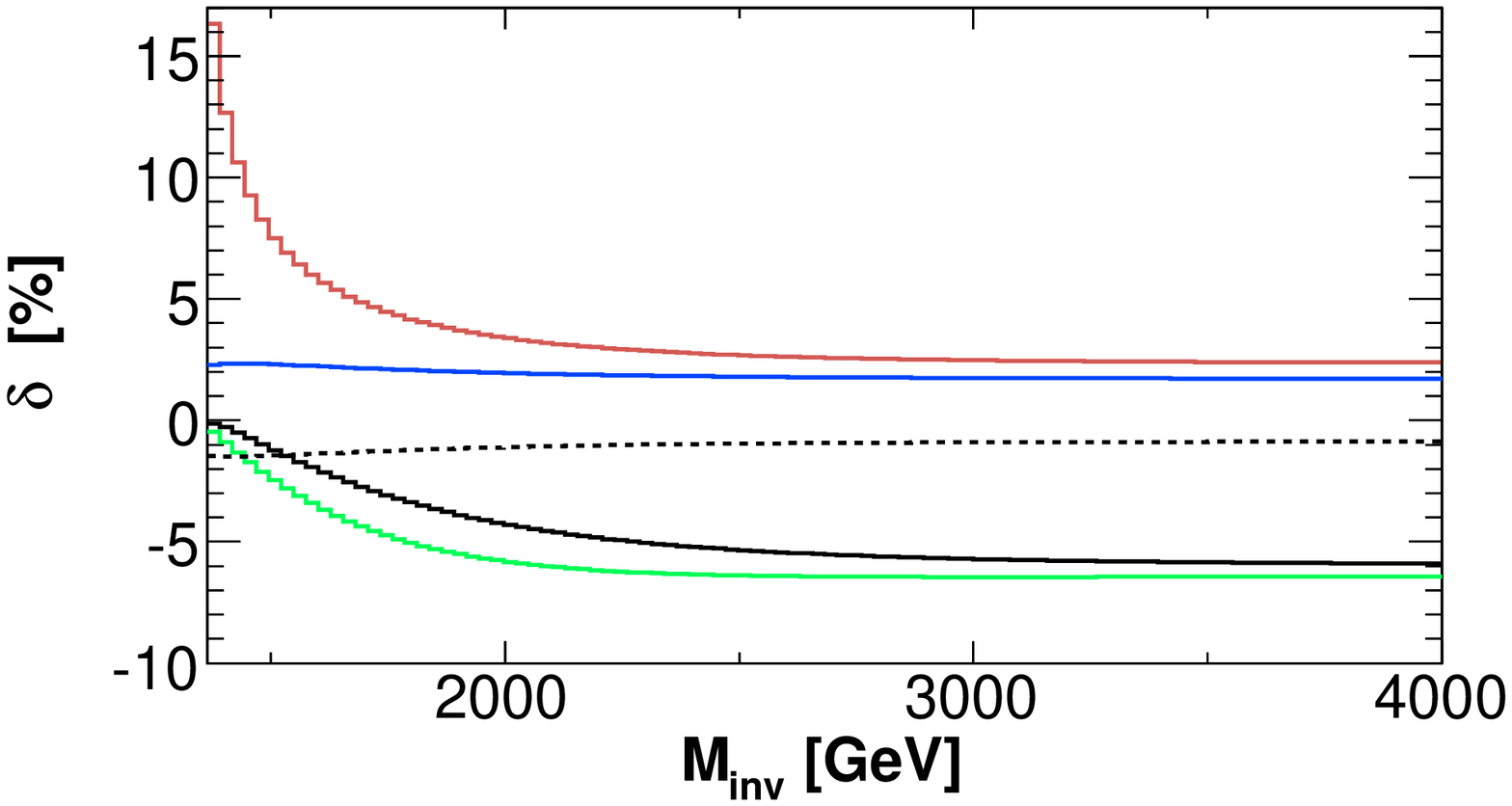, width=7.9cm}
\epsfig{file= 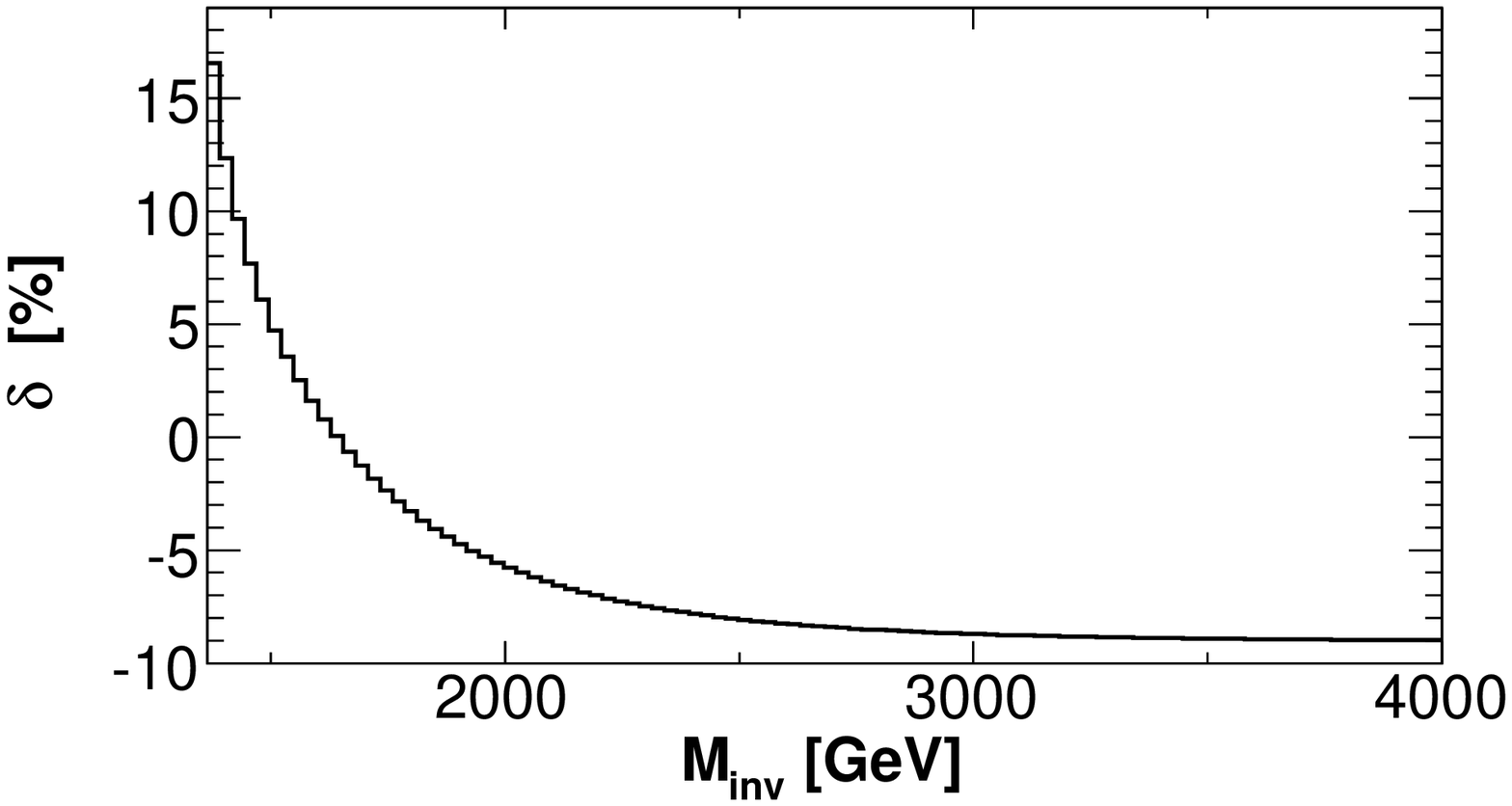, width=7.9cm}
\underline{SU1}\\
\epsfig{file= 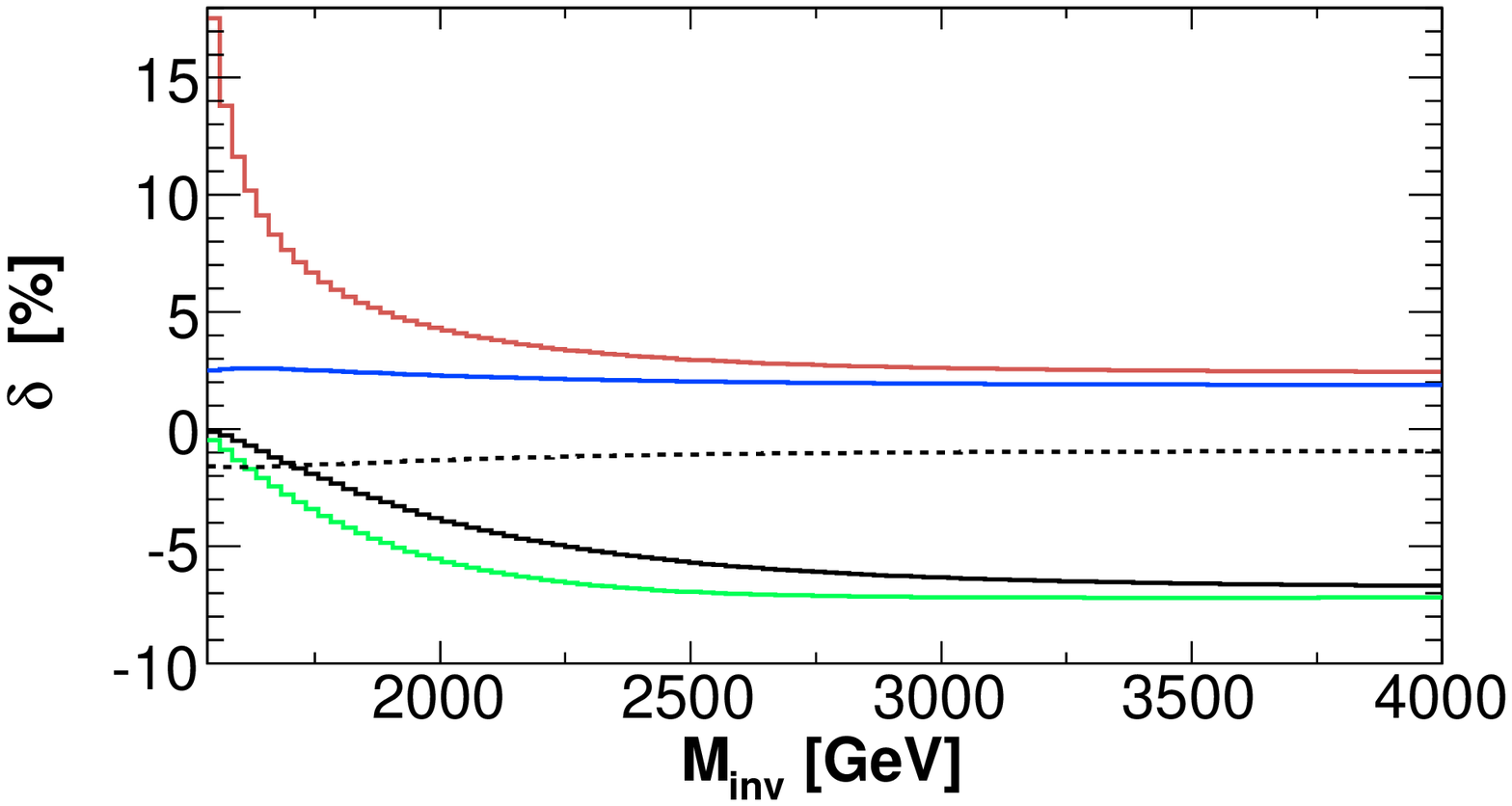, width=7.9cm}
\epsfig{file= 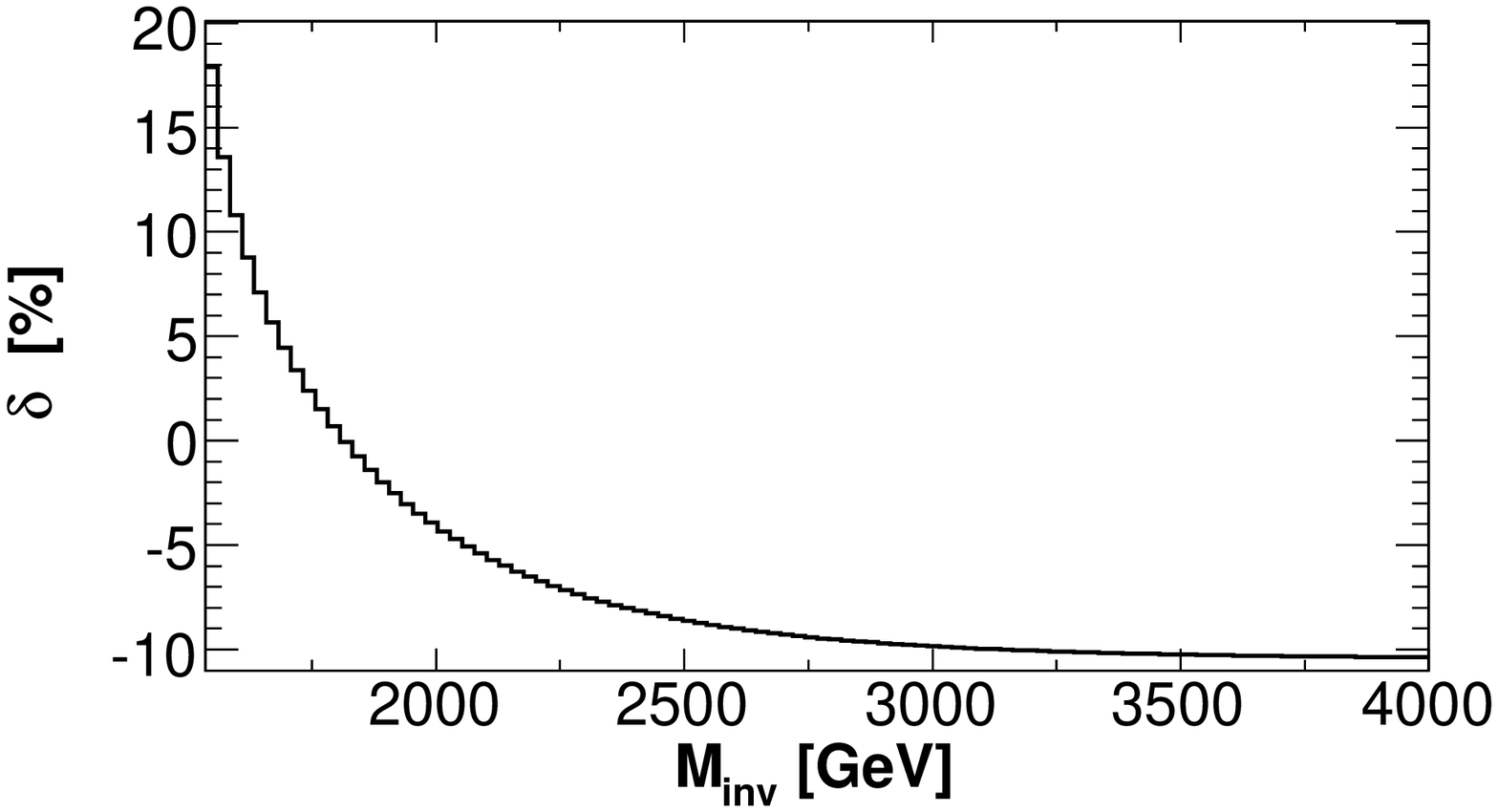, width=7.9cm}
\underline{SU4}\\
\epsfig{file= 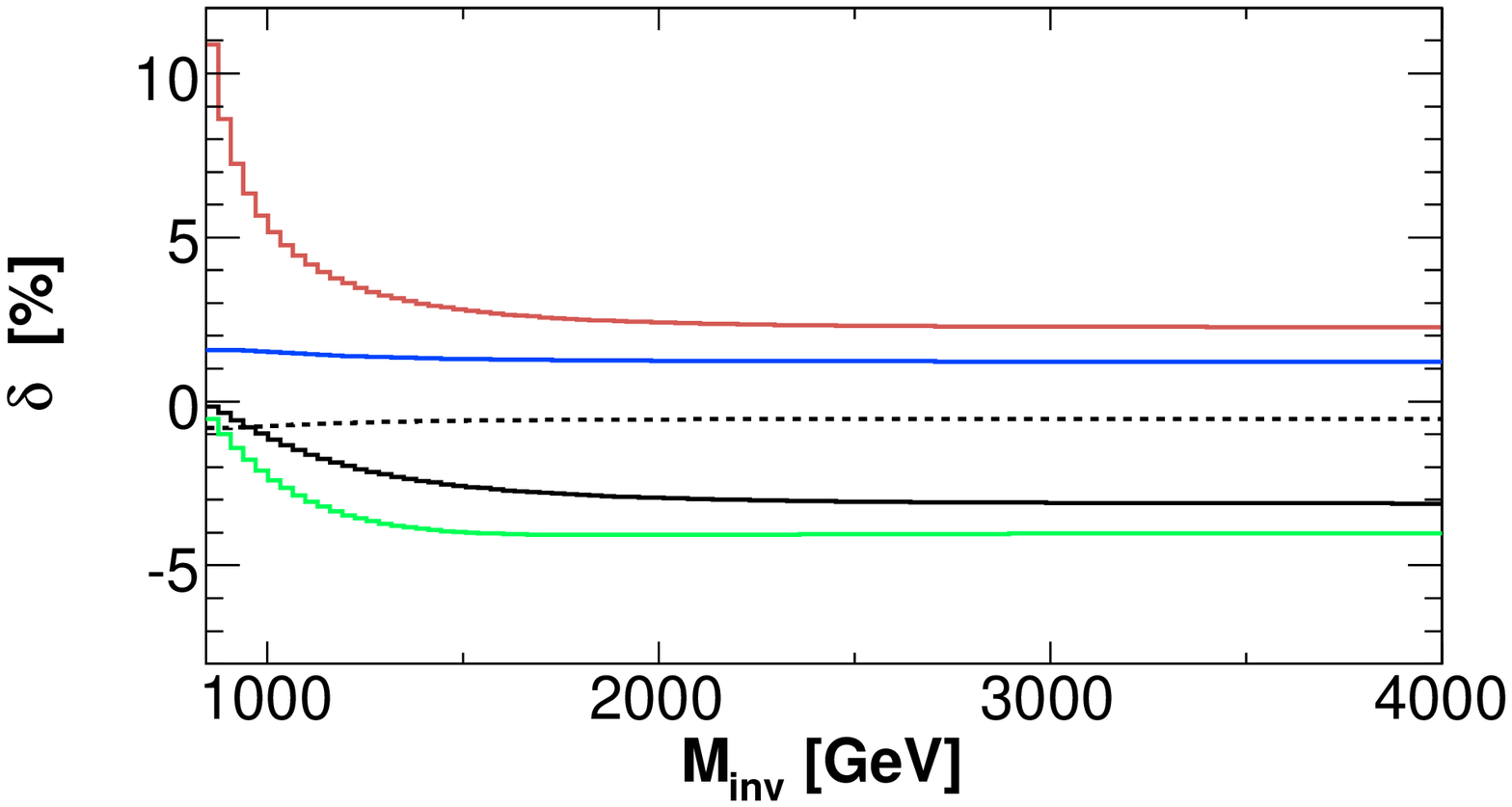, width=7.9cm}
\epsfig{file= 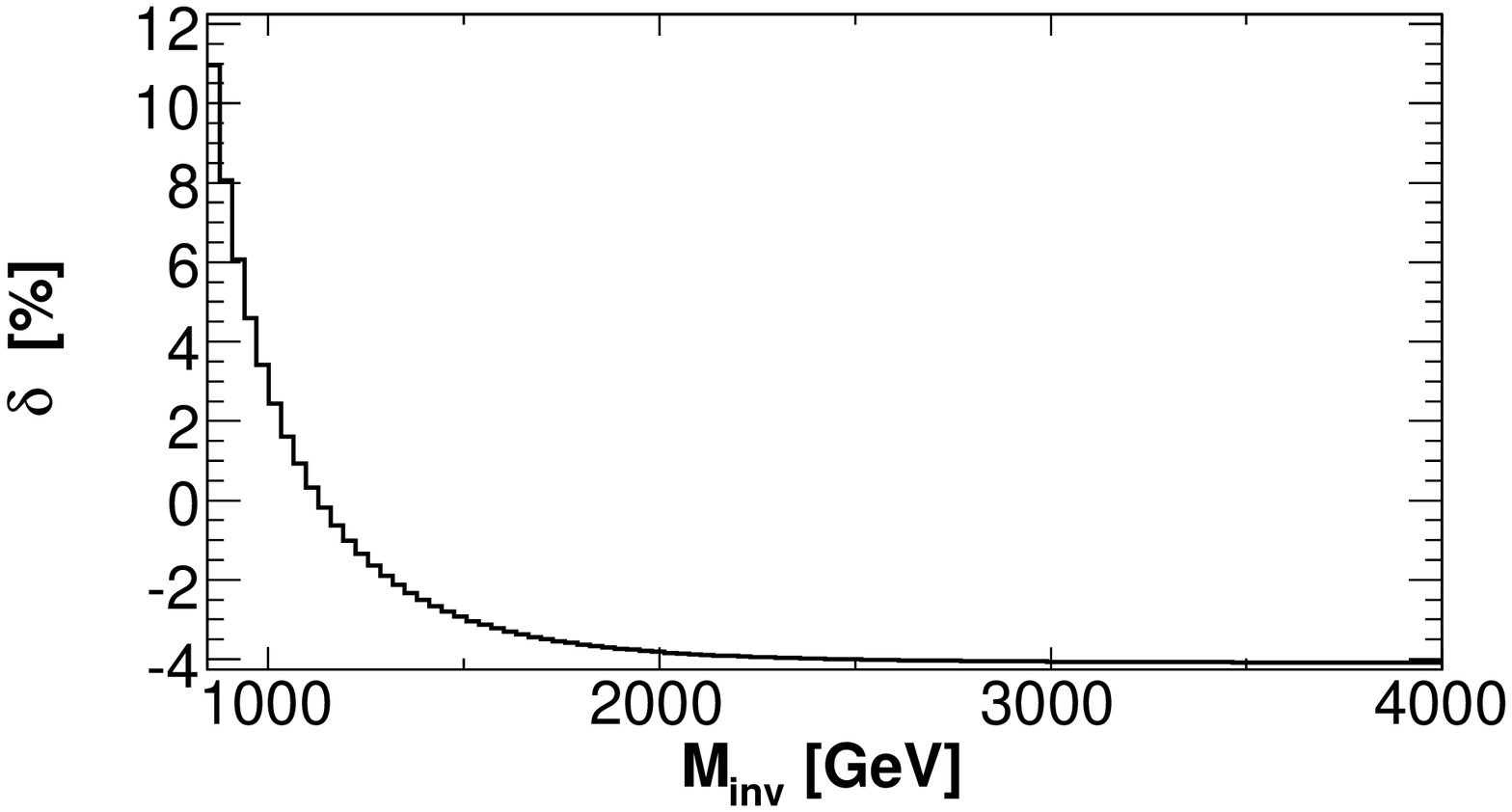, width=7.9cm}
\caption{Cumulative invariant mass distribution for $PP \to \tilde{u}^L\tilde{u}^{L*}X$ 
in different SUSY scenarios.
Notations as in Fig.~\ref{Fig:IC_SQ}.  }
\label{Fig:IC}
\end{figure}

\newpage

\begin{figure}
\centering
\underline{SPS1a$'$}\\
\epsfig{file= PLOT/UP2/SPS1a_IM_Ch.eps, width= 7.9cm}
\epsfig{file= PLOT/UP2/SPS1a_IM_Per.eps, width=7.9cm}
\underline{SPS5}\\
\epsfig{file= 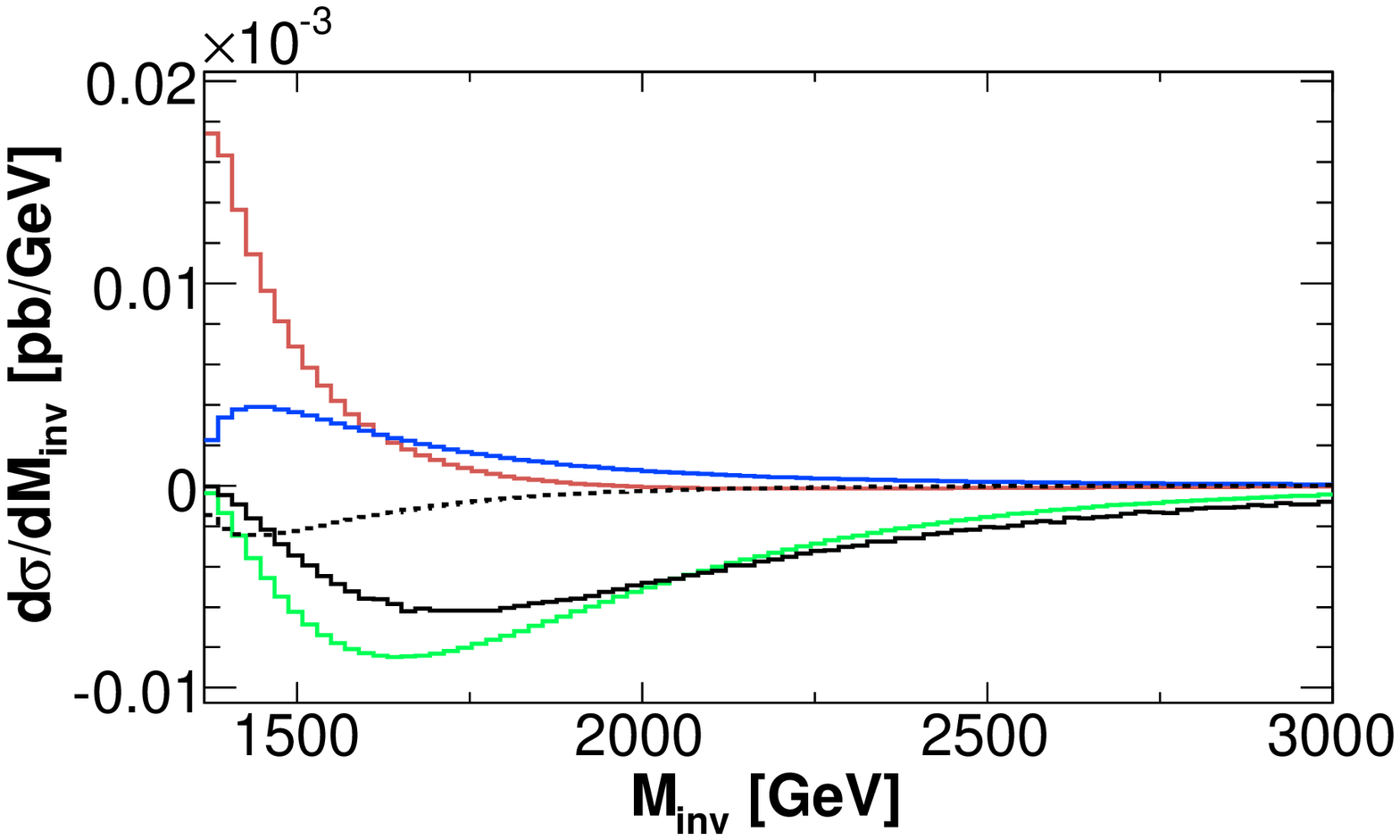, width= 7.9cm}
\epsfig{file= 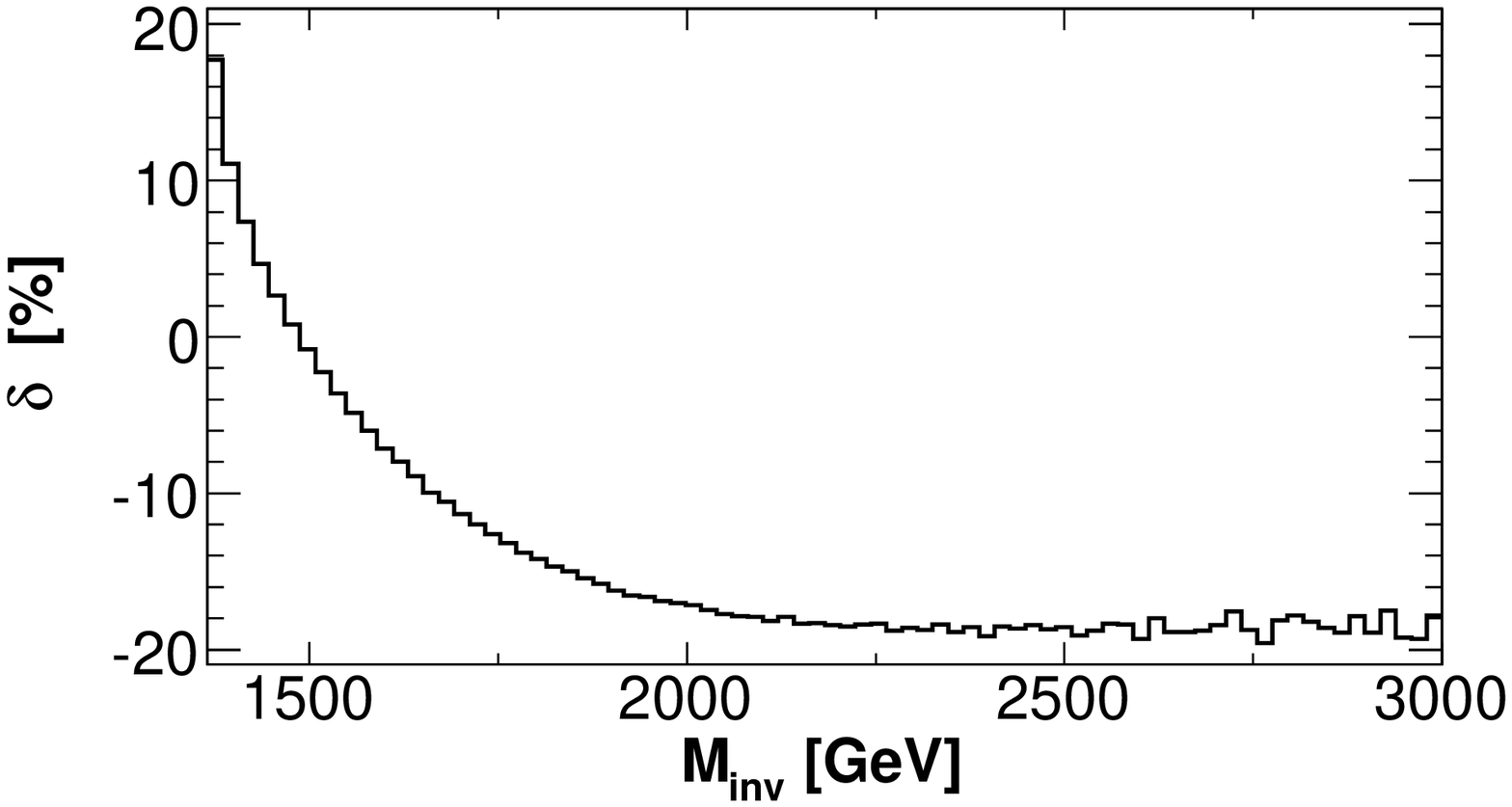, width=7.9cm}
\underline{SU1}\\
\epsfig{file=  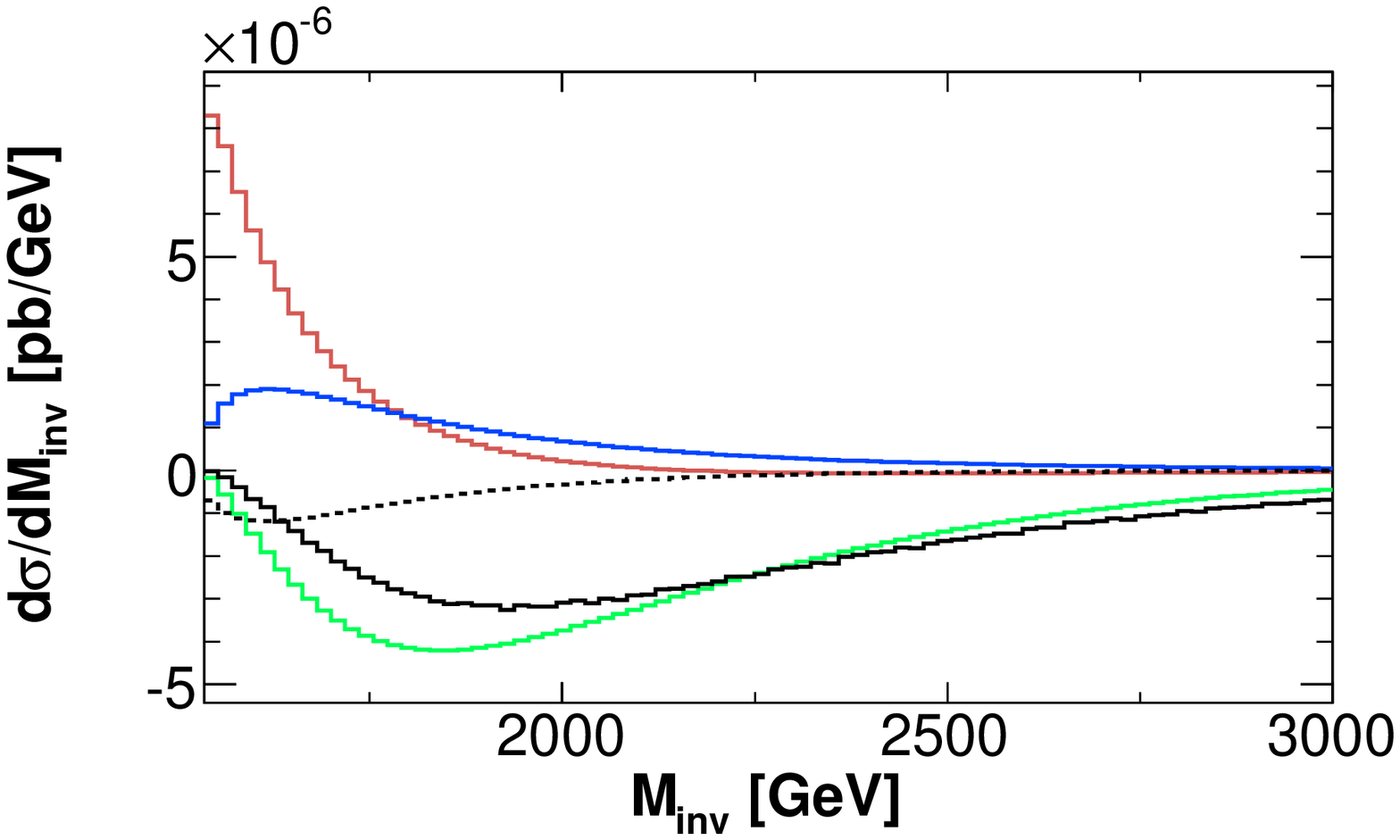, width= 7.9cm}
\epsfig{file= 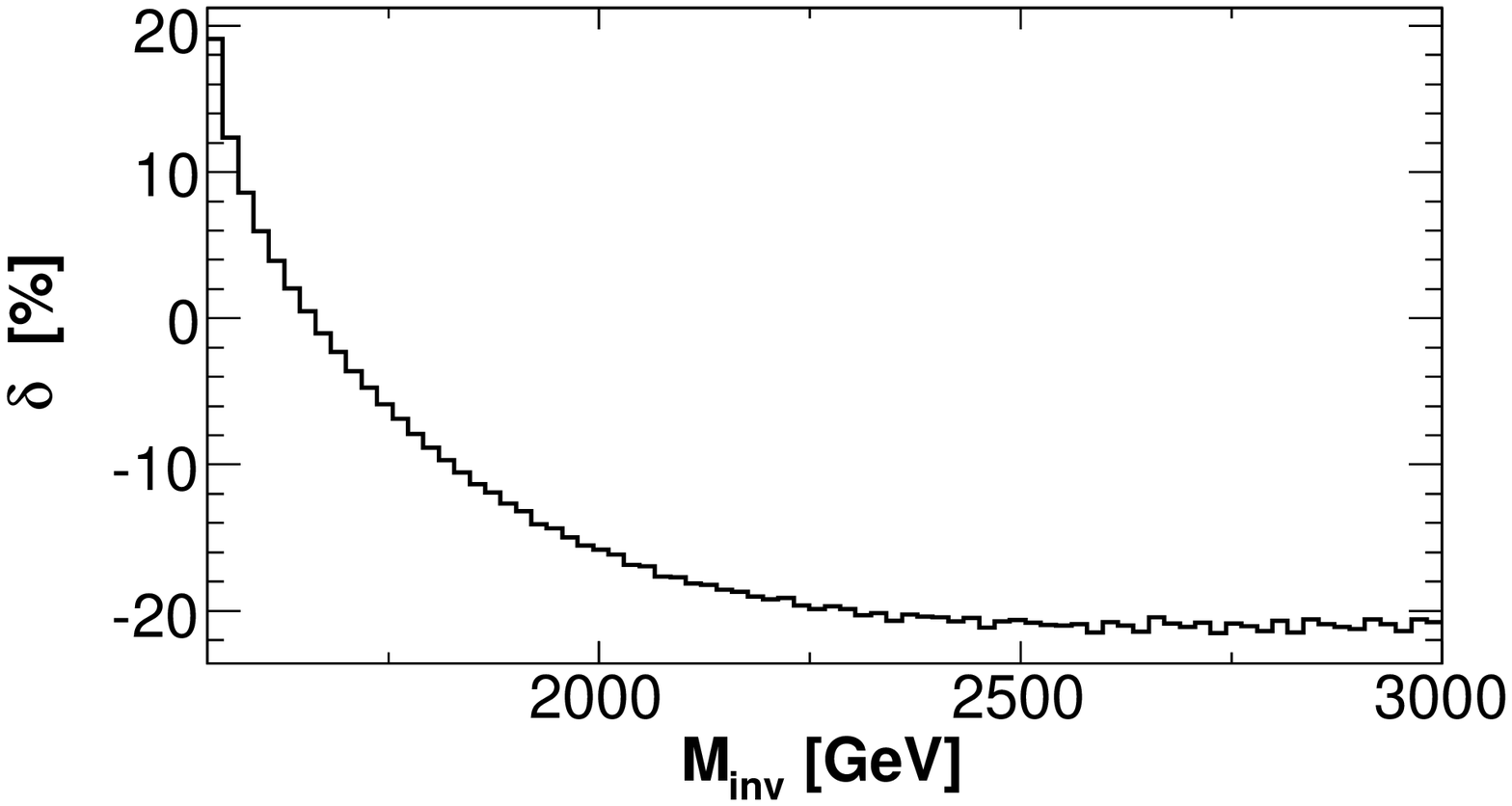, width=7.9cm}
\underline{SU4}\\
\epsfig{file= 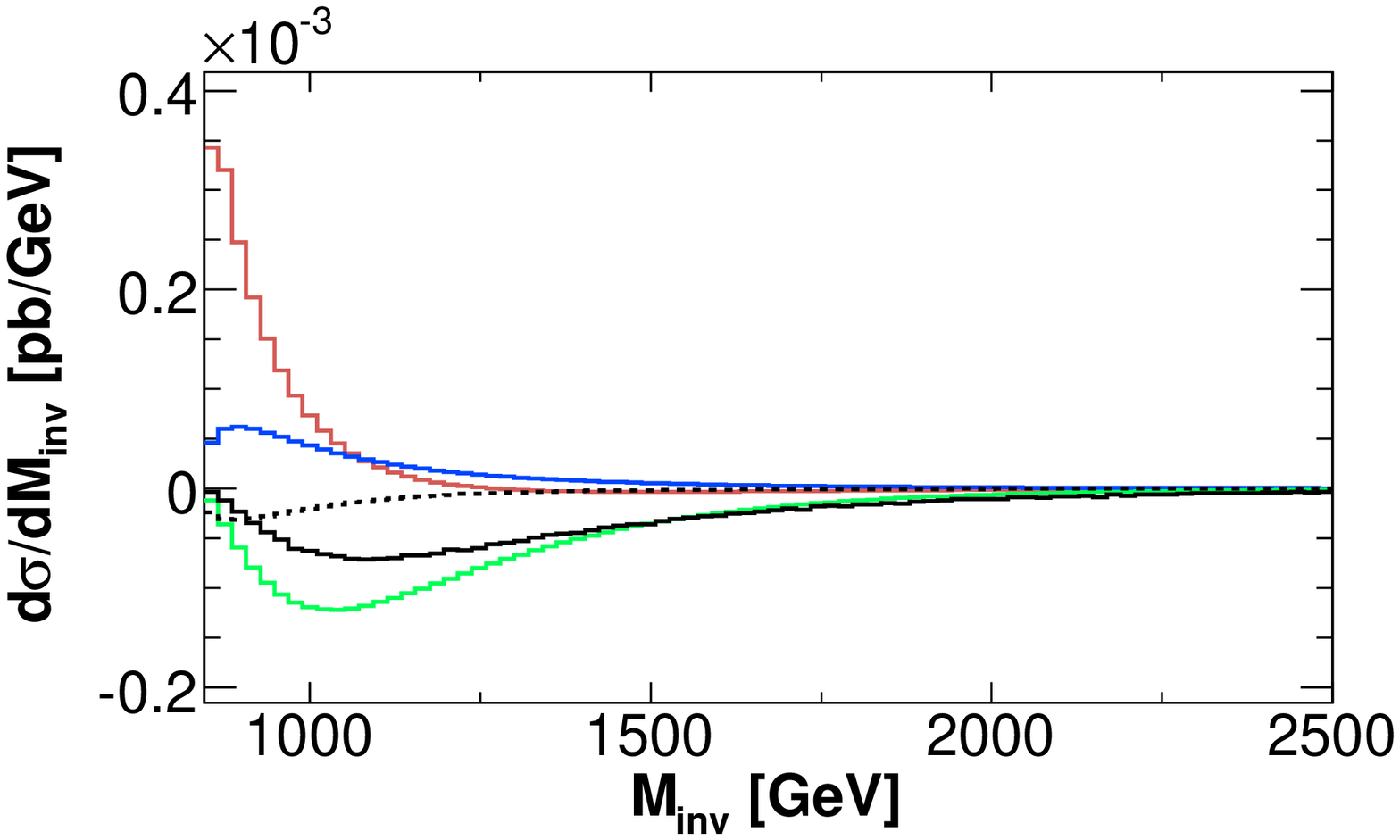, width= 7.9cm}
\epsfig{file= 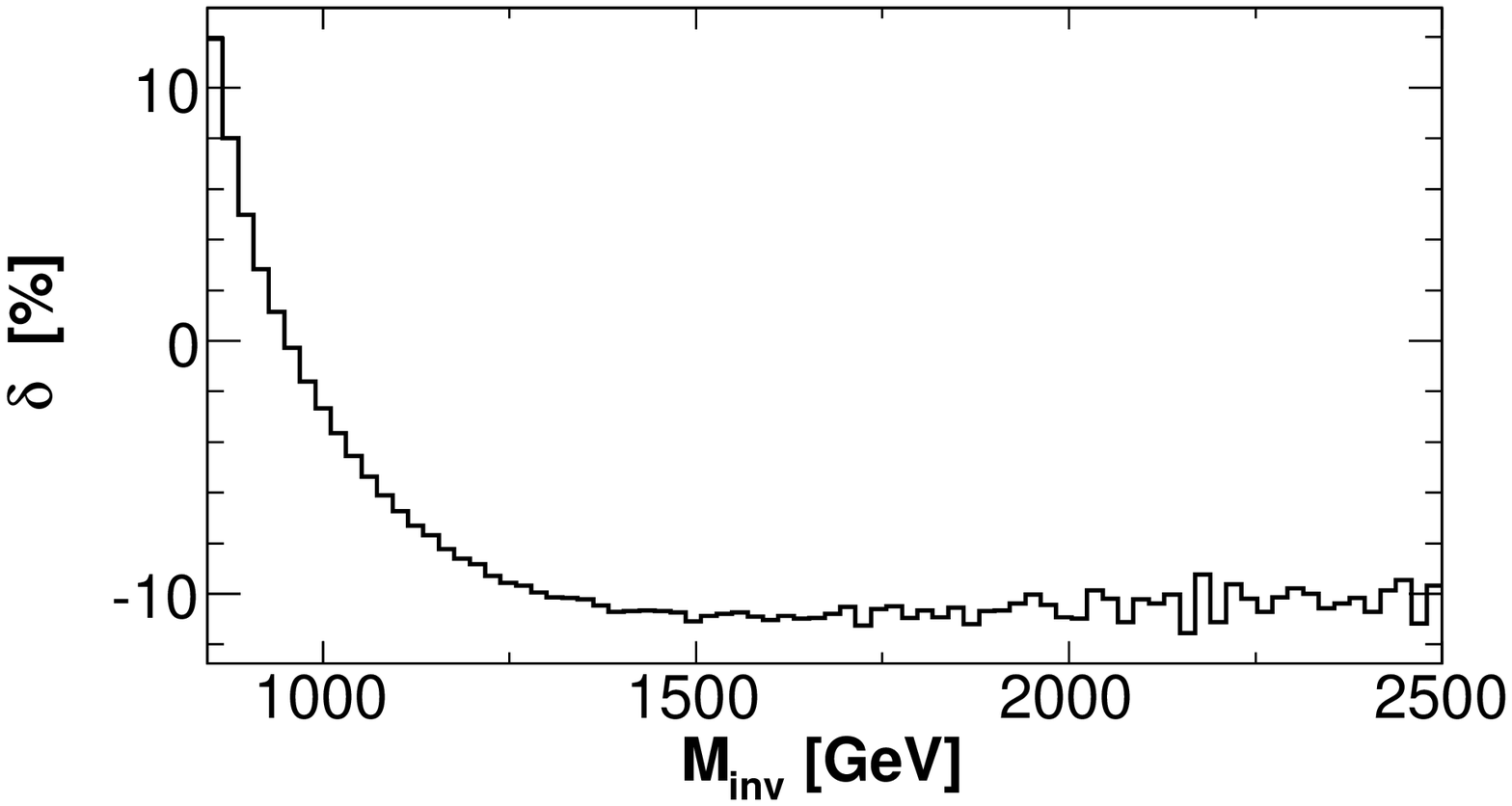, width=7.9cm}
\caption{
Invariant mass distribution for $PP \to \tilde{u}^L\tilde{u}^{L*}X$
in different SUSY scenarios.
Notations as in Fig.~\ref{Fig:IM_SQ}.  }
\label{Fig:IM}
\end{figure}

\newpage

\begin{figure}
\centering
\underline{SPS1a$'$}\\
\epsfig{file=PLOT/UP2/SPS1a_PT_Ch.eps, width= 7.9cm}
\epsfig{file= PLOT/UP2/SPS1a_PT_Per.eps, width=7.9cm}
\underline{SPS5}\\
\epsfig{file= 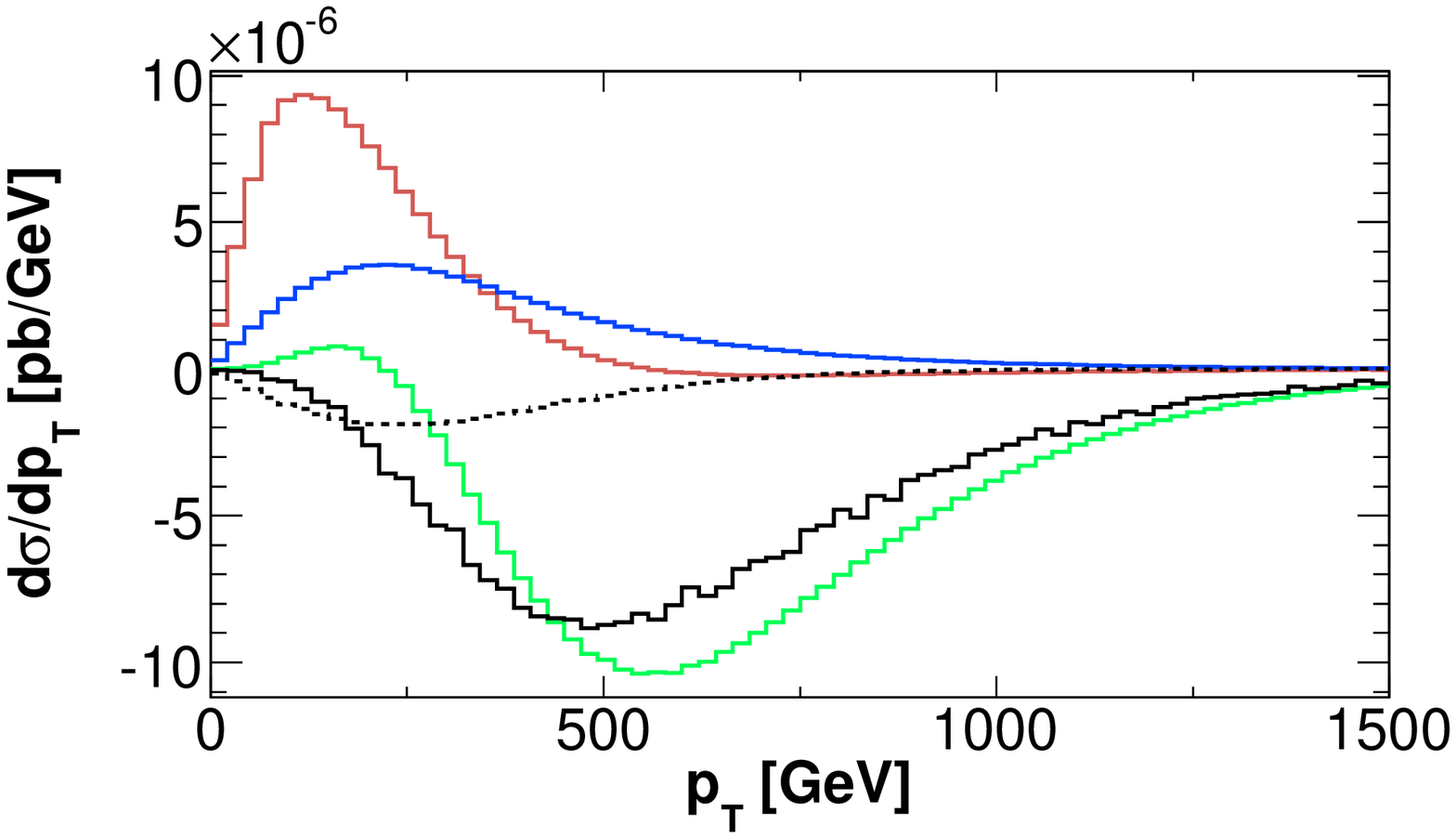, width= 7.9cm}
\epsfig{file= 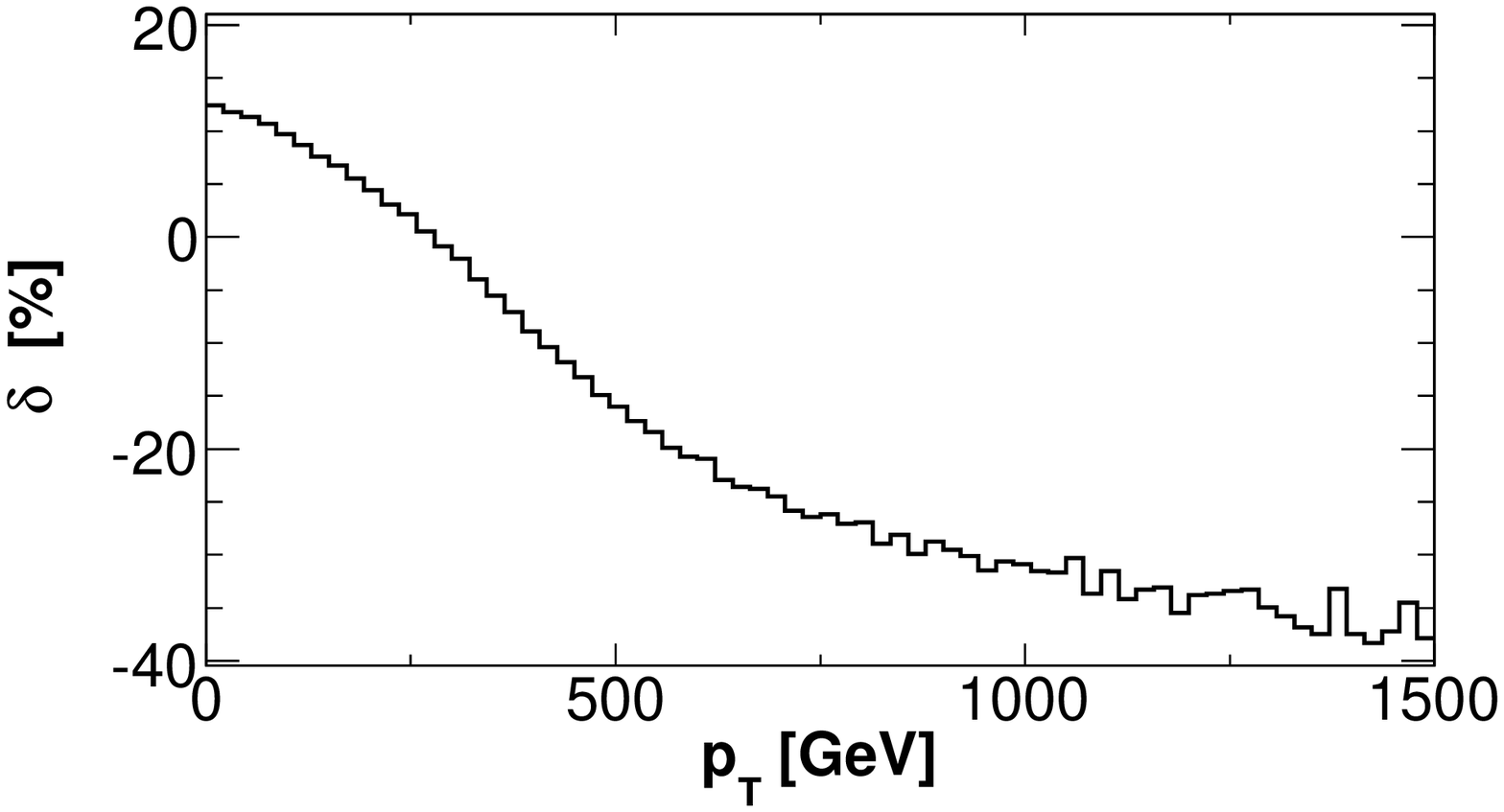, width=7.9cm}
\underline{SU1}\\
\epsfig{file= 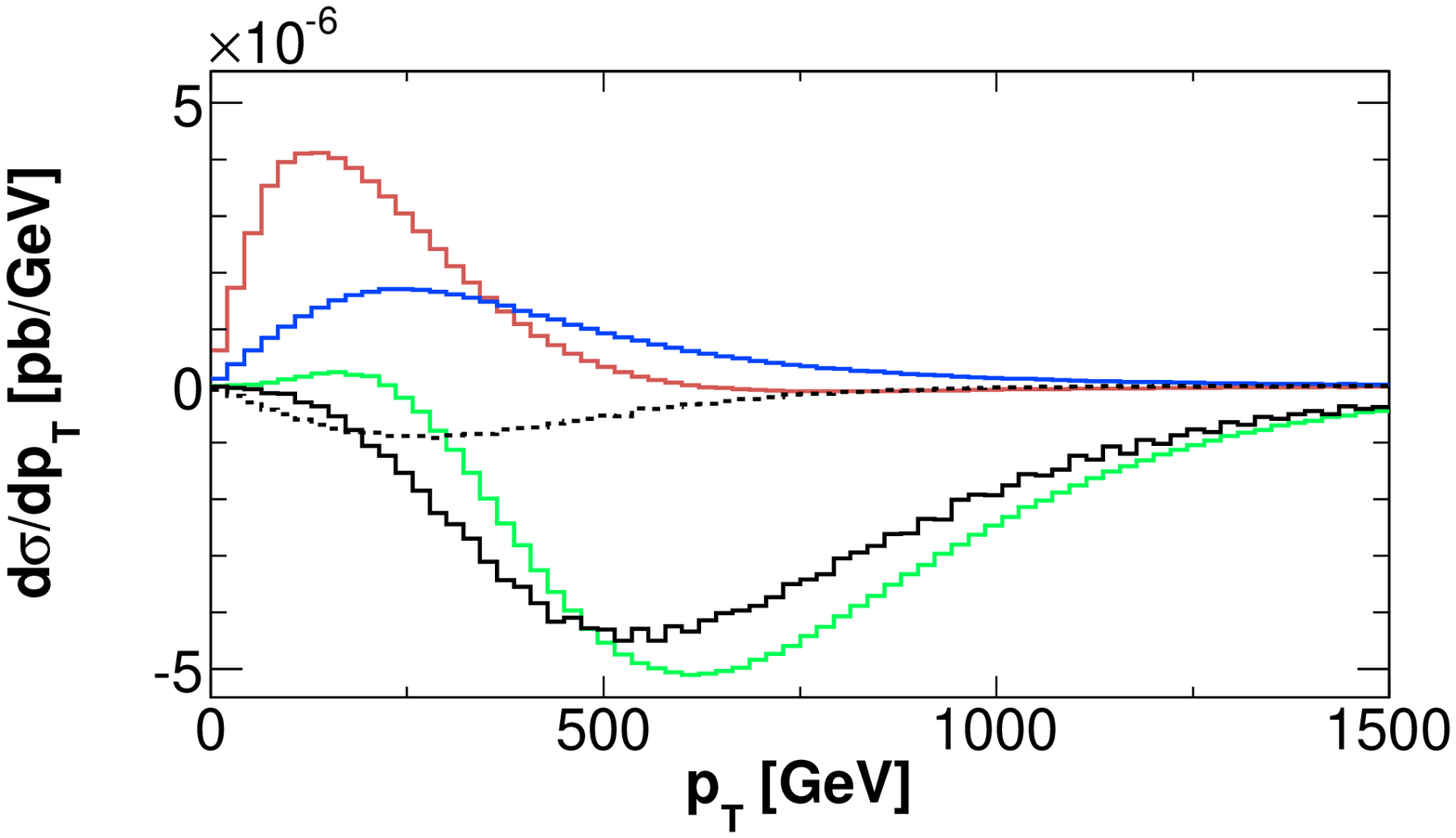, width= 7.9cm}
\epsfig{file=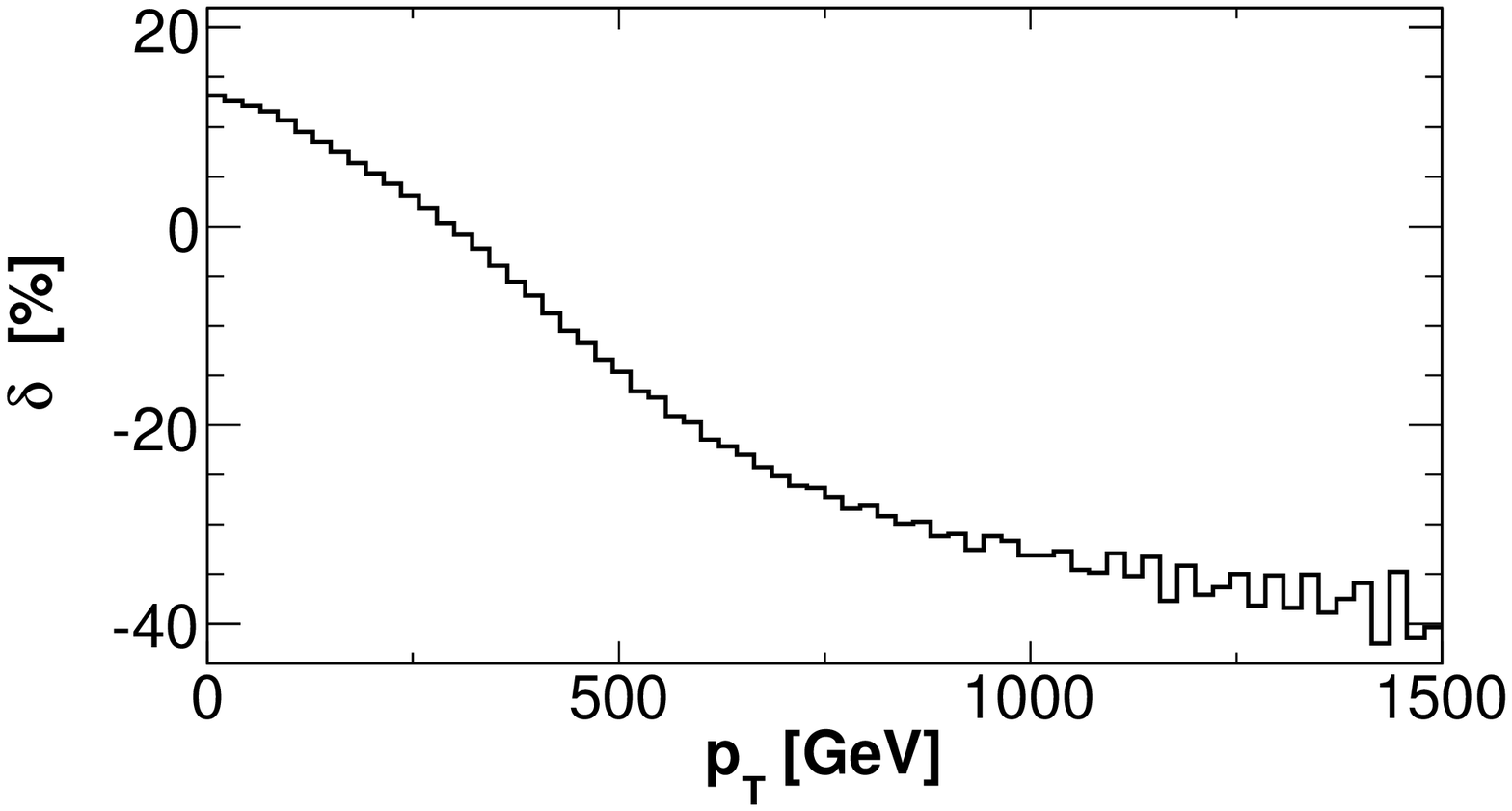, width=7.9cm}
\underline{SU4}\\
\epsfig{file= 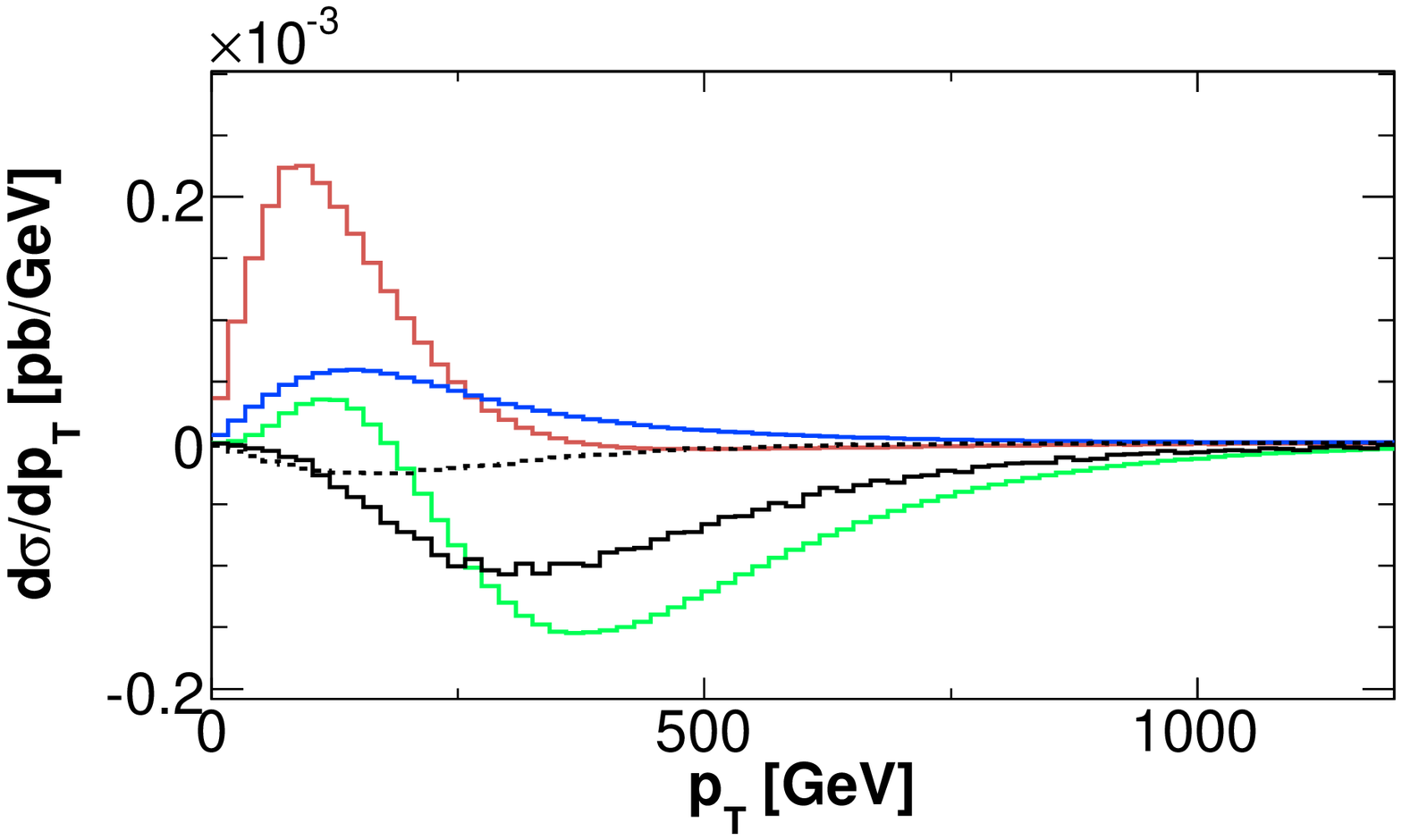, width= 7.9cm}
\epsfig{file= 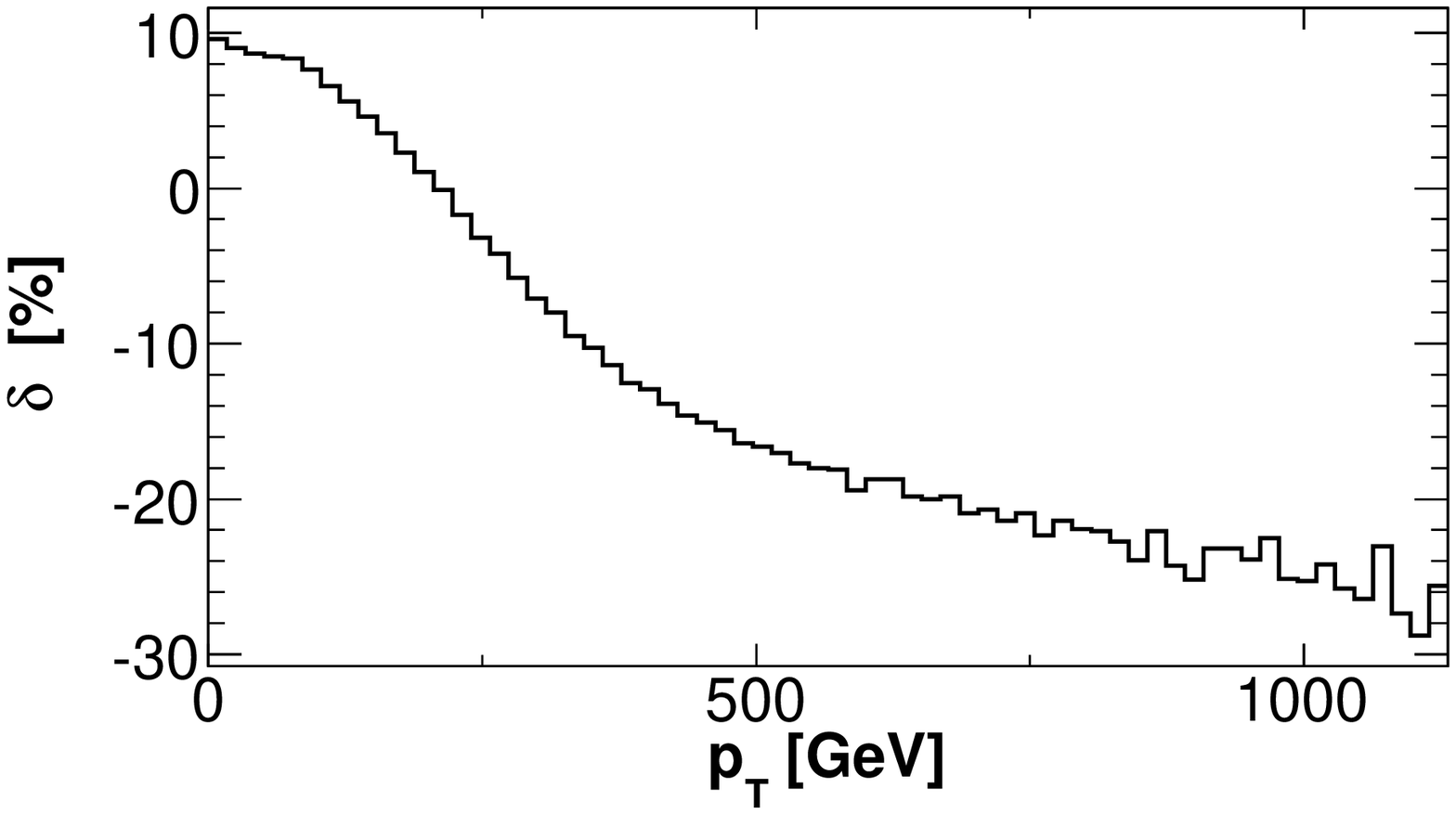, width=7.9cm}
\caption{Transverse momentum distribution of the process $PP \to \tilde{u}^L\tilde{u}^{L*}X$ 
in different SUSY scenarios. 
Notations as in Fig.~\ref{Fig:PT_SQ}.  }
\label{Fig:PT}
\end{figure}

%\newpage

%\begin{figure}
%\centering
%%
%\underline{SPS1a$'$}\\
%\epsfig{file= PLOT/UP2/SPS1a_YY_Ch.eps, width= 7.9cm}
%\epsfig{file= PLOT/UP2/SPS1a_YY_Lo.eps, width=7.9cm}
%
%\underline{SPS5}\\
%\epsfig{file=PLOT/UP2/SPS5_YY_Ch.eps, width= 7.9cm}
%\epsfig{file= PLOT/UP2/SPS5_YY_Lo.eps, width=7.9cm}
%
%\underline{SU1}\\
%\epsfig{file= PLOT/UP2/SU1_YY_Ch.eps, width= 7.9cm}
%\epsfig{file= PLOT/UP2/SU1_YY_Lo.eps, width=7.9cm}
%
%\underline{SU4}\\
%\epsfig{file=PLOT/UP2/SU4_YY_Ch.eps, width= 7.9cm}
%\epsfig{file= PLOT/UP2/SU4_YY_Lo.eps, width=7.9cm}
%
%\caption{Rapidity distribution of the process $PP \to \tilde{u}^L\tilde{u}^{L*}X$ 
%in different SUSY scenarios. 
%Notations as in Fig.~\ref{Fig:YY_SQ}.  }
%\label{Fig:YY}
%\end{figure}

\newpage

\begin{figure}
\centering
\underline{SPS1a$'$}\\
\epsfig{file= 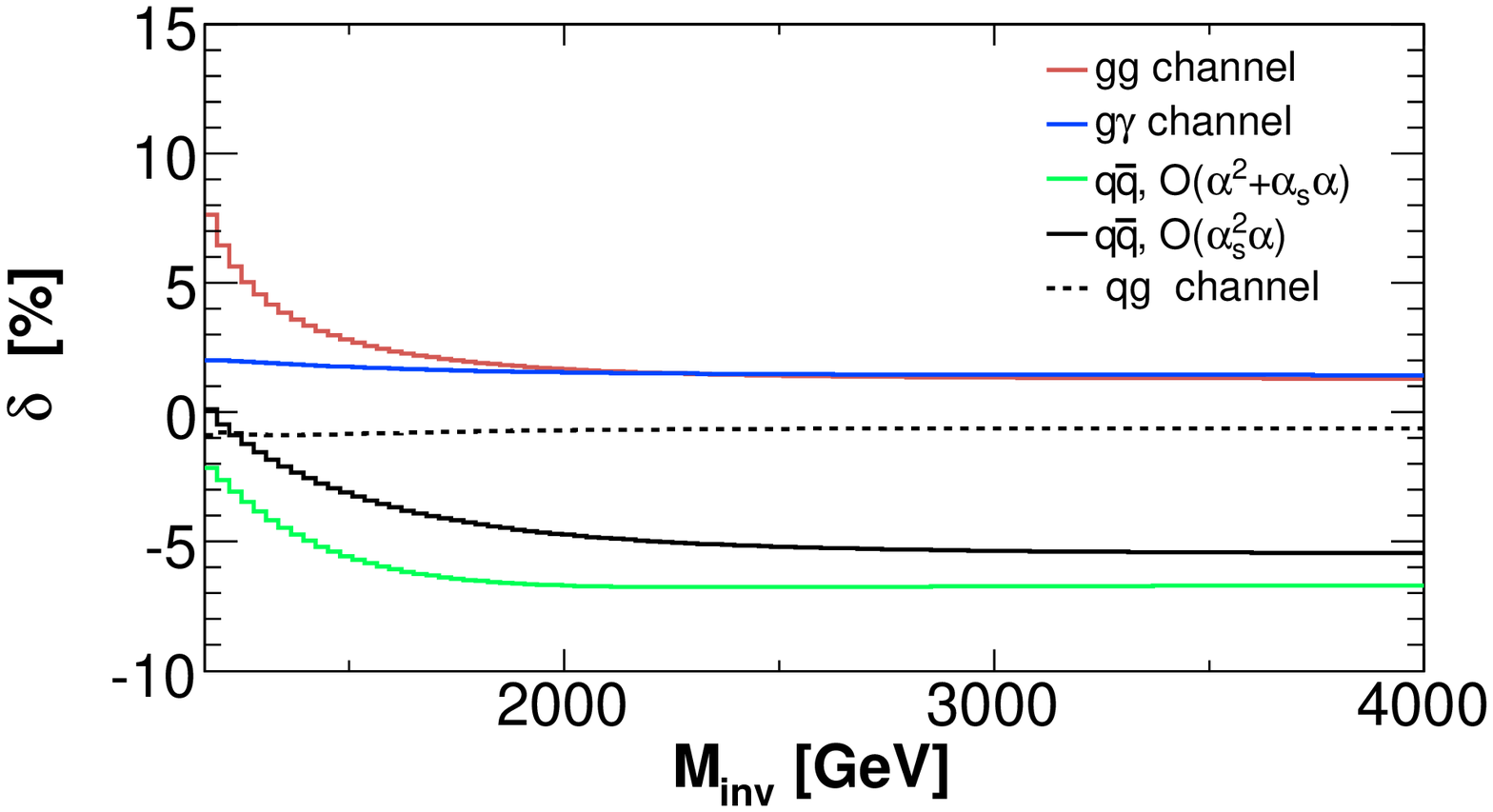, width=7.9cm}
\epsfig{file= 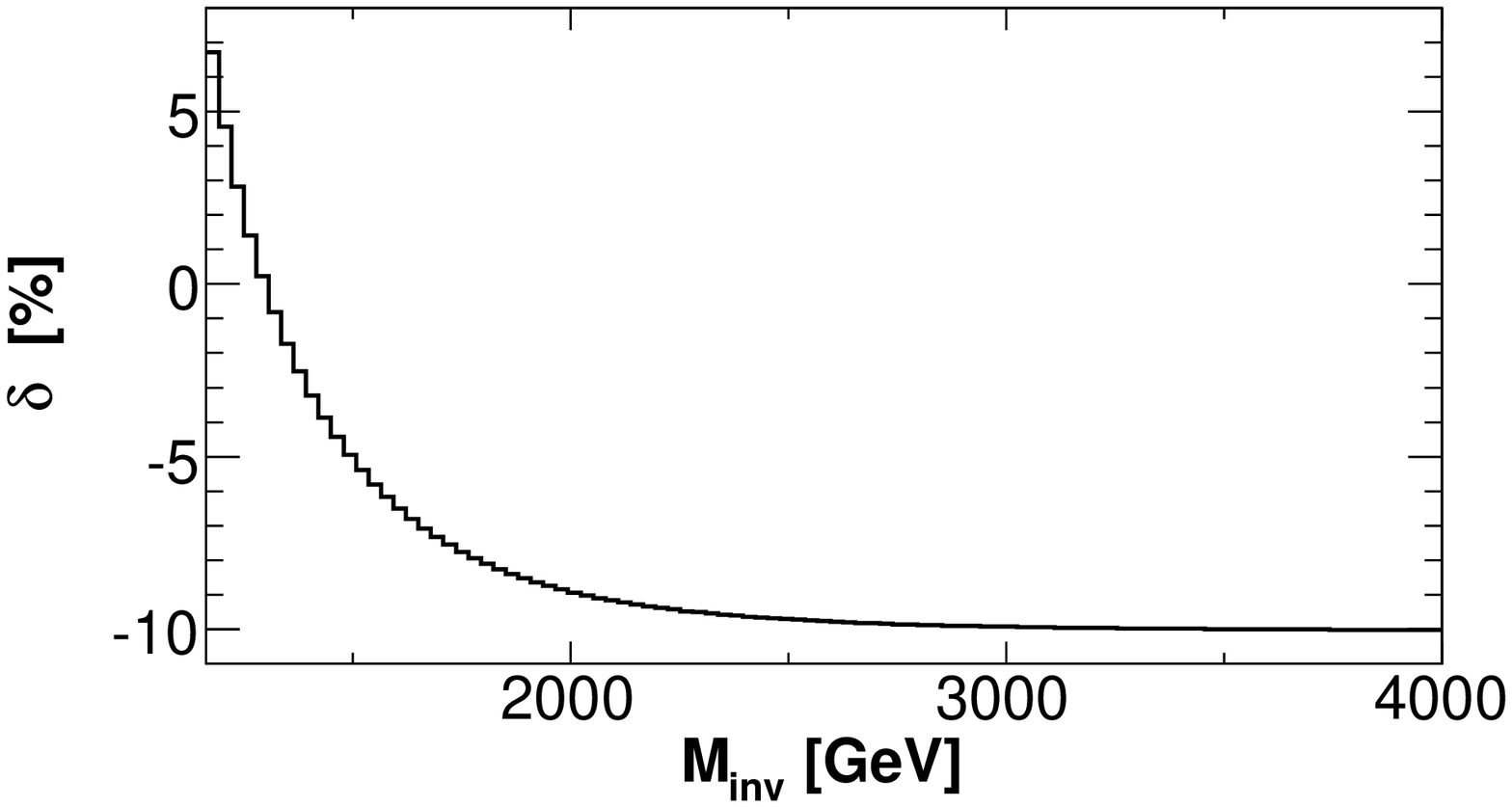, width=7.9cm}
\underline{SPS5}\\
\epsfig{file= 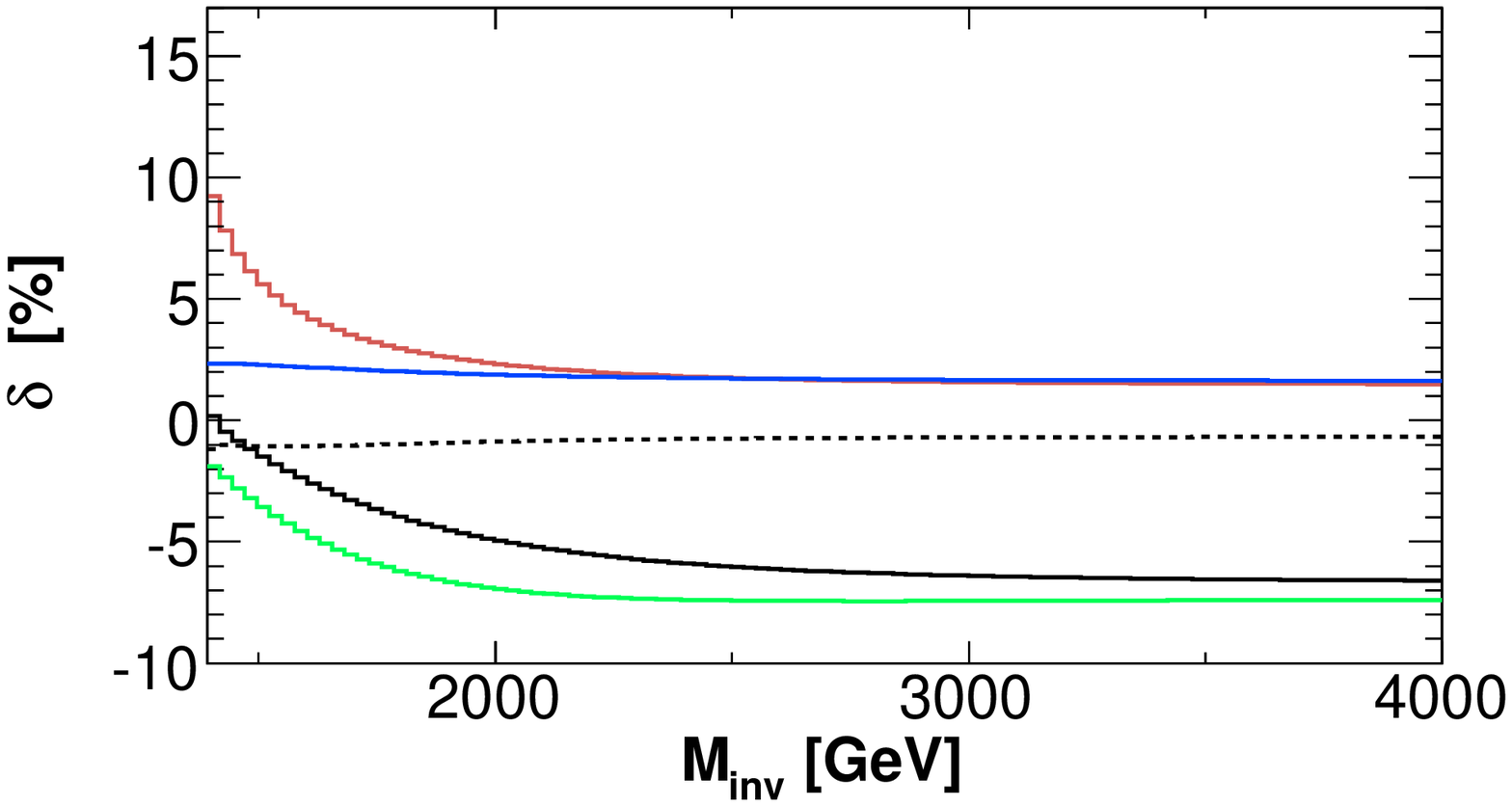, width=7.9cm}
\epsfig{file= 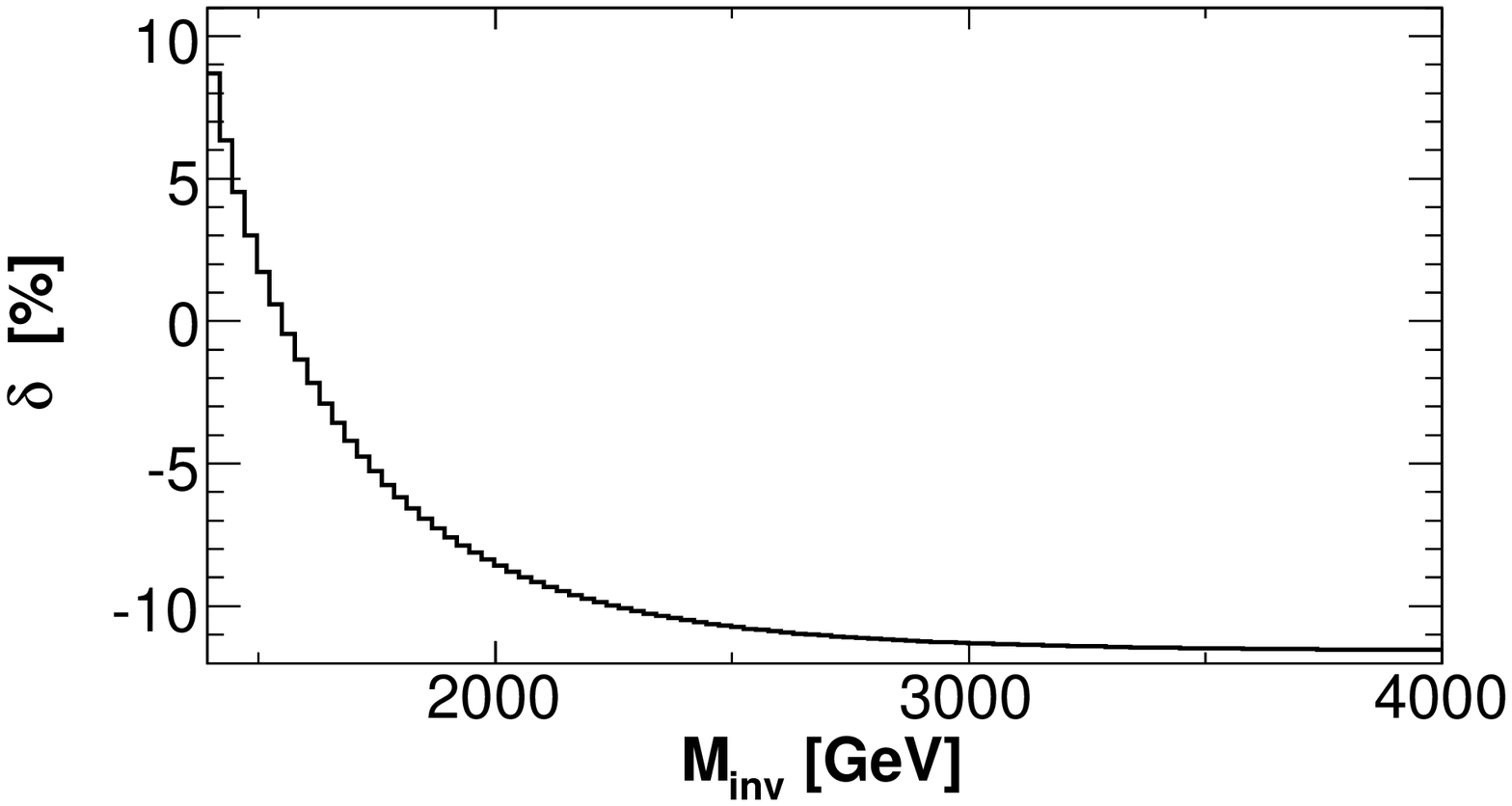, width=7.9cm}
\underline{SU1}\\
\epsfig{file= 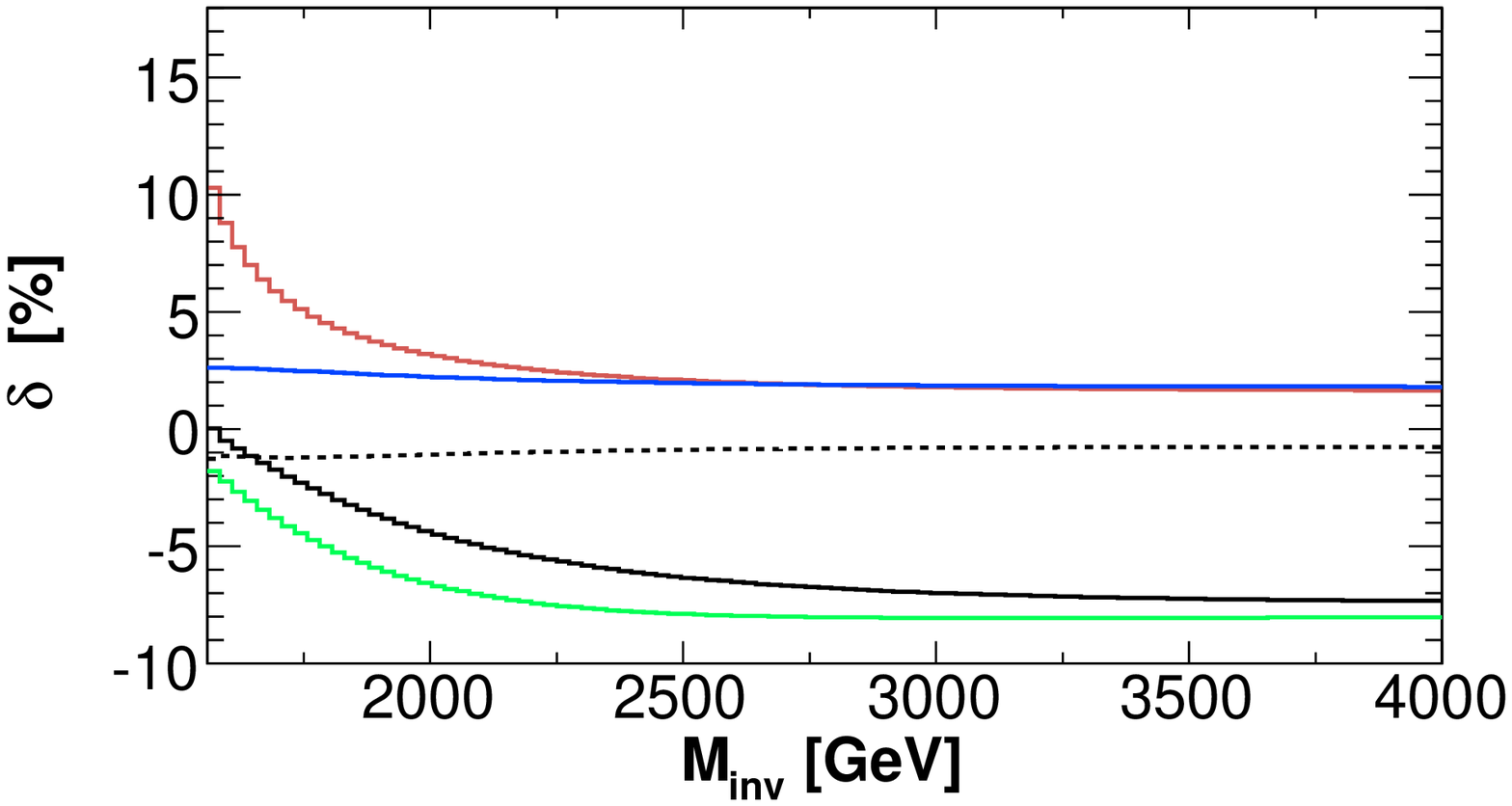, width=7.9cm}
\epsfig{file= 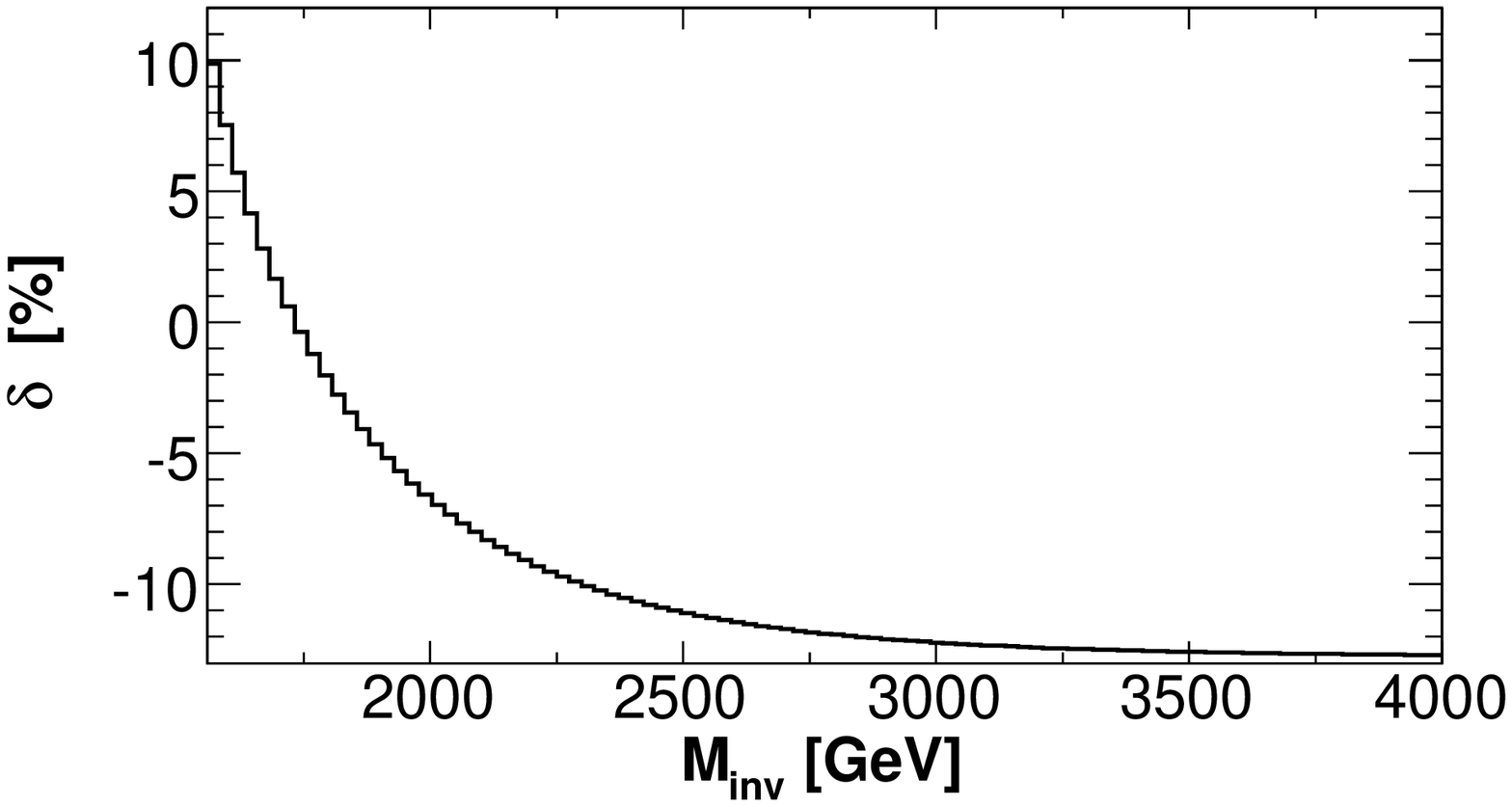, width=7.9cm}
\underline{SU4}\\
\epsfig{file= 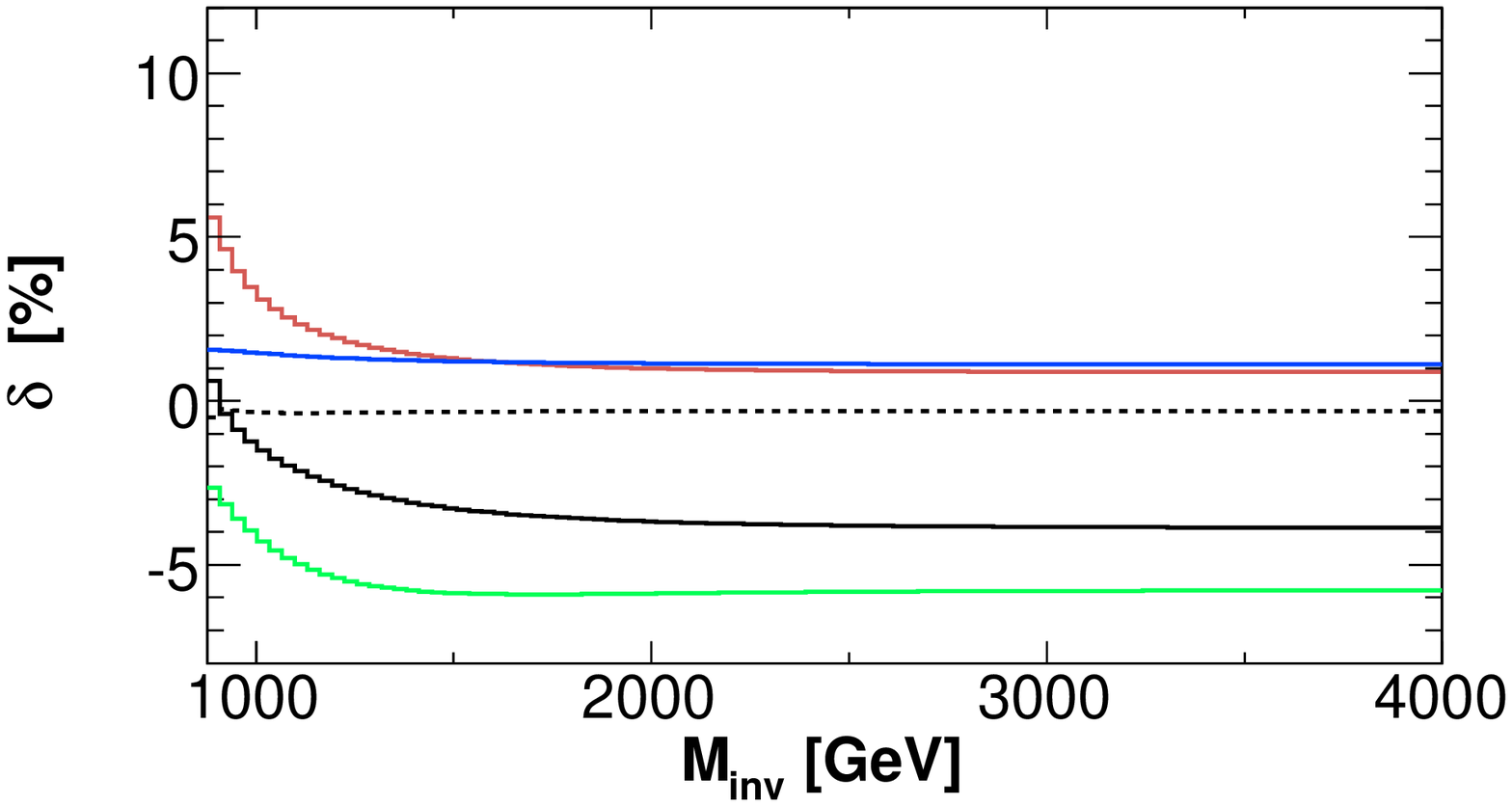, width=7.9cm}
\epsfig{file= 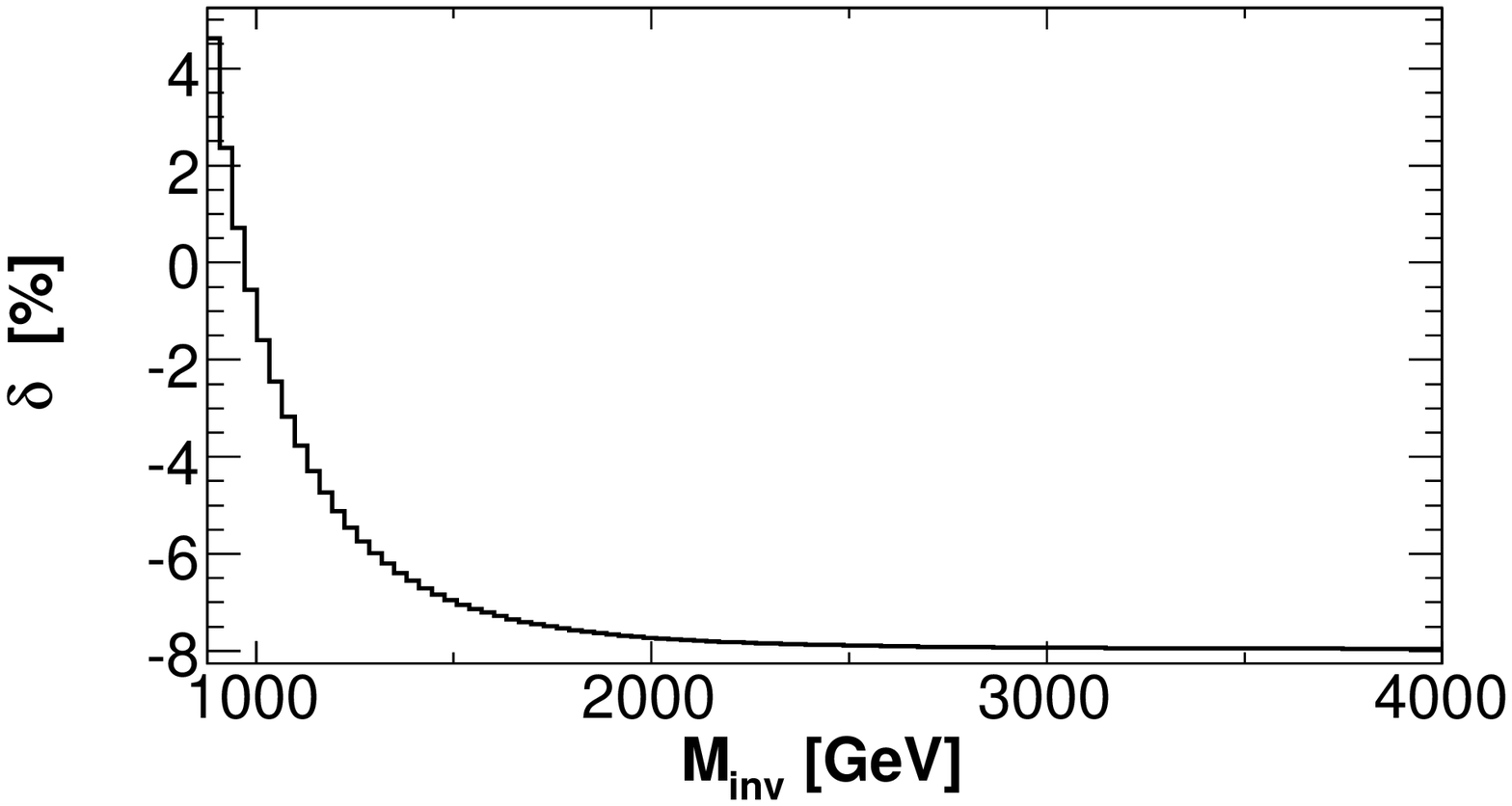, width=7.9cm}
\caption{Same as Fig.~\ref{Fig:IC}, but with the kinematical cuts 
defined in section~\ref{SSec:Disc2}. }
\label{Fig:ICcut}
\end{figure}

\newpage

\begin{figure}
\centering
\hspace{-8.2cm}
\epsfig{file= 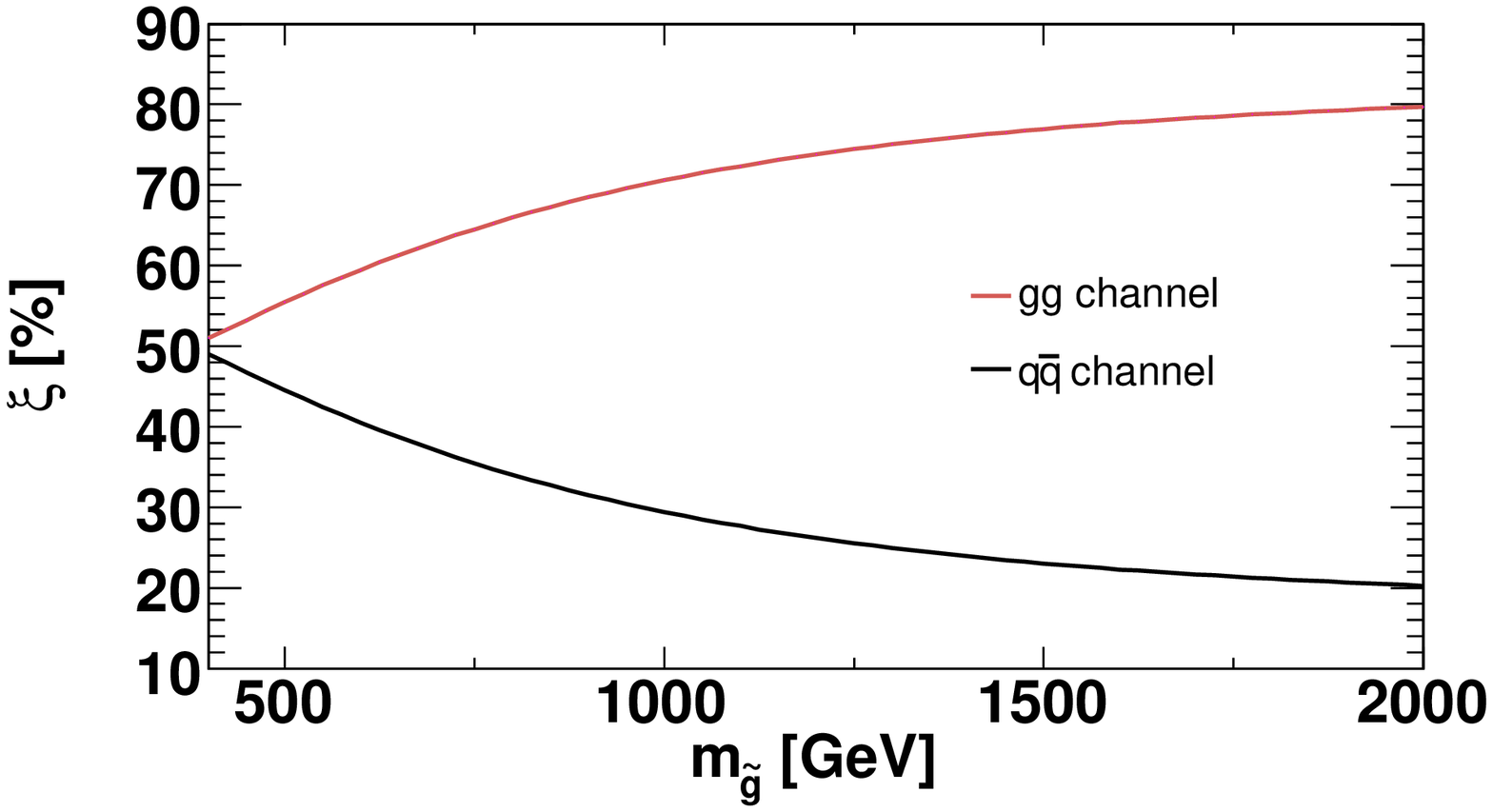, width=7.9cm}\\
\epsfig{file= 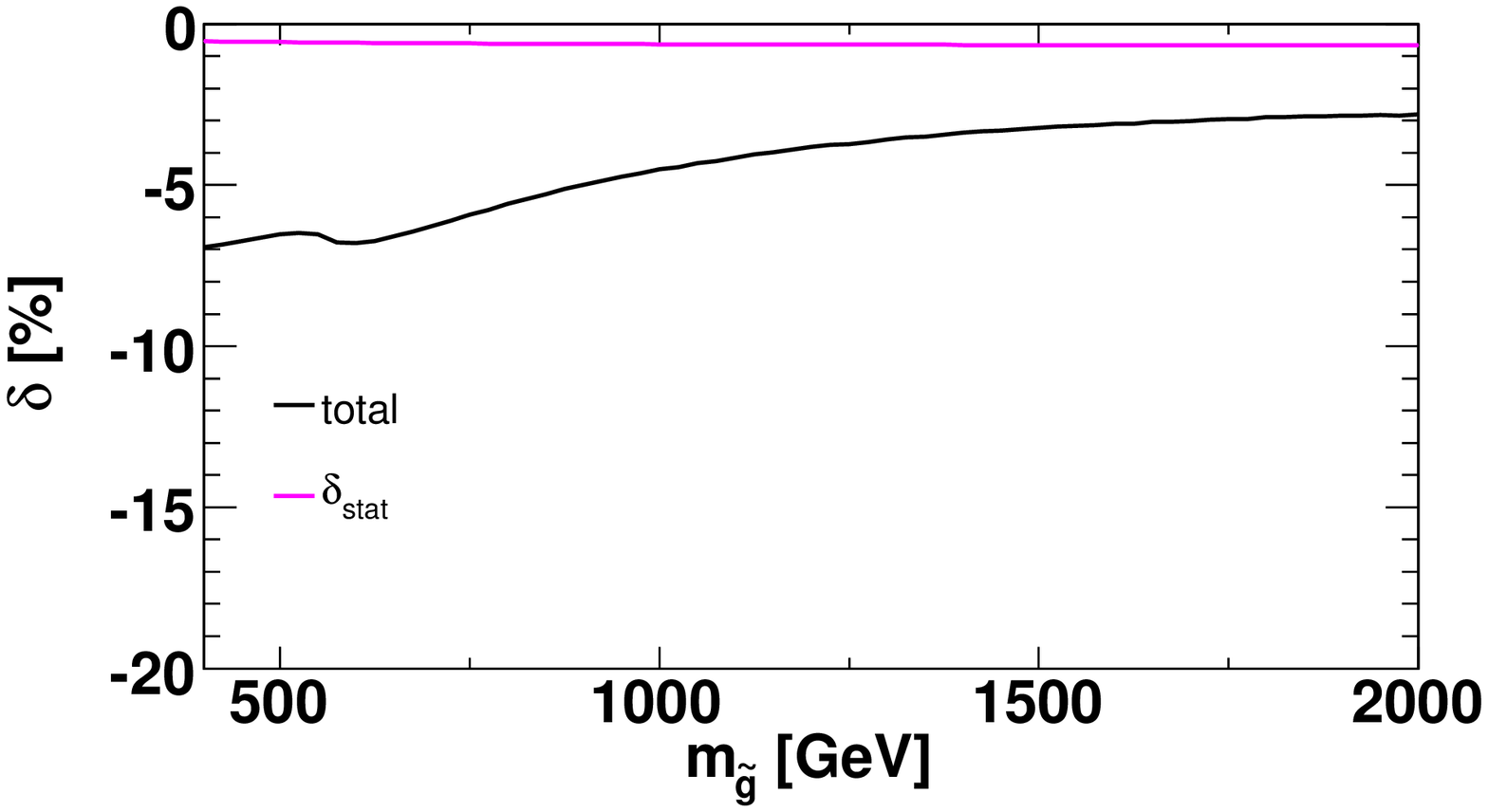, width= 7.9cm}
\epsfig{file=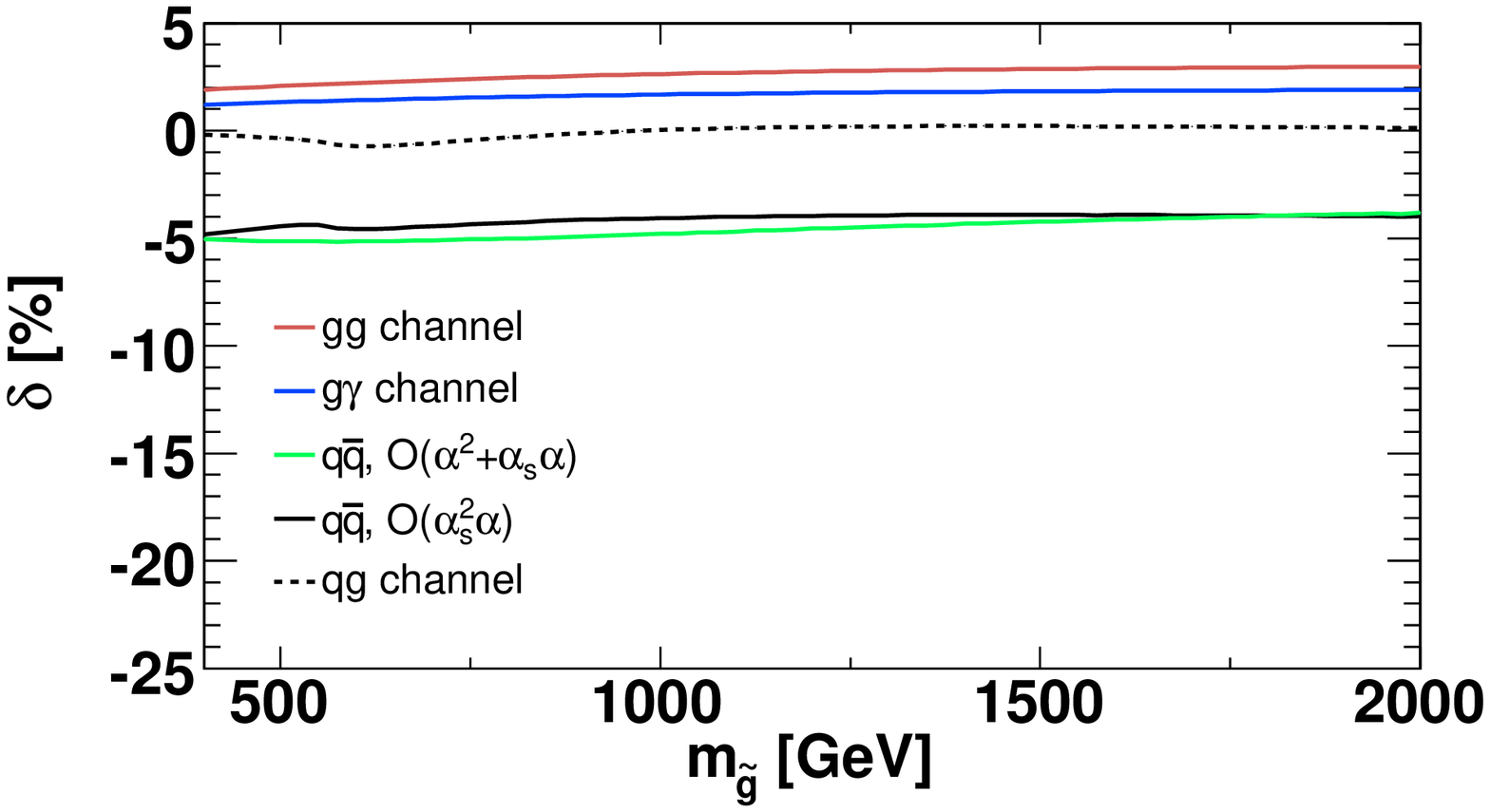, width=7.9cm}
\caption{Gluino mass dependence of the total (lower left) and of the individual (lower right)
EW contributions to the total cross section for $PP \to \tilde{u}^L\tilde{u}^{L*}X$. 
The upper panel
shows the relative yield of the two channels that contribute at LO.}
\label{Fig:Scan3}
\end{figure}

\end{document}